\documentclass[12pt,sans,showpacs,showkeys]{revtex4-2}
\usepackage{physics}
\usepackage{graphicx}
\usepackage{dcolumn}
\usepackage{bm}
\usepackage[colorlinks,citecolor=blue,urlcolor=blue,linkcolor=blue]{hyperref}
\usepackage[mathlines]{lineno}
\usepackage{amsmath,amsfonts,amsthm,amscd,amssymb,latexsym,mathrsfs}

\begin{document}

\title{Photon-Gravity Coupling in Schwarzschild Spacetime}

\author{Masoud Molaei}
 \email{masoud.molaei@sharif.ir}
\affiliation{
 Department of Physics, Sharif University of Technology, Tehran 11365-9161, Iran.}%




\date{\today}

\begin{abstract}
A canonical formalism for quantum electrodynamics in curved spacetime is developed. This formalism enables a systematic investigation of photons in the Schwarzschild gravitational field, yielding novel results as well as refining previous results that were predicted by heuristic methods. The claim that "the gravitational redshift is a shift in the sharp frequencies of the photons for all frequencies of the spectrum" is proved. It is shown the gravitational decoherence is due to photon-gravity coupling and observer-dependent quantum electrodynamics in curved spacetime phenomena. The proper value of the photon gravitational interferometric relative phase shift is calculated and its full quantum-general relativistic nature is demonstrated. It is shown its observation will falsify the validity of Newtonian gravity and the extension of the Einstein equivalence principle beyond a single point (even in the weak uniform gravitational field.)

\end{abstract}


\maketitle


\section{Introduction}\label{Intro}
Quantum field theory and Einstein’s theory of gravity are two main theories and cornerstones of modern theoretical physics. Both theories have been studied significantly and independently on both theoretical and experimental sides. However, the quantum-gravity interface has been studied much less on both theoretical and experimental sides. In this paper, we investigate the influence of gravitational field on the photons as the quantum particle of the electromagnetism field.
\\

The influence of a background gravitational field of a source on the propagation of electromagnetic waves when the sources are black holes and electromagnetic fields are considered to be linear perturbations on the background such that their gravitational back reactions on the background are of second order and hence negligible to linear order in the perturbation, is the subject of black hole perturbation theory. This problem for Maxwell's equations on Schwarzschild and Kerr black holes was solved long ago by using the Newman-Penrose formalism. The equations, known as Tueckolsky equations, are useful for finding quasi-normal modes etc.~\cite{chandra}. However, due to the complexity of equations and solutions, it is challenging to investigate all physical aspects of electromagnetic waves in the gravitational field by using them. Therefore, it is common to investigate the influence of gravitational fields on photons by toy models, hand-waving methods, and effective descriptions, like assigning mass (based on mass-energy equivalence)~\cite{weinberg1972gravitation, einstein1911influence, MTW, pound1960apparent, zych2012general, hilweg2017gravitationally}, or clock (based on Shapiro time delay)~\cite{zych2012general} to photons or solving and analyzing the Klein-Gordon equation~\cite{zych2012general, hilweg2017gravitationally, exirifard2021towards, rodriguez2023introduction, bruschi2021spacetime, bruschi2023gravitational, bruschi2014spacetime, mieling2022measuring}. All of these approaches contrast the physical nature of photons which are massless spin-1 particles. 

This paper aims to construct a physical formalism of quantum electrodynamics in curved spacetime to investigate the influence of the gravitational field on photons. We will derive the gravitational effects on photons in a straightforward and unified way within this framework of quantum field theory in curved spacetime. We stress that our results are based only on Einstein's gravitation theory and
quantum field theory framework, with no assumptions about any new physical principles. Let us emphasize that the
obtained results are valid beyond the weak field approximation for the gravitational field (unless stated otherwise). Hence, they allow one to study the photons in \underline{strong gravity}.

The paper is organized into several sections, Section \ref{II} discusses a formulation of classical electrodynamics in curved spacetime based on the Riemann-Silberstein vectors and density tensor, while in section \ref{III} a canonical formalism for QED in CST is developed. Section \ref{IV} uses a class of solutions of Maxwell’s equations in Schwarzschild spacetime and the QED in CST formalism to obtain ladder operators. A proper physical approach to gravitational redshift is set in section \ref{V}. In section \ref{VI} by use of an extension of the locality principle, a measurement scheme for fundamental observers is developed. Propagation of photons in Schwarzschild spacetime and wave function distortion are investigated in \ref{VII}. Interferometry with photons and gravitational decoherence of superposition in Schwarzschild spacetime are studied in section \ref{VIII}.

This work is mainly based on the works of B. Mashhoon in electromagnetism in curved spacetime and the works of I. Bialynicki-Birula and Z. Bialynicka-Birula in the photon wave mechanics.  In this paper, we ignored the spin-gravity coupling~\cite{Frolov, Lambiase} and spin-orbit coupling~\cite{Andersson} (gravitational spin Hall effect), but it is possible to extend the analysis and consider the corrections due to such effects. In some sections, the analogy of the gravitational field and a hypothetical optical medium is discussed, this analogy is not a premise of the formulation, final results, and conclusions. Throughout the paper, we interchangeably use gravitational decoherence and gravitational decoherence of superposition.

In this paper, the units $G = c =\hbar= 1$, unless specified otherwise. The signature of the metric is $+2$ and Greek indices run from $0$ to $3$, while Latin indices run from $1$ to $3$.
\section{Classical Electrodynamics in curved spacetime}\label{II}
The Lagrangian of electromagnetic field which is not coupled to any kind of physical field other than the gravitational field is the covariant Lagrangian of free electromagnetic field in curved spacetime and its action is given by:
\begin{equation}\label{a1}
    S_{\text{Maxwell}}=\frac{-1}{4}\int {f^{\mu\nu} f_{\mu\nu}}\sqrt{-g} \, \dd{x^{4}},
\end{equation} 
where $f_{\mu\nu}$ is anti-symmetric bi-tensor. The field equations of motion for this Lagrangian are
\begin{equation}\label{a2}
\nabla _{\mu}f^{\mu\nu}=0, \qquad \nabla _{[\lambda}f_{\mu\nu]} =0.
\end{equation}
where  $\nabla_{\mu}$ is covariant derivative. Using, $ {\sqrt{-g}}T_{\alpha\beta}=-2\fdv{\cal{L}}{g^{\alpha\beta}}$, the energy-momentum tensor of the Lagrangian is:
\begin{equation}\label{a4}
    T_{\alpha\beta}=\frac{1}{4}(f_{\alpha\mu}f_{\beta}\\^{\mu}-\frac{1}{4}f_{\mu\nu}f^{\mu\nu}g_{\alpha\beta}).
\end{equation}
By use of the Hodge operator, the Hodge dual of the Faraday tensor is specified as
\begin{equation}\label{a5}
    {^{*}}f^{\mu\nu}=\frac{i}{2\sqrt{-g}}\epsilon^{\mu\nu}\\_{\alpha\beta}f^{\alpha\beta}.
\end{equation}
The self-dual and anti-self-dual parts of $f^{\mu\nu}$ are specified as
\begin{align}\label{a6}
   {^{\pm}} f^{\mu\nu}=f^{\mu\nu}\pm \frac{i}{2\sqrt{-g}}\epsilon^{\mu\nu\alpha\beta}f_{\alpha\beta},
\end{align}

To write the equation of motions based on ${^{\pm}}f^{\mu\nu}$, the reality of $\nabla_{\mu}$ and self-duality and anti-self-duality of $ {^{+}}f_{\alpha\mu}$ are used. The field equations of motion~\eqref{a2} based on ${^{\pm}} f^{\mu\nu}$ are written as 
\begin{equation}\label{a9}
\nabla_{\mu}\,{^{\pm}}f^{\mu\nu}=0 .
\end{equation}
Therefore field equation for self-dual and anti-self-dual parts ${^{-}}f^{\mu\nu}$ and ${^{+}}f^{\mu\nu}$ become separated. The action based on the self-dual and anti-self-dual sector of the field takes the following form:
\begin{equation}
    S_{\text{Maxwell}}=\frac{-1}{16}\int ({^{+}}{f^{\mu\nu}\, {^{+}}f_{\mu\nu}}+{^{-}}{f^{\mu\nu} \,{^{-}}f_{\mu\nu}})\sqrt{-g} \, \dd{x^{4}},
\end{equation} 
and energy-momentum tensor density based on ${^{\pm}}f^{\mu\nu}$ takes following form 
\begin{equation}\label{a10}
    T_{\alpha\beta}=\frac{1}{8}\,{^{+}}f_{\alpha\mu}\,\,{^{-}}f_{\beta}\\^{\mu}\, \, .
\end{equation}
To derive this formula the general Identity: 
\begin{equation}
-\frac{1}{2}g_{\mu\nu}A_{\alpha\beta}B^{\alpha\beta}={^{*}}A_{\mu\alpha}{^{*}}B^{\alpha}\\_{\nu}+B_{\mu\alpha}A^{\alpha}\\_{\nu}\,,
\end{equation}
is used. By use of $\Gamma^{\mu}_{\nu\mu}=\partial_{\nu}\ln{\sqrt{-g}}$, field equation can be written in the following form
\begin{equation}\label{a13}
\partial_{\mu}(\sqrt{-g}\,\,\, {^{\pm}}f^{\mu\nu})=0 .
\end{equation}
It is convenient to define density tensors, 
\begin{equation}\label{a14}
F{^{\pm}}^{\mu\nu}=\sqrt{-g}\,\,\, {^{\pm}}f^{\mu\nu} \, ,
\end{equation}
 we call it, the Riemann-Silberstein density tensor. Based on the Riemann-Silberstein density tensor, the field equations~\eqref{a9} take the following form
\begin{equation}\label{a15}
    \partial_{\mu}F{^{\pm}}^{\mu\nu}=0 .
\end{equation}
We define the contravariant dual of the Riemann-Silberstein density tensor as:
\begin{equation}\label{a16}
   F^{\pm}_{\mu\nu}=\frac{1}{\sqrt{-g}} \,g_{\mu\alpha}\,g_{\nu\beta}\, F^{\pm\alpha\beta}={^{\pm}}f_{\mu\nu}
\end{equation}
Define Riemann-Silberstein vectors as
\begin{equation}\label{a17}
    \mathbf{S}^{\pm}_{i}=F{^{\pm}}^{0i}, \qquad \mathbf{F}^{\pm}_{i}=\,{^{\pm}}f_{i0} \, .
\end{equation}
The Faraday field tensor has $6$ physical degrees of freedom, therefore, the components of $\mathbf{S}^{\pm}$ and  $\mathbf{F}^{\pm}$ are dependent. To find the relation between  $\mathbf{S}^{\pm}$ and  $\mathbf{F}^{\pm}$ it makes essential to use the fact that the matrices $(g_{\mu \nu})$ and $(g^{\mu \nu})$ are inverse of each other. The explicit expression of this fact, after separating temporal and spatial coordinates, results in relations that can be used to show that $(\gamma_{ij})$ and its dual $(\hat{\gamma}^{ij})$,
\begin{equation}\label{a18}
\gamma_{ij} := g_{ij} - \frac{g_{0i}\,g_{0j}}{g_{00}}\,, \qquad \hat{\gamma}^{ij} := g^{ij} - \frac{g^{0i}\,g^{0j}}{g^{00}}\,,          \end{equation}
are inverses of $(g^{ij})$ and $(g_{ij})$, respectively; that is, $g^{ik}\,\gamma_{kj} = \delta^i_j$ and  
$g_{ik}\,\hat{\gamma}^{kj} = \delta_i^j$~\cite{bini2018spinning, bini2012spacetime}. The quantities in Equation~\eqref{a18} also appear in the (1+3) timelike threading and the (3+1) spacelike slicing of spacetime~\cite{bini2012spacetime}; It is possible to show

\begin{equation}\label{a22}
\mathbf{F}^{\pm}_{l}={^{\pm}}\boldsymbol{\kappa}^{-1}_{li}\,\,\mathbf{S}^{\pm}_{i}\, ,
\qquad \qquad
    \mathbf{S}^{\pm}_{l}={^{\pm}}\boldsymbol{\kappa}_{li}\,\,\mathbf{F}^{\pm}_{i}\, ,
\end{equation}

where
\begin{align}
   & {^{\pm}}\boldsymbol{\kappa}^{-1}_{li}=-\frac{g_{il}}{\sqrt{-g}g^{00}}\pm i\frac{g^{0j}}{g^{00}}\epsilon_{ijl}\, \\
    & {^{\pm}}\boldsymbol{\kappa}_{li}=- \frac{g^{il}{\sqrt{-g}}}{g_{00}}\pm i\frac{1}{g_{00}}\epsilon^{ijl} g_{0j}\, .
\end{align}

Maxwell equations~\eqref{a15} can be written based on  $\mathbf{S}^{\pm}$ and $\mathbf{F}^{\pm}$  as:
\begin{align}\label{a27}
     \partial_{\mu}F{^{\pm}}^{\mu0}=0 &\Rightarrow \nabla\cdot \mathbf{S}^{\pm}=0 \, ,
\\ \label{a28}
      \partial_{\mu}F{^{\pm}}^{\mu i}=0 &\Rightarrow\pm\, i\,\frac{\partial \mathbf{S}^{\pm}}{\partial t} = \nabla \times \mathbf{F}^{\pm} \, .
\end{align}
The field equations for different helicity states at linear order perturbation treatment are completely decoupled. Taking the divergence of Equation~\eqref{a28} implies $\partial_t(\nabla \cdot \mathbf{S}^{\pm}) = 0$, which means that once the Equation~\eqref{a27} is satisfied at any given time, then it is valid for all time. The equation~\eqref{a28} is known as the Dirac equation for photons in a gravitational field. These equations are equivalent to the standard source-free Maxwell's equations in a hypothetical optical medium~\cite{bini2018spinning, mashhoon1973scattering, mashhoon1974electromagnetic, mashhoon1974can, mashhoon1975influence, kopeikin2002gravitomagnetic, skrotskii1957influence, plebanski1960electromagnetic, de1971gravitational, volkov1971propagation,bialynicki2013role}.
This hypothetical optical medium is conformally invariant and its optical properties are given by~\cite{bini2018spinning},
\begin{equation}\label{a29}
\epsilon_{ij} = \mu_{ij} = -\sqrt{-g}\,\frac{g^{ij}}{g_{00}}\,, \qquad G_i = - \frac{g_{0i}}{g_{00}}\,.
\end{equation} 
It is important to note there is no such exotic material in nature. The self-dual and anti-self-dual part of Faraday tensor, ${{^{\pm}}f^{\mu\nu}}$ and ${{^{\pm}}f_{\mu\nu}}$ based on  $\mathbf{S}^{\pm}$ and  $\mathbf{F}^{\pm}$ are found to be:
\begin{equation}\label{a32}
    {^{\pm}}f_{0i}=-\mathbf{F}^{\pm}_{i} \, , \qquad   {^{\pm}}f_{ij}= \pm i \epsilon_{ijk}\mathbf{S}^{\pm}_{k} \, ,
\end{equation}
\begin{equation}\label{a33}
    {^{\pm}}f^{0i}=\frac{1}{\sqrt{-g}}\mathbf{S}^{\pm}_{i}\, ,  \qquad   {^{\pm}}f^{ij}= \frac{\mp i}{\sqrt{-g}} \epsilon_{ijk}\mathbf{F}^{\pm}_{k}\, .
\end{equation}
In general, if some solutions are found for the set of Equations~\eqref{a27} and~\eqref{a28}, a solution for Maxwell's equation in curved spacetime is found.The reality of $f^{\mu\nu}$ imply, ${^{-}} f^{\mu\nu}=\bar{{^{+}} f^{\mu\nu}} $ which means $ {^{+}} f^{\mu\nu}$ and ${^{-}} f^{\mu\nu}$ are complex conjugate of each other. We define $f_{0i} = -E_i$, $f_{ij} = \epsilon_{ijk}\,B_k$, $\sqrt{-g}\,f^{0i} = D_i$ and $\sqrt{-g}\,f^{ij} = \epsilon_{ijk}\,H_k$. It is straightforward to show

\begin{equation}
   \mathbf{S}^{\pm}_{i}= D_{i}\pm i B_{i}\, , \qquad \mathbf{F}^{\pm}_{i}=E_{i}\pm i H_{i}\, .
\end{equation}

Due to the complex conjugate duality between the self-dual and anti-self-dual sectors for the case of a real Faraday tensor (physically consistent one), we consider only the self-dual sector and obtain the anti-self-dual sector by complex conjugation. For another argument for the adequacy of one Reimann-Silberstein vector to describe photon see~\cite{Reply}. By rewriting $ \mathbf{S}_{i}= \mathbf{S}^{+}_{i}\,,\mathbf{F}_{i}=\mathbf{F}^{+}_{i} \,$ 
and $ F^{\mu\nu}=F^{+\mu\nu}, F_{\mu\nu}=F^{+}_{\mu\nu},$ the relations between ${\mathbf{S}}$ and ${\mathbf{F}}$ is  expressed by ${\mathbf{S}}=\boldsymbol{\kappa}\,{\mathbf{F}} $ and its inverse, where $\boldsymbol{\kappa}={^{+}}\boldsymbol{\kappa}$.    
The equation of motion can be written only based on the self-dual part as,
 \begin{equation}\label{a41}
     \partial_{\mu}F^{\mu\nu}=0 \quad\Rightarrow \qquad  \nabla\cdot\mathbf{S}=0\, , \quad i \pdv{\mathbf{S}}{t}=\nabla\times \mathbf{F}.
 \end{equation}
The energy-momentum density tensor takes the following form based on the Riemman-Silberstein density tensor:
\begin{equation}\label{a43}
    \sqrt{-g}T_{\alpha}\\^{\beta}=\frac{1}{8}F_{\alpha\mu}\,\bar{F}^{\beta\mu}\, .
\end{equation}
Its expansion based on the Riemman-Silberstein vector is 
\begin{align}\label{a44}
    &\sqrt{-g}T_{0}\,^{0}=\frac{-1}{8} \mathbf{F}_{i}\,\,\bar{\mathbf{S}}_{i}\, ,
    \\ \label{a45}
    &\sqrt{-g}T_{0}\,^{i}=\frac{i}{8} \mathbf{F}_{j}\,\bar{\mathbf{F}}_{k}\, \epsilon_{jki}\, ,
    \\
    \label{a46}
    &\sqrt{-g}T_{i}\,^{0}=\frac{-i}{8} \mathbf{S}_{j}\,\bar{\mathbf{S}}_{k}\,\epsilon_{jki} \, , 
    \\
    \label{a47}
    &\sqrt{-g}T_{i}\,^{j}=\frac{1}{8} (\mathbf{S}_{j}\bar{\mathbf{F}}_{i}-\mathbf{S}_{k}\,\bar{\mathbf{F}}_{k}\,\delta_{ij}),
\end{align}
the repeated index shows summation. These are consistent with the result of \cite{mashhoonEMW}.

\section{Quantum Electrodynamics in Curved Spacetime}\label{III}
In this section, a new method for canonical quantization of electromagnetic fields in curved spacetime is developed. To construct the quantum field theory of electromagnetic field in curved spacetime, a canonical formulation of electromagnetic field in curved spacetime is needed. To construct a canonical formulation, canonical variables, canonical equations, and generalized Poisson brackets are introduced in this section. Then by use of the Dirac quantization procedure, the commutators are obtained. In this section, spacetime is considered to be stationary, or conformally stationary (static and conformally static are obtained as special cases by putting $g_{0i}=0$ ) and the field variables are considered at a particular time $t$. It is important to mention that our approach for developing a canonical formulation for the electromagnetic field in curved spacetime owes to~\cite{QED}, where a canonical formulation of electromagnetic field in optical media in Minkowski spacetime was developed.    
\subsection{Canonical Form of Maxwell's Equations in Curved Spacetime}\label{IIIA}
Expanding Equation~\eqref{a27} and~\eqref{a28} base on $\mathbf{E},\, \mathbf{D},\,\mathbf{B},\,\mathbf{H}$ gives,
\begin{equation}\label{b1}
    \Dot{\mathbf{D}}(\mathbf{x})=\nabla\times \mathbf{H}(\mathbf{x})\, , \qquad \Dot{\mathbf{B}}(\mathbf{x})=-\nabla\times \mathbf{E}(\mathbf{x})\, ,
\end{equation}
\begin{equation}\label{b2}
    \nabla\cdot \mathbf{D}(\mathbf{x})=0 \, ,\qquad \nabla\cdot \mathbf{B}(\mathbf{x})=0\, .
\end{equation}
The Lagrangian in terms of $\mathbf{E},\, \mathbf{D},\,\mathbf{B},\,\mathbf{H}$ takes the following form
\begin{equation}\label{b3}
    \mathcal{L}=\frac{1}{2}(D_{i}E_{i}-H_{i}B_{i})\, .
\end{equation}
Based on this Lagrangian, it can be written 
\begin{equation}\label{b4}
    \mathbf{D}(\mathbf{x})=\pdv{\mathcal{L}\bigl(\mathbf{E}(\mathbf{x}),\mathbf{B}(\mathbf{x})\bigl)}{\mathbf{E}(\mathbf{x})} , \quad
    \mathbf{H}(\mathbf{x})=-\pdv{\mathcal{L}\bigl(\mathbf{E}(\mathbf{x}),\mathbf{B}(\mathbf{x})\bigl)}{\mathbf{B}(\mathbf{x})}.
\end{equation}
It is not hard to show that these relations are equivalent to Equations~\eqref{a22}.  
If $-4\sqrt{-g}T_{0}^{0}$ be considered as electromagnetic density Hamiltonian and denote it by $\mathcal{H}$. It takes the following form based on $\mathbf{E},\, \mathbf{D},\,\mathbf{B},\,\mathbf{H}$ 
\begin{equation}\label{b5}
    \mathcal{H}=\frac{1}{2}\mathbf{F}_{i}\,\,\bar{\mathbf{S}}_{i}=\frac{D_{i}E_{i}+B_{i}H_{i}}{2}\, .
\end{equation}
Expansion of $T_{0}^{0}$ component in Equation~\eqref{a4} and use of Lagrangian density definition in Equation~\eqref{a1} gives
\begin{equation}\label{b6}
    \mathcal{H}=\mathbf{D}(\mathbf{x})\cdot \mathbf{E}(\mathbf{x})-\mathcal{L} \, .
\end{equation}
 The set of Equations~\eqref{b1} and~\eqref{b4} form a complete set of equations that can specify the value of field equation based on the initial value of $\mathbf{D(\mathbf{x})}$ and $\mathbf{B(\mathbf{x})}$. The set of Equations~\eqref{b2} are axillary conditions, once they are satisfied at the initial time (or any given time), then it is valid for all
time. This equation resembles the equation of Hamilton in classical mechanics. We treat $ \mathbf{D}(\mathbf{x})$ and $\mathbf{B}(\mathbf{x})$ as independent variables and $\mathbf{E}(\mathbf{x})$ and 
$\mathbf{H}(\mathbf{x})$ as functions of $ \mathbf{D}(\mathbf{x})$ and $\mathbf{B}(\mathbf{x})$ by use of Equation~\eqref{b4}. The transformation from variables $\mathbf{E}(\mathbf{x})$ and $\mathbf{B}(\mathbf{x})$ which the Lagrangian is defined to variables $\mathbf{D}(\mathbf{x})$ and $\mathbf{B}(\mathbf{x})$ 
is obtained by following the Legendre transformation 
\begin{align}\label{b7}
   \mathcal{H}(\mathbf{x}) =\mathbf{D}(\mathbf{x})\cdot \mathbf{E}\bigl(\mathbf{D}(\mathbf{x}),\mathbf{B}(\mathbf{x})\bigl)-\mathcal{L}\bigl(\mathbf{E}\bigl(\mathbf{D}(\mathbf{x}),\mathbf{B}(\mathbf{x})\bigl),\mathbf{B}(\mathbf{x})\bigl) .
\end{align}
The resulting Hamiltonian is equivalent to what we defined as Hamiltonian density in equation~\eqref{b5}.
This Hamiltonian explicitly in terms of $\mathbf{D}(\mathbf{x})$ and $\mathbf{B}(\mathbf{x})$ is
\begin{align}\label{b8}
    \mathcal{H}\bigl(\mathbf{D}(\mathbf{x}),\mathbf{B}(\mathbf{x})\bigl)=\frac{1}{2}\mathbf{D}(\mathbf{x})\cdot \mathbf{E}\bigl(\mathbf{D}(\mathbf{x}),\mathbf{B}(\mathbf{x})\bigl)+\frac{1}{2}\mathbf{B}(\mathbf{x})\cdot \mathbf{H}\bigl(\mathbf{D}(\mathbf{x}),\mathbf{B}(\mathbf{x})\bigl)\, .
\end{align}
$\mathbf{E}(\mathbf{x})$ and $\mathbf{H}(\mathbf{x})$ can be expressed as function of derivative of $\mathcal{H}$ as 
\begin{align}\label{b9}
    \mathbf{E}\bigl(\mathbf{D}(\mathbf{x}),\mathbf{B}(\mathbf{x})\bigl)=\pdv{\mathcal{H}(\mathbf{x})}{\mathbf{D}(\mathbf{x})} ,  \\ \label{b9-1} \mathbf{H}\bigl(\mathbf{D}(\mathbf{x}),\mathbf{B}(\mathbf{x})\bigl)=\pdv{\mathcal{H}(\mathbf{x})}{\mathbf{B}(\mathbf{x})} .
\end{align}
By use of Hamiltonian of the field that is space integral of the Hamiltonian density ${H}=\int \mathcal{H}(\mathbf{x}) \dd x^{3}\, ,$ and use of the functional derivative, Equations~\eqref{b9} and~~\eqref{b9-1} written as,
\begin{equation}\label{b11}
    \mathbf{E}\bigl(\mathbf{D}(\mathbf{x}),\mathbf{B}(\mathbf{x})\bigl)=\fdv{{H}(\mathbf{x})}{\mathbf{D}(\mathbf{x})} , \quad \mathbf{H}\bigl(\mathbf{D}(\mathbf{x}),\mathbf{B}(\mathbf{x})\bigl)=\fdv{{H}(\mathbf{x})}{\mathbf{B}(\mathbf{x})} .
\end{equation}
Similarly, equations~\eqref{b1} written as,
\begin{align}\label{b12}
    \Dot{\mathbf{D}}(\mathbf{x})=\nabla\times \fdv{{H}(\mathbf{x})}{\mathbf{B}(\mathbf{x})}\, ,\quad \Dot{\mathbf{B}}(\mathbf{x})=-\nabla\times \fdv{{H}(\mathbf{x})}{\mathbf{D}(\mathbf{x})}\, .
\end{align}
This set of equations resembles the Hamiltonian system of equations. 


\subsection{Generalized Poisson Brackets for Maxwell Fields}\label{IIIB}

The generalized Poisson brackets for~\eqref{b12} Hamiltonian system are developed and discussed in (Appendix~\ref{GPB}) take the following form
\begin{align}\label{b18-1}
    \{\mathbf{F}\bigl(\mathbf{D}(\mathbf{x}),\mathbf{B}(\mathbf{x})\bigl),\mathbf{G}\bigl(\mathbf{D}(\mathbf{y}),\mathbf{B}(\mathbf{y})\bigl)\}
   =\left[\pdv{\mathbf{F}(\mathbf{x})}{D_{i}(\mathbf{x})}\pdv{\mathbf{G}(\mathbf{x})}{B_{j}(\mathbf{x})}-\pdv{\mathbf{F}(\mathbf{x})}{B_{i}(\mathbf{x})}\pdv{\mathbf{G}(\mathbf{y})}{D_{j}(\mathbf{y})}\right]\epsilon_{ijk}\partial_{k}\delta(\mathbf{x}-\mathbf{y})\,.
\end{align}
Therefore the Equations~\eqref{b1} take the following form based on the generalized Poisson brackets
\begin{equation}\label{b16}
    \dot{\mathbf{D}}(\mathbf{x})=-\{{H},\mathbf{D}(\mathbf{x})\}\, , \qquad  \dot{\mathbf{B}}(\mathbf{x})=-\{{H},\mathbf{B}(\mathbf{x})\}\, ,
\end{equation}
and, Poisson brackets for fields $\mathbf{D}(\mathbf{x})$ and $\mathbf{B}(\mathbf{x})$ are
\begin{equation}\label{b19}
    \{B_{i}(\mathbf{x}),D_{j}(\mathbf{y})\}=-\{D_{j}(\mathbf{x}),B_{i
}(\mathbf{y})\}=\epsilon_{ijn}\partial_{n}\delta (\mathbf{x}-\mathbf{y}) \, ,
\end{equation}
\begin{equation}\label{b20}
     \{D_{i}(\mathbf{x}),D_{j}(\mathbf{y})\}=0=\{B_{i}(\mathbf{x}),B_{j}(\mathbf{y})\}\, .
\end{equation}
It is important to notice, that they are well-defined on points where the metric is non-singular. As it is pointed out in~\cite{bialynicki2013role} these Poisson brackets are universal because they contain no physical constant. In addition, gravitational quantities are absent in this Poisson brackets. These validate the choice of $\mathbf{B}(\mathbf{x})$ and $\mathbf{D}(\mathbf{x})$ (and not $\mathbf{E}(\mathbf{x})$ and $\mathbf{H}(\mathbf{x})$ ) as canonical variables.  
It is not hard to show that these equations imply
\begin{equation}\label{b21}
     \{\nabla\cdot \mathbf{D}(\mathbf{x}),D_{j}(\mathbf{y})\}=0=\{\nabla\cdot \mathbf{D}(\mathbf{x}),B_{j}(\mathbf{y})\} \, ,
\end{equation}
\begin{equation}\label{b22}
     \{\nabla\cdot \mathbf{B}(\mathbf{x}),D_{j}(\mathbf{y})\}=0=\{\nabla\cdot \mathbf{B}(\mathbf{x}),B_{j}(\mathbf{y})\}\, .
\end{equation}
Generalized Poisson bracket for $\mathbf{S}_{i}$ can be written by use of its components Poisson bracket as
\begin{align}\label{b23}
    \bigl\{ \mathbf{S}_{i}(t,\mathbf{x}),\bar{\mathbf{S}}_{j}(t,\mathbf{y}) \bigl\}&=2 i\epsilon_{ijk}\partial_{k}\delta(\mathbf{x}-\mathbf{y})\, , \\  \bigl\{ \mathbf{S}_{i}(t,\mathbf{x}),\mathbf{S}_{j}(t,\mathbf{y}) \bigl\}&=0\, .
\end{align}
Generalized Poisson brackets for $\mathbf{F}$ are 
\begin{align}\label{b24}
    \bigl\{ \mathbf{F}_{i}(t,\mathbf{x}),\bar{\mathbf{F}}_{j}(t,\mathbf{y}) \bigl\}&=2 i\boldsymbol{\kappa}^{-1}_{ik}(\mathbf{x})\bar{\boldsymbol{\kappa}}^{-1}_{jl}(\mathbf{y})\epsilon_{kln}\partial_{n}\delta(\mathbf{x}-\mathbf{y})\\
   \bigl\{ \mathbf{F}_{i}(t,\mathbf{x}),\mathbf{F}_{j}(t,\mathbf{y}) \bigl\}&=0\, .
  \end{align}
Calculation of Poisson brackets of other combinations of $\mathbf{F}$ and $\mathbf{S}$ and their complex conjugate are straightforward.


\subsection{Quantization and Commutators}\label{IIIC}
To obtain the commutation relation from Poisson brackets, we use the standard Dirac procedure$\,\,\,\frac{ \{ \, \, \, \, \, \, , \,  \, \, \, \, \, \}}{i\hbar} \Rightarrow   \bigl[ \, \, \, \,  ,  \, \, \, \, \bigl]\, 
$.
Consequently, the commutation relations for combination of $\mathbf{S}$ and $\mathbf{S}^{\dag}$ are found to be,
\begin{align}\label{b28}
    \bigl[ \hat{\mathbf{S}}_{i}(t,\mathbf{x}),\hat{\mathbf{S}}^{\dag}_{j}(t,\mathbf{y})\bigl]&=-2\hbar\epsilon_{ijn}\partial_{n}\delta(\mathbf{x}-\mathbf{y}) \, ,\\  
      \bigl[ \hat{\mathbf{S}}_{i}(t,\mathbf{x}),\hat{\mathbf{S}}_{j}(t,\mathbf{y})\bigl]&= 0
      \, .
\end{align}
Other commutation relations can be obtained by use of $\boldsymbol{\kappa}^{-1}_{ik}(\mathbf{x})$ or by Dirac procedure for other generalized Poisson brackets. 
It is noteworthy to mention such commutation relations just for the case of closed FRW spacetime (which is conformally static) were obtained in~\cite{mashhoonEMW} by a different approach.


\section{Quantum Electrodynamics in  Schwarzschild Spacetime}\label{IV}
The Schwarzschild metric in the isotropic form~\cite{mashhoon1973scattering, weinberg1972gravitation} is
\begin{equation}\label{c1}
    \dd s^{2}=-\biggl(\frac{\mathcal{B}_{-}(r)}{\mathcal{B}_{+}(r)}\biggl)^2\dd t^{2} +\mathcal{B}^{\,4}_{+}(r)\bigl(\dd x^{2}+\dd y^{2}+\dd z^{2}\bigl),
\end{equation}
 Here $\mathcal{B}_{\pm}(r)=1\pm\frac{MG}{2r}$ and $r$ is the radius $r^{2}=x^{2}+y^{2}+z^{2}$ which shows the spatial distance from the singularity point. It is important to note that $g_{00}$ starts from unity at $r = \infty$ far from the source and decreases monotonically with decreasing isotropic radial coordinate $r$ until it diverges at the horizon $r = M/2$ in isotropic coordinates. We recall that $r_{S} = M + r + M^2/(4\, r)$, where $r_{S}$ is the standard Schwarzschild radial coordinate~\cite{mashhoon1973scattering}. The contravariant form of the metric is,
\begin{align}\label{c2}
    \dd s^{2}=-\biggl(\frac{\mathcal{B}_{+}(r)}{\mathcal{B}_{-}(r)}\biggl)^2 \pdv[2]{}{t}+\mathcal{B}^{\,-4}_{+}(r)\bigl(\pdv[2]{}{x}+\pdv[2]{}{y}+\pdv[2]{}{z}\bigl).
\end{align}
It is not hard to show 
\begin{equation}\label{c6}
    \boldsymbol{\kappa}_{ij}(\mathbf{x})=\varkappa_{(\mathbf{x})} \,\delta_{ij} \, , \qquad \boldsymbol{\kappa}^{-1}_{ij}(\mathbf{x})=\varkappa^{-1}_{(\mathbf{x})}\,\delta_{ij} \, ,
\end{equation}
where $\varkappa_{(\mathbf{x})}=(1+\frac{MG}{2r})^{3}(1-\frac{MG}{2r})^{-1}$.


\subsection{Solutions of Maxwell's Equations in Schwarzschild Spacetime}
As Schwarzschild Spacetime is static, we can assume that $\mathbf{F}^{\pm} $ and hence $\mathbf{S}^{\pm} $ depend upon time as $\exp(-i\omega t)$. That is, the field equation in~\eqref{a28} takes the form
\begin{equation}
     {\pm}\,\omega\, \mathbf{S}^{\pm} = \nabla \times \mathbf{F}^{\pm}\, .
\end{equation}
By use of relation $\mathbf{S}^{\pm }={^{\pm}}\boldsymbol{\kappa}\,\mathbf{F}^{\pm} $ and the form of $^{\pm}\boldsymbol{\kappa}$ in Schwarzschild spacetime, we obtain
\begin{equation}
     {\pm}\,\omega\,\varkappa_{(\mathbf{x})}\, \mathbf{F}^{\pm} = \nabla \times \mathbf{F}^{\pm}\, .
\end{equation}
It is straightforward to show this equation has a set of solutions in the following form,
\begin{equation}\label{c7}
\mathbf{F}^{\pm}(t,\mathbf{x}) = A_{\pm}(\omega,\hat{\mathbf{k}})\, \hat{\ell}_{\pm}(\hat{\mathbf{k}}) \,e^{-i\omega\,t + i\,\omega \int^{\mathbf{x}} \varkappa_{(\mathbf{x}')} \hat{\mathbf{k}} \cdot dx'}\, ,
\end{equation} 
at the zero order in $\omega^{-1}$ to make our analysis more precise, we consider $\omega_{0}\ll\omega$ and $\omega_{0}$ is related to the curvature of background (we can replace this condition with $\nabla\times\hat{\ell}_{\pm}(t,\mathbf{x})=0 $ condition). $A_{+}(\omega,\hat{\mathbf{k}})$ and $A_{-}(\omega,\hat{\mathbf{k}})$ are constant complex amplitudes for positive and negative helicity radiation, respectively, and $\hat{\ell}_{\pm}=\hat{\ell}_1\pm i\hat{\ell}_2$ where $\hat{\ell}_1$ and $\hat{\ell}_2 $ are two unit vectors which with $\hat{\mathbf{k}}$ form a right-oriented orthonormal basis in space (Appendix.~\ref{Pol}).
We can use dimensional analysis (Appendix.~\ref{Acc}) to set a value to $\omega_{0}$,
\begin{equation}\label{c7-1}
    \omega_{0}=\frac{2\pi \,g_{(r)}}{c}\, ,
\end{equation}
where $g_{(r)}$, is the value of the gravitational acceleration measured by fundamental observers (Appendix.~\ref{Acc}). A similar scale also appears in the discussion of the locality principle~\cite{mashhoon2017nonlocal}. 

By use of relations $\mathbf{S}^{\pm}={^{\pm}}\boldsymbol{\kappa}\,\mathbf{F}^{\pm} $, we obtain
\begin{equation}\label{c8}
\mathbf{S}^{\pm}(t,\mathbf{x}) = A_{\pm}(\omega,\hat{\mathbf{k}})\,
\varkappa_{(\mathbf{x})}\, 
\hat{\ell}_{\pm}(\hat{\mathbf{k}}) \,e^{-i\omega\,t + i\,\omega \int^{\mathbf{x}} \varkappa_{(\mathbf{x}')} \hat{\mathbf{k}} \cdot \dd x' }\, .
\end{equation} 
 To check the $\nabla\cdot\mathbf{S}^{\pm}=0$ condition, we use the relation $\mathbf{S}^{\pm}=\mp\omega^{-1}\nabla\times\mathbf{F}^{\pm} $, therefore, this condition holds for this type of solution. 
 
 Based on these solutions, the self-dual sector has a general solution in Schwarzschild in the form of:
\begin{align}\label{c9}
    \mathbf{S}^{}(t,\mathbf{x}) =\int\frac{\dd^3\mathbf{k}}{(2\pi)^{3/2}}\varkappa_{(\mathbf{x})}
\hat{\ell}(\mathbf{\hat{k}})\bigl(A_{+(\omega,\hat{\mathbf{k}})} e^{-i\omega t+i\omega\int^{\mathbf{x}}\varkappa_{(\mathbf{x}')}\hat{\mathbf{k}}\cdot\dd x'}  +\bar{A}_{-(\omega,\hat{\mathbf{k}})}e^{+i\omega t-i\omega \int^{\mathbf{x}} \varkappa_{(\mathbf{x}')} \hat{\mathbf{k}}\cdot\dd x'}\bigl),
\end{align}
where integration over $\int\dd^3\mathbf{k}=\int_{{0}}^{+\infty}\omega^{2}\,\dd\omega\int\dd\hat{\mathbf{k}}^2$ means integration over wave vector space or integration over angular frequency and closed 2-dimensional space of unit vectors. To be more precisely, Due to the limit in the solutions for low frequencies,
\begin{equation}\label{c10}
    A_{\pm(\omega,\hat{\mathbf{k}})}=0 \quad \text{if}  \quad \omega\leq \omega_{1}\quad \text{where}  \quad \omega_{1} \gg \omega_{0}\,.
\end{equation}

This solution~\eqref{c9} is valid even in a strong gravitational field as soon as the condition~\eqref{c10} be satisfied. The solution~\eqref{c9} in the asymptotic region far from the source where the gravitational field is absent takes the form of solutions of Maxwell equations in Minkowski spacetime which are constructed by adding plane wave solutions.
As it is apparent from these solutions \textit{the photon wave vector $\vec{\mathbf{\mathbf{k}}}$ in the asymptotic region far from the source where the gravitational field is absent is replaced by $\varkappa_{(\mathbf{x})}\,\vec{\mathbf{\mathbf{k}}}$, in regions the gravitational field is nonzero.} Therefore, the gravitational field couples to the wave vector through $\varkappa_{(\mathbf{x})}$. We refer to this coupling as \textit{photon-gravity coupling}. It is important to note there is no frequency-gravity (energy-gravity) coupling.
\subsection{QED Field Operators in Schwarzschild Spacetime}
By use of Dirac second quantization Procedure, we replace 
\begin{equation}\label{c10-1}
 A_{\lambda(\omega,\hat{\mathbf{k}})}   \rightarrow \sqrt{\hbar c} \, \hat{a}_{\lambda(\omega,\hat{\mathbf{k}})}\, , \quad \bar{A}_{\lambda(\omega,\hat{\mathbf{k}})}   \rightarrow \sqrt{\hbar c} \, \hat{a}^{\dag}_{\lambda(\omega,\hat{\mathbf{k}})}\, .
\end{equation}
and promote~\ref{c9} solution to field operators
\begin{align}\label{c11}
    \hat{\mathbf{S}}(t,\mathbf{x})=\sqrt{\hbar c}\int\frac{\dd^3{\mathbf{k}}}{(2\pi)^{3/2}}\varkappa_{(\mathbf{x})} \hat{\ell}(\mathbf{\hat{k}}) \bigl(\hat{a}_{+(\omega,\hat{\mathbf{k}})} e^{-i\omega t+i\int^{\mathbf{x}} \varkappa_{(\mathbf{x}')} {\mathbf{k}}\cdot\dd x'}+\hat{a}^{\dag}_{-(\omega,\hat{\mathbf{k}})}e^{+i\omega t-i \int^{\mathbf{x}} \varkappa_{(\mathbf{x}')}{\mathbf{k}}\cdot\dd x'}\bigl) ,
\end{align}

where $\hat{a}_{+(\omega,\hat{\mathbf{k}})}$ ($\hat{a}_{-(\omega,\hat{\mathbf{k}})}$)  annihilate right (left) helicity photon with frequency $\omega$ in the direction $\hat{\mathbf{k}}$ and $\hat{a}^{\dag}_{+(\omega,\hat{\mathbf{k}})}$ ($\hat{a}^{\dag}_{-(\omega,\hat{\mathbf{k}})}$) create right (left) helicity photon with frequency $\omega$ in the direction $\hat{\mathbf{k}}$. 
If we impose the following commutation relations for creation and annihilation operators 
\begin{align}\label{c13}
    [\hat{a}_{\lambda,(\omega,\hat{\mathbf{k}})}, \hat{a}^{\dag}_{\lambda',(\omega,\hat{\mathbf{k'}})}]=\hbar^{-1}\omega^{-1} \delta_{\lambda\lambda'}\delta(\omega-\omega')\delta^{2}(\hat{\mathbf{k}}-\hat{\mathbf{k'}}) =\hbar\omega \delta_{\lambda\lambda'}\delta^{3}({\mathbf{k}}-{\mathbf{k'}}) \, ,
\end{align}
    \begin{equation}\label{c14}
    [\hat{a}_{\lambda,(\omega,\hat{\mathbf{k}})}, \hat{a}_{\lambda',(\omega,\hat{\mathbf{k'}})}]=0\, ,
\end{equation}

It is possible to show (see Appendix~\ref{consis}),

\begin{equation}\label{c15-1}
    [ \hat{\mathbf{S}}_{i}(t,\mathbf{x}),\hat{\mathbf{S}}_{j}^{\dag}(t,\mathbf{y})]=-2\hbar\epsilon_{ijk}\pdv{}{x^k}\delta(\mathbf{x}-\mathbf{y})\, .
\end{equation}
So our quantization procedure is consistent and the field operators' commutation relations and creation-annihilation operators' commutation relations are equivalent.
As the photon wave function in momentum representation, and not in position representation, plays the fundamental role~\cite{bialynicki2020three}, we work in momentum space. Due to its convenience, we chose the Heisenberg picture. Therefore, the operators are spacetime-dependent and states are constructed by the operation of creation and annihilation operators on the vacuum state. 

To make our analysis more precise, we first set a proper physical approach to the redshift of photons in Schwarzschild spacetime.

\section{Gravitational Redshift in Schwarzschild Spacetime}\label{V}
Gravitational redshift is the change in the frequency of electromagnetic waves from emitter to receiver as they propagate in the gravitational field~\cite{einstein1911influence, MTW}. To set the proper approach to gravitational redshift, consider two fundamental observers Alice and Bob in different radii around astronomical mass, in a configuration that can signal directly to each other. To conduct a redshift experiment, we suppose both observers have an identical discharge tube of dilute gas or vapor of a simple atom. Bob as the receiver also has other apparatus to measure the frequency of emission lines of his own and Alice's sample. The gravitational redshift happens when Alice uses her radiation source and sends her sample emission spectra line to Bob. Bob uses his detector to detect Alice's emission spectra lines and compare their frequency with his own sample emission spectra line. To conduct a theoretical analysis of Bob's measurements, we consider the Hamiltonian of a simple atom sample in a gravitational field to compute the frequency of the lines in spectra. The Schrödinger equation of the single atom in the fundamental frame is,
\begin{equation}\label{c38}
    \dv{\ket{\psi_{0}}}{\tau}=\mathcal{H}\ket{\psi_{0}} \, ,
\end{equation}
where $\mathcal{H}$ is the Hamiltonian of the single simple atom in the inertial frames in Minkowski spacetime. Using $ \dd \tau = \chi_{(\mathbf{x})}^{-1} \, \dd t $, where $\chi_{(\mathbf{x})}=\frac{(1+\frac{MG}{2r_{(\mathbf{x})}})}{(1-\frac{MG}{2r_{(\mathbf{x})}})}$, it is possible to rewrite the Schrödinger equation based on coordinate time as
\begin{equation}\label{c39}
    {\chi_{(\mathbf{x})}}\dv{\ket{\psi_{0}}}{t}=\mathcal{H}\ket{\psi_{0}} \, .
\end{equation}
We can rewrite the Hamiltonian of Alice and Bob's source in isotropic coordinates in the following form:
\begin{equation}\label{r1}
    H_{A}=\chi_{A}^{-1}\, \mathcal{H}, \qquad \qquad  H_{B}=\chi_{B}^{-1}\, \mathcal{H}.
\end{equation}
For simplicity, we consider just ground state $E_{g}$ and first excited state $E_{1}$ so there is just one sharp line in the spectrum in our analysis. The line in Alice and Bob's spectrum at their rest frames has the following energy and frequency relative to time coordinate,
\begin{align}\label{r3}
    \Delta E_{A,A}=(E_{1}-E_{g})\,\chi_{A}^{-1}=\chi_{A}^{-1} \Delta E\Rightarrow  f_{A,A}=\frac{\Delta E}{\hbar} \chi_{A}^{-1},
\end{align}
\begin{align}\label{r5}
    \Delta E_{B,B}=(E_{1}-E_{g})\,\chi_{B}^{-1}=\chi_{B}^{-1} \Delta E\Rightarrow f_{B,B}=\frac{\Delta E}{\hbar}\chi_{B}^{-1} .
\end{align}
Here we used~\cite{earman1980gravitational} notation where the first index refers to the emitter and the second refers to the receiver.
Therefore, the problem of redshift reduces to calculations of $f_{A,B}$. 
From the previous section analysis, the photon in the gravitational field can be described by an effective Hamiltonian $H_{_{\text{Photon}}}$ such that:
\begin{equation}\label{r7}
  \pdv{\ket{\Psi}}{t}=H_{_{\text{Photon}}}\ket{\Psi} ,
\end{equation}
where $\ket{\Psi}$ is a single photon state and $t$ is coordinate time, due to the gravitational field statics, the Hamiltonian is symmetric under time translations so photon energy is invariant and does not change as they propagate in the gravitational field. This is compatible with the existence of the solutions in~\eqref{c7} and~\eqref{c8} form and the calculation in black hole perturbation theory~\cite{chandra}. Therefore, the static
gravitational field does not change the frequency of photons and electromagnetic radiation that Alice sends to Bob:
\begin{equation}\label{r8}
    f_{A,B}=f_{A,A},
\end{equation}
and
\begin{equation}\label{r9}
    1+z = \frac{f_{A,B}}{f_{B,B}}={\chi_{A}}^{-1}{\chi_{B}}.
\end{equation}
Expansion of $\chi$ in $r$ gives $\chi(r)=1+\frac{MG}{r}+\cdots$, Consequently, the redshift in the weak field limit is,
\begin{equation}\label{r10}
    z = \frac{f_{A,B}-f_{B,B}}{f_{B,B}} \simeq -\frac{MG}{r_{A}}+\frac{MG}{r_{B}}\, = V{(r_{A})}-V{(r_{B})}\, ,
\end{equation}
where $V{(r)}$ is Newtonian potential. It is not hard to show when Alice and Bob are in a configuration such that the gravitational field can be considered uniform, Equation~\eqref{r10} can be rewritten as, 
\begin{equation}\label{r11}
    z =\frac{1}{c^2}\mathbf{g}_{(r_{0})}\cdot (\mathbf{X}_{A}-\mathbf{X}_{B}) \, ,
\end{equation}
where $r_0$ is some reference point in the uniform gravitational field region, $\mathbf{g}_{(r_{0})}$ is the gravitational acceleration (see Appendix.~\ref{Acc}), and $\mathbf{X}_{A/B}$ is the position vector of Alice or Bob relative to $r_{0}$. These results are in agreement with experimental results~\cite{pound1960apparent,vessot1980test} and astrophysical observational evidence~\cite{greenstein1971effective}. 

Therefore, \textit{the gravitational redshift problem is explainable based on considering the influence of the gravitational field on the emitter and the receiver setups and without considering any frequency shift for the radiation as it propagates in the static gravitational field.} Other similar analyses of gravitational redshift can be found in~\cite{okun2000interpretation, okun2000photons, mashhoon1988complementarity}.

In this section, we only analyzed the observed frequency shift for photons with sharp frequency from Alice to Bob. After this digression, we return the formalism we have developed in section~\ref{IV} and try to develop a measurement scheme for fundamental observers for analyzing realistic photons with spectrum width in gravitational field. 


\section{Generalized Locality Principle and QED measurement scheme for the Fundamental Observers in Schwarzschild Spacetime}\label{VI}
To construct a scheme for QED measurements of the fundamental observers at different points in Schwarzschild spacetime, field operators in the fundamental frames must be obtained. In this regard, we develop a tentative generalization of \textit{Locality Principle}. The locality principle and measurements in SR and GR are discussed in~\cite{mashhoon2017nonlocal} and references cited therein. Based on this principle, \textit{the projection of various tensorial classical quantities on the orthonormal tetrad frame of an observer can be physically interpreted as the classical measurement of these quantities by the observer.}
\subsection{Fundamental Frames in Schwarzschild Spacetime}
To develop a measurement scheme, we first consider the tetrads of fundamental observers at rest outside the Schwarzschild black hole in isotropic coordinates:
\begin{align}\label{c20}
     e_{\hat{0}}\,^{\mu}=\frac{(1+\frac{MG}{2r})}{(1-\frac{MG}{2r})}\,\delta_{0}^{\mu}\, ,\qquad
      e_{\hat{i}}\,^{\mu}=\bigl( 1+\frac{MG}{2r}\bigl)^{-2}\,\delta_{i}^{\mu}\,,
\end{align}
where the spatial axes of the observer’s frame are primarily along the background coordinate axes. Introduce the dual coframe as
\begin{align}\label{c22}
    e^{\hat{0}}\,_{\mu}=\frac{(1-\frac{MG}{2r})}{(1+\frac{MG}{2r})}\, \delta^{0}_{\mu}\,,\qquad
     e^{\hat{i}}\,_{\mu}=\bigl( 1+\frac{MG}{2r}\bigl)^{2}\,\delta^{i}_{\mu}\,.
\end{align}
By use of this tetrad field, we can compute the local coordinate $(X^{\hat{0}},X^{\hat{i}})$ about fundamental observers at rest outside the Schwarzschild black hole in isotropic coordinate and obtain the vectorial and tensorial value of physical fields by projecting by the tetrads (locality principle). Orthonormal tetrad fields satisfy the relation
\begin{equation}\label{c24}
g_{\mu\nu}=\eta_{\hat{a}\hat{b}}\,e^{\hat{a}}\\_{\mu}\,e^{\hat{b}}\\_{\nu}\, .
\end{equation} 
Based on this tetrad, we have the following relations between the coordinate segments 
\begin{equation}\label{c26}
    \dd X^{\hat{a}}=e^{\hat{a}}\\_{\mu}\, \dd x^{{\mu}}\, .
\end{equation}
By use of the explicit form of tetrad fields, we have
\begin{equation}\label{c27}
    \dd X^{\hat{0}}=\frac{(1-\frac{MG}{2r})}{(1+\frac{MG}{2r})}\, \dd t \, , \qquad  \dd X^{\hat{i}}=\left( 1+\frac{MG}{2r}\right)^{2} \dd x^{i} .
\end{equation}
If we consider observer proper time $\tau$, it is not hard to show  $X^{0}=\tau$. We can choose the initial time such that $ \tau=\frac{(1-\frac{MG}{2r})}{(1+\frac{MG}{2r})} t $.

 It is worth mentioning that by use of tetrad fields, we have calculated the first approximation to the Fermi coordinate of fundamental observers in the Schwarzschild outer region. In better words, we established an approximate Fermi normal coordinate system with coordinates $(X^{\hat{0}},X^{\hat{i}})$
along the fundamental orthonormal tetrad frame
adapted to this observer~\cite{synge1960relativity}. The observer permanently occupies the spatial origin of his Fermi system; thus, $(X^{\hat{0}}, 0, 0, 0)$ are the Fermi coordinates of the fundamental observer and $X^{\hat{0}}$ is its proper time. Concerning this observer and in terms of approximate Fermi coordinates established along its world line, it is possible to obtain the values of observer measurements by obtaining the values of physical quantities in the Fermi coordinate. The values of physical quantities in the Fermi coordinate are obtained by projection by tetrad field. Higher order approximations to Fermi coordinate are obtained in~\cite{MashRosh} and reference cited therein.


\subsection{Generalized Locality Principle}
 By considering the general field operator in the form of~\eqref{c11}, we try to find the transformed value of the phase part, we can rewrite the phase part in the form of
\begin{align}\label{c29}
i\Phi_{\text{phase}}(t,{\mathbf{x}})=i\int_{{(0,\mathbf{x}_{0})}}^{(t,\mathbf{x})}k_{\mu}(\mathbf{x}')\,\dd x'^{\mu}=i\int_{{(0,\mathbf{X}_{0})}}^{(\tau,\mathbf{X})}K_{\hat{a}}(\mathbf{X}')\,\dd X'^{\hat{a}}\, ,
\end{align}
where $k_{\mu}(\mathbf{x})=(-\omega,\omega \, \varkappa_{(\mathbf{x})}\,\hat{\mathbf{k}}) $ and we used the nominal relation $x^{\mu}(X^{\nu})$ (it is important to note we do not need an explicit form of this relation for our calculations). By using tetrad explicit relations, we get
\begin{equation}\label{c30}
    K_{\hat{a}}(\mathbf{x})=e_{\hat{a}}\\^{\mu}(\mathbf{x}) k_{\mu}(\mathbf{x})=(-\Omega(\mathbf{x}),\Omega(\mathbf{x})\hat{\mathbf{K}}),
\end{equation}
where $\Omega(\mathbf{x})=\chi_{(\mathbf{x})}\,\omega$ and $\chi_{(\mathbf{x})}=\frac{(1+\frac{MG}{2r_{(\mathbf{x})}})}{(1-\frac{MG}{2r_{(\mathbf{x})}})}, $ that is the redshifted factor at the point $\mathbf{x}$. It is important to mention that the $\mathbf{x}$ in $\Omega(\mathbf{x})$ refers to the fact that it is relative to the proper time of the static observer at the point $\mathbf{x}$. Therefore, $\Omega(\mathbf{x})$ for an observer at point $\mathbf{x}$ is not a space dependent number.

We consider \textit{The Generalized Locality Principle} to obtain the field operators in the fundamental frames.
\\
\\
\textit{\textbf{Generalized Locality Principle:} The projection of the Riemann-Silberstein density tensor field operator by tetrad frame $\{e^{\hat{a}}_{\mu}\}$, is equivalent to a combination of a projection-dependent phase change $e^{-i\Theta(\mathbf{k},e^{\hat{a}}_{\mu(\mathbf{x})})}$ on polarization vector, and a projection-dependent unitary transformation $\hat{U}_{(e^{\hat{a}}_{\mu(\mathbf{x})})}$ on annihilation and creation operators.}
\\
$\hat{U}_{(e^{\hat{a}}_{\mu(\mathbf{x})})}$ act on $\hat{a}_{\lambda(\omega,\hat{\mathbf{k}})}$ as
\begin{equation}\label{c31}
\hat{U}^{\dag}_{(e^{\hat{a}}_{\mu(\mathbf{x})})}\,\hat{a}_{\lambda(\omega,\hat{\mathbf{k}})}\,\hat{U}_{(e^{\hat{a}}_{\mu(\mathbf{x})})}=\hat{a}_{\lambda(\Omega,\hat{\mathbf{K}})}\, ,
\end{equation}
\\
where $(\omega,\hat{\mathbf{k}})$ and $(\Omega,\hat{\mathbf{K}})$ are related by Equation~\eqref{c30}. As we choose the spatial axis of the fundamental frame along the coordinate axis, no aberration happens for wave vectors due to projection. Due to the statics of observers in two coordinates (Isotropic and fundamental), it  is reasonable to suppose that there is no change in the vacuum state, therefore:
\begin{equation}\label{c32}
    \hat{a}_{\lambda(\omega,\hat{\mathbf{k}})}\ket{0}=0\, ,\qquad  \hat{a}_{\lambda(\Omega,\hat{\mathbf{K}})}\ket{0}=0\, .
\end{equation}
The above vacuum can be considered as the QED analog of the Boulware vacuum~\cite{Singleton, Mukhanov}. 
\\
It is important to mention to make the measurement scheme for fundamental and non-fundamental frames more precise, a theoretical model of photon detectors like Unruh-DeWitt detectors in the gravitational field and an extension of Glauber's measurement theory in curved spacetime must be developed. We leave it to further work to establish a complete measurement theory in a gravitational field.
\\
\\
\subsection{QED Field Operators in Fundamental Frames in Schwarzschild Spacetime}
Putting all things together and considering the field operators in the fundamental frames only in single point $\mathbf{X}=0$ (due to the limitation of the locality principle), we get

\begin{align}\label{c35}
 \hat{\Tilde{\mathbf{S}}}(\tau,\mathbf{X}=0)=\sqrt{\hbar c} \int \frac{\dd^3{\mathbf{K}}}{(2\pi)^{3/2}}n{(\mathbf{X}=0)}\, 
e^{-i\Theta(\mathbf{k},e^{\hat{a}}_{\mu(\mathbf{x})})}\,\hat{\ell}(\mathbf{\hat{k}}) \bigl(\hat{a}_{+(\Omega,\hat{\mathbf{K}})} e^{-i\Omega_{(\mathbf{X}=0)} \tau}+\hat{a}^{\dag}_{-(\Omega,\hat{\mathbf{K}})}e^{+i\Omega_{(\mathbf{X}=0)} \tau}\bigl),
\end{align}

By use of Dirac delta properties, redshift relation, and $\hat{U}_{e^{\hat{a}}_{\mu(\mathbf{x})}}$, the commutators in the fundamental frames are
\begin{align}\label{c37}
    [\hat{a}_{\lambda,(\Omega,\hat{\mathbf{K}})}, \hat{a}^{\dag}_{\lambda',(\Omega',\hat{\mathbf{K}}')}]&=\hbar\Omega \delta_{\lambda\lambda'}\delta^{3}(\mathbf{\mathbf{K}}-\mathbf{\mathbf{K'}}), \\ [\hat{a}_{\lambda,(\Omega,\hat{\mathbf{K}})}, \hat{a}_{\lambda',(\Omega',\hat{\mathbf{K}'})}]&=0 .
\end{align}
In the rest of this paper we ignore phase change $e^{-i\lambda\Theta(\mathbf{k},e^{\hat{a}}_{\mu(\mathbf{x})})}\,$, due to ignorance of helicity. 
Now we have enough tools to systematically investigate the influence of the Schwarzschild gravitational field on photons that propagate between Fundamental observers. 
\section{Pure Photon in Schwarzschild spacetime and wave function Distortion}\label{VII}
To investigate photon propagation in Schwarzschild spacetime, we consider two fundamental observers, Alice and Bob in different radii around Schwarzschild mass in such a configuration that they can signal directly to each other and the spatial axes of the observer’s frame are primarily along the background coordinate axes. Both observers have identical sources of photons. Bob as the receiver also has other apparatus to detect the wave function of Alice’s source and compare it with his own source wave function. Alice uses her photon source and sends a photon to Bob. Bob uses his detector to observe Alice’s photon and compare it with the photon generated by his own source.  

To conduct a theoretical analysis of Bob’s measurements we use the formalism we have developed in the previous sections, the generalized locality principle, redshift operation, and redshift relations between coordinate and fundamental observer frequencies $\Omega_{A/B}=\chi_{A/B}\,\omega$ where $\chi_{A/B}=\frac{(1+\frac{MG}{2r_{A/B}})}{(1-\frac{MG}{2r_{A/B}})} $. Commutation relations in each frame are 
\begin{align}\label{c40}
    [\hat{a}_{\lambda,(\Omega,\hat{\mathbf{K}})}, \hat{a}^{\dag}_{\lambda',(\Omega',\hat{\mathbf{K}}')}]&=\hbar\Omega \delta_{\lambda\lambda'}\delta^{3}(\mathbf{\mathbf{K}}-\mathbf{\mathbf{K}}'), \\  [\hat{a}_{\lambda,(\Omega,\hat{\mathbf{K}})}, \hat{a}_{\lambda',(\Omega',\hat{\mathbf{K}}')}]&=0.
\end{align}
In this section, we consider the state of photons generated by sources in their frames. For simplicity, we consider that our photon has positive helicity and its wave function is $\psi(\Omega,\mathbf{K})$. Therefore, the creation operator for this wave function is~\cite{bialynicki2019photons}:
\begin{equation}\label{c41}
    \hat{a}^{\dag}_{\psi}= \int \frac{\dd^3{\mathbf{K}}}{K}\, 
\left(\psi(\Omega,\mathbf{K})\hat{a}^{\dag}_{+(\Omega,\hat{\mathbf{K}})}\right) ,
\end{equation}
where $\Omega$ is the frequency measured by the local observer. Operation of this operator on a vacuum state gives:   
\begin{equation}\label{c42}
    \ket{\psi}= \hat{a}^{\dag}_{\psi}\ket{0}= \int \frac{\dd^3{\mathbf{K}}}{K}\, 
\left(\psi(\Omega,\mathbf{K})\hat{a}^{\dag}_{+(\Omega,\hat{\mathbf{K}})}\right) \ket{0}\, .
\end{equation}
The photon states form Hilbert space, endowed with the inner product
\begin{equation}\label{c43}
    \bra{\phi}\ket{\psi}=\int \frac{\dd^3{\mathbf{K}}}{K}\, 
\left(\phi^{*}(\Omega,\mathbf{K})\psi(\Omega,\mathbf{K})\right)\, .
\end{equation}
By use of this inner product, the normalization condition for $\psi(\Omega,\mathbf{K})$ is $\bra{\psi}\ket{\psi}=1$. For simplicity, we strongly confine the wave function profile $\psi(\Omega,\mathbf{K})$ in the direction of propagation, therefore we can replace $\psi(\Omega,\mathbf{K})\rightarrow\psi(\Omega)$ and ignore integration over $\mathbf{\hat{K}}$ we also omit the $+$ sign as they do not have any effect in our final result, therefore
\begin{equation}\label{c45}
    \ket{\psi}=\int_{{0}}^{+\infty}\,\frac{\dd\Omega}{\Omega} \, 
\left(\psi(\Omega)\hat{a}^{\dag}_{(\Omega,\hat{\mathbf{k}}_{1})} \right)\ket{0}\, .
\end{equation}
The normalization becomes $\int\frac{\dd{\Omega} }{\Omega} \abs{  \psi(\Omega)}^{2}=1$. The creation and annihilation operators in the fundamental frames have the following commutation relations
\begin{align}\label{c47}
    [\hat{a}_{\lambda,(\Omega,\hat{\mathbf{k}})}, \hat{a}^{\dag}_{\lambda,(\Omega',\hat{\mathbf{k}})}]&=\hbar\Omega \delta_{\lambda\lambda'}\delta(\Omega-\Omega'), \\  [\hat{a}_{\lambda,(\Omega,\hat{\mathbf{k}})}, \hat{a}_{\lambda',(\Omega',\hat{\mathbf{k}})}]&=0.
\end{align}


\subsection{Photon Propagation through Gravitational Field from Alice to Bob}
We consider Alice to produce wave function  $\psi_{A}$ by her source and send it to Bob. To analyze the photon propagation as we work in the Heisenberg picture, we propagate the creation operator $\hat{a}^{\dag}_{\psi_{A}}$. By operation of redshift operator $\hat{a}^{\dag}_{(\omega_{},\hat{\mathbf{k}}_{1})}=\hat{U}_{(\chi_{A})} \hat{a}^{\dag}_{(\Omega_{A},\hat{\mathbf{K}}_{1})} \hat{U}^{\dag}_{(\chi_{A})}=\hat{a}^{\dag}_{(\chi^{-1}_{A}\Omega_{A},\hat{\mathbf{k}}_{1})}$, we rewrite creation operators in isotropic coordinate. As the creation operators propagate in spacetime they get phase and become  
\begin{equation}\label{c48}
 \rightarrow e^{-i\chi^{-1}_{A}\Omega_{A} \,t+i\chi^{-1}_{A}\Omega_{A}\int_{\mathbf{x}_{A}}^{\mathbf{x}_{B}}\varkappa_{(\mathbf{x})}\hat{\mathbf{k}}_{1}\cdot\dd x}\hat{a}^{\dag}_{(\chi^{-1}_{A}\Omega_{A},\hat{\mathbf{k}}_{1})}, 
\end{equation}
at Bob's location. To find the creation operators in Bob's Lab, we use redshift relation $\Omega_{B}=\chi_{B}\,\omega $ and operation of redshift operator $\hat{a}^{\dag}_{(\Omega_{B},\hat{\mathbf{K}}_{1})}=\hat{U}^{\dag}_{(\chi_{B})} \hat{a}^{\dag}_{(\omega_{},\hat{\mathbf{k}}_{1})} \hat{U}_{(\chi_{B})}=\hat{a}^{\dag}_{(\chi_{B}\omega_{},\hat{\mathbf{k}}_{1})}$, we obtain  Alice creation operators in Bob frame as 
\begin{equation}\label{c49}
  \rightarrow e^{-i\chi_{B}\chi^{-1}_{A}\Omega_{A} \,\tau_{A}+i\chi_{B}\chi^{-1}_{A}\Omega_{A}\int_{\mathbf{x}_{A}}^{\mathbf{x}_{B}}\varkappa_{(\mathbf{x})}\hat{\mathbf{k}}_{1}\cdot\dd x}\hat{a}^{\dag}_{(\chi_{B}\chi^{-1}_{A}\Omega_{A},\hat{\mathbf{K}}_{1})}. 
\end{equation}
Putting these together, Alice creation operator $\hat{a}_{\psi_{A}}$ at Bob frame is
\begin{align}\label{c50}
\hat{a}_{\psi_{A,B}}^{\dag}=\int_{{0}}^{+\infty}\frac{\dd\Omega_{A}}{\Omega_{A}} 
\bigl(\psi(\Omega_{A}) e^{-i\beta\Omega_{A}\tau_{B}+i\beta\Omega_{A}\int_{\mathbf{x}_{A}}^{\mathbf{x}_{B}}\varkappa_{(\mathbf{x})}\hat{\mathbf{k}}_{1}\cdot\dd x}\hat{a}^{\dag}_{(\beta\Omega_{A},\hat{\mathbf{K}}_{1})} \bigl),
\end{align}
where the combination of redshift factors $\chi_{B}\,\chi^{-1}_{A}$ at Alice and Bob frame is denoted by $\beta$. Alice's photon state in Bob's frame is
\begin{equation}\label{c51}
    \ket{\psi_{A}}_{B}=\hat{a}^{\dag}_{\psi_{A,B}}\ket{0}.
\end{equation}
\subsection{Bob's Measurements, Fidelity, Distortion, and  Purity}
To investigate variation in Alice's photon state Bob must compare it with his own photon wave function, generated by his photon source
\begin{align}\label{c52}
\ket{\psi_{B}}_{B}=\hat{a}_{\psi_{B,B}}^{\dag}\ket{0} =\int_{{0}}^{+\infty}\,\frac{\dd\Omega_{B}}{\Omega_{B}} \, 
\left(\psi(\Omega_{B})e^{-i\Omega_{B} \,\tau_{B}}\,\hat{a}^{\dag}_{(\Omega_{B},\hat{\mathbf{K}}_{1})} \right)\ket{0}.
\end{align}
Considering overlap of two state $\ket{\psi}$ and $\ket{\phi}$ definition $\Delta={\bra{\phi}\ket{\psi}}$, the overlap of $\{\ket{\psi_{A}}_{B},\ket{\psi_{B}}_{B}\}$ is 
\begin{align}\label{c53}
\Delta={{_B}\bra{\psi_{B}}\ket{\psi_{A}}_{B}}={\int_{{0}}^{+\infty}\,\frac{\dd\Omega_{}}{\Omega_{}} \, 
\bar{\psi}(\beta\,\Omega_{})\,\psi(\Omega_{}) \,e^{i\beta\,\Omega_{}\int_{\mathbf{x}_{A}}^{\mathbf{x}_{B}}\varkappa_{(\mathbf{x})}\hat{\mathbf{k}}_{1}\cdot\dd x}} .
\end{align}
To drive this relation we have used the commutation relations. Based on~\cite{bialynicki2020three} the standard definition of fidelity in momentum space, $\mathcal{F}_{m}$, applied to two pure states of the photon in momentum space is
\begin{equation}\label{c54}
   \mathcal{F}_{m}=\frac{\abs{\bra{\phi}\ket{\psi}}^{2}}{\bra{\phi}\ket{\phi}\bra{\psi}\ket{\psi}} =\abs{\Delta}^{2}.
\end{equation}
By applying this formula to Bob's measurement results we get
\begin{align}\label{c55}
   \mathcal{F}_{m}= \abs{{_{B}}\bra{\psi_{B}}\ket{\psi_{A}}_{B}}^{2}=\abs{\int_{{0}}^{+\infty}\,\frac{\dd\Omega_{}}{\Omega_{}} 
\bar{\psi}(\beta\,\Omega_{})\,\psi(\Omega_{}) \,e^{i\beta\,\Omega_{}\int_{\mathbf{x}_{A}}^{\mathbf{x}_{B}}\varkappa_{(\mathbf{x})}\hat{\mathbf{k}}_{1}\cdot\dd x}}^{2}.
\end{align}
As Fidelity is the measure of the similarity of two states, this equation shows the distortion of a photon by gravitational field as sent from Alice to Bob. It is important to note that the distinct influence of gravitational field on the Alice and Bob sources and photon-gravity coupling, both change the shape of the wave function of Alice's photon as it is sent from Alice to Bob relative to Bob's photon wave function. This motivates us to make our analysis more systematic, we define the following types of distortions:\\
\\
\textit{\textbf{Active Distortion (Phase Distortion):} Distortion due to the photon-gravity coupling as it propagates in spacetime.}\\
\\
\textit{\textbf{Passive Distortion (Shift Distortion):} Distortion due to the distinct effects of the gravitational field on the sources of the photons.}
\\

The main conclusion is that the total gravitational distortion of photons arises from a \textit{nonlinear} combination of active and passive distortions.

It is interesting like~\cite{bruschi2021spacetime} to ask: \textit{does the coherence of a pure state change due to gravity?} 
To answer this question we consider the purity $\gamma$ of a quantum state $\hat{\rho}$ in momentum space, $ \gamma_{m}(\hat{\rho}_{m})=\text{Tr}[\hat{\rho}_{m}^2]$, which is limited to $0\leq \gamma_{m}\leq 1$, and $\gamma_m=1$ for pure states. To calculate the trace, we use the definition $ \text{Tr}\hat{\mathcal{A}}=\int \frac{\dd^3{\mathbf{K}}}{K} \bra{K} \hat{\mathcal{A}}\ket{K}$, where $\ket{K}= \hat{a}^{\dag}_{(\Omega,\hat{\mathbf{K}})}\ket{0}$. By use of commutation relations and Alice's photon state in his frame and Bob's frame, we get:
\begin{equation}
    \gamma_{m}\bigl(\ket{\psi_A}_{A}\bigl)=1\, , \qquad   \gamma_{m}\bigl(\ket{\psi_A}_{B}\bigl)=1\, .
\end{equation}
Therefore, a pure photon remains pure as it propagates through the gravitational field from Alice to Bob, and active and passive distortions do not affect its purity.

\subsection{ Fidelity for Gaussian-like Photons and Coherent Photons}
We consider the Gaussian-like waveform, $\psi_{G(\Omega)}= \frac{1}{\sqrt{\sqrt{2\pi}\sigma}} \sqrt{\Omega} \exp{-\frac{(\Omega-\Omega_{0})^{2}}{4\sigma^{2}}},$ and suppose $\Omega_{0}\gg \sigma$, therefore, we can replace $\int_{0}^{\infty}$ by $\int_{-\infty}^{+\infty}$ in integral. Therefore, we have, 
\begin{align}\label{c61}
 \Delta={{_B}\bra{\psi_{G\,B}}\ket{\psi_{G\,A}}_{B}}= \sqrt{\frac{2\beta}{1+\beta^{2}}} \, e^{-\frac{(1-\beta)^{2}}{(1+\beta^{2})}\frac{\Omega_{0}^{2}}{4\sigma^{2}}-\frac{\beta^{2}}{1+\beta^{2}}\sigma^{2}\delta t^{2}} \, 
e^{i\frac{(1+\beta)\beta}{(1+\beta^{2})}\Omega_{0}\delta t},
\end{align}
where $\delta t=\int_{\mathbf{x}_{A}}^{\mathbf{x}_{B}}\varkappa_{(\mathbf{x})}\hat{\mathbf{k}}_{1}\cdot\dd x$, is the time delay between instant Alice send photon and instant it received by Bob (see Appendix.~\ref{PT}). Fidelity for a Gaussian-like photon is,
\begin{equation}\label{c62}
    \mathcal{F}_{m}= \left({\frac{2\beta}{1+\beta^{2}}}\right)\, e^{-\frac{(1-\beta)^{2}}{(1+\beta^{2})}\frac{\Omega_{0}^{2}}{2\sigma^{2}}-\frac{2\beta^{2}}{1+\beta^{2}}\sigma^{2}\delta t^{2}}\, .  
\end{equation}
The fidelity of a coherent photon (see Appendix.~\ref{coh}) that is constructed by Gaussian-like waveform with average $\overline{N}$ photons, is
\begin{align}\label{c63}
    \mathcal{F}_{c}=\exp\bigl\{-2\overline{N}\bigl[1-\,(\frac{2\beta}{1+\beta^{2}})^{\frac{1}{2}} \, \cos{\bigl(\frac{(1+\beta)\beta}{(1+\beta^{2})}\Omega_{0}\delta t\bigl)}  e^{-\frac{(1-\beta)^{2}}{(1+\beta^{2})}\frac{\Omega_{0}^{2}}{4\sigma^{2}}-\frac{\beta^{2}}{(1+\beta^{2})}\sigma^{2}\delta t^{2}} \, \bigl]\bigl\}\,.
\end{align}
It is important to note that, \textit{distinct effects of the gravitational field on sources and detectors} causes asymmetry between Fidelity as well as phase difference of photon wave functions ($\varphi$ in $\Delta=\abs{\Delta}\, e^{i\varphi}$) when Alice and Bob change their roles as sender and detector while their positions are prefixed. Therefore, Fidelity (distortion) as well as the phase difference are observer-dependent concepts.

As we limited our analysis to momentum space, we calculated momentum space fidelity but in principle, it is also possible to calculate position space fidelity $\mathcal{F}_{p}$ which is distinct from momentum space fidelity $\mathcal{F}_{m}$~\cite{bialynicki2020three}.

It is enlightening to mention the active distortion (phase distortion) in the Schwarzschild gravitational field has recently been calculated perturbatively and in the null Fermi coordinate of radiation in~\cite{exirifard2021towards, exirifard2022gravitational}, by analyzing Klein-Gordon and electromagnetic vector potential models of photon and passive distortion (shift distortion) in the Schwarzschild gravitational field has been investigated and discussed in detail in ~\cite{rodriguez2023introduction, bruschi2014spacetime, bruschi2021spacetime, bruschi2023gravitational}, by analyzing Klein-Gordon toy model of photon in (1+1) spacetime.

\subsection{Gravitational Redshift as a Shift in the Sharp Frequencies of the Spectrum}

In a series of papers~\cite{rodriguez2023introduction, bruschi2014spacetime, bruschi2021spacetime, bruschi2023gravitational}, it is tried to analyze the influence of gravitational redshift on photons (Passive Distortion) based on the Klein-Gordon model of photons. Some interesting conclusions were drawn, especially pointed out: "It is not possible to obtain gravitational redshift in the form of a linear shift of the spectrum of sharp frequencies $\omega$ as a unitary operation acting on the set of field modes $\{\hat{a}_{\omega}\}$ alone. This result corroborates the claim that the gravitational redshift cannot be interpreted simply as a shift in the sharp frequencies of
the photons for all frequencies of the spectrum"~\cite{rodriguez2023introduction, bruschi2023gravitational}. Consequently, they try to solve this conundrum by introducing realistic photons and investigating the physical effects that are caused by gravitational redshift.
\\

Indeed, based on our formalism (QED in CST),\textit{ We have shown gravitational redshift manifests as a linear shift of the spectrum of sharp frequencies $\omega$, which can be represented as a unitary operation $\{\hat{U}_{(\chi)}\}$ acting on the set of field modes $\{\hat{a}_{\omega}\}$. This result proves the claim that the gravitational redshift is a shift in the sharp frequencies of the photons for all frequencies of the spectrum.}
\\

In this work, we take a different point of view from that of~\cite{rodriguez2023introduction, bruschi2014spacetime, bruschi2021spacetime, bruschi2023gravitational}. In this sense, since we focus on the local gravitational influence at Alice's and Bob's locations, we do not trace the evolution of the wave packet as done in the literature~\cite{rodriguez2023introduction, bruschi2014spacetime, bruschi2021spacetime, bruschi2023gravitational}. Our sharp-frequency operators are transformed unitarily under the effects of redshift because of the different commutation relations (appearance of $\omega$ or $\Omega$ next to Dirac delta functions) and normalization condition of the wave function in momentum space (appearance of $\omega^{-1}$ or $\Omega^{-1}$ in integrals). It is important to note these factors were introduced to make the inner product, Lorentz invariant in flat spacetime~\cite{bialynicki2020three}. Therefore, although~\cite{rodriguez2023introduction, bruschi2014spacetime, bruschi2021spacetime, bruschi2023gravitational} analysis of passive distortion seemingly gives different results, according to the fact that all the effects are mapped in the transformation of the photon profiles, their work will give identical physically measurable quantities as ours in related limit.

\section{Single Photon Interferometry in Schwarzschild Spacetime}\label{VIII}

\begin{figure}[h!tbp]
    \centering
    \includegraphics[scale=0.4]{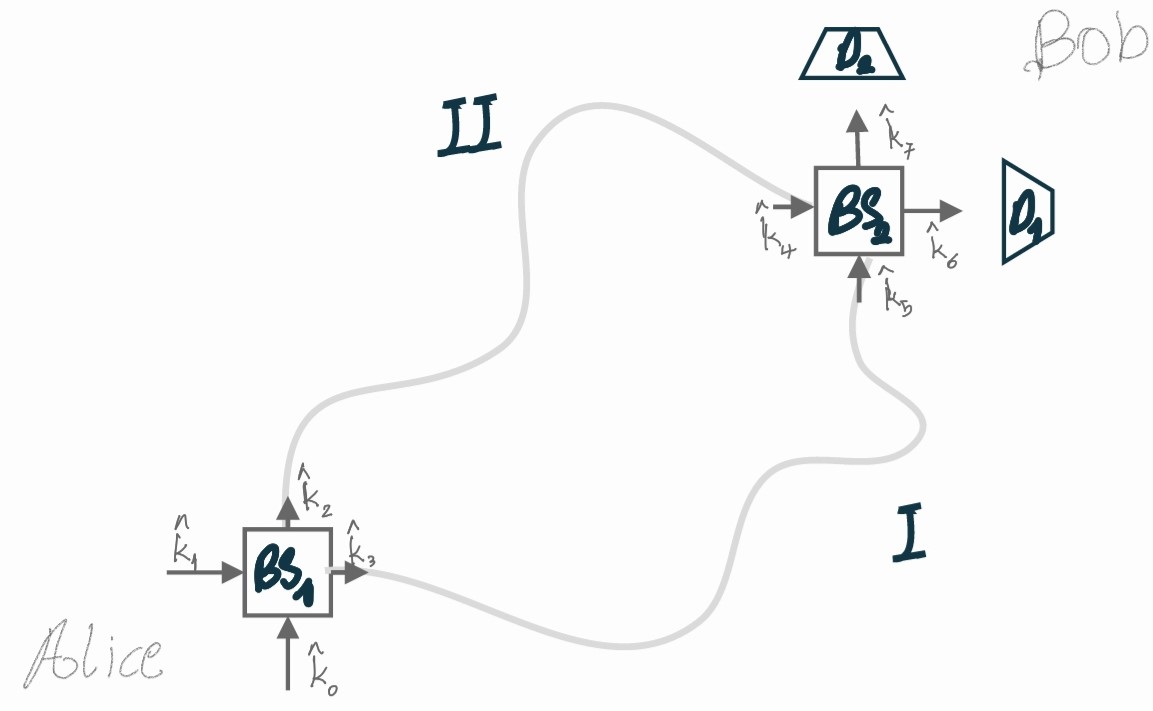}\caption{Schematic diagram of general Mach-Zehnder Interferometer in Schwarzschild spacetime.}
\end{figure}

This section analyzes the general form of the Interferometry experiment in Schwarzschild spacetime within the context of QED in CST. Special cases are proposed and analyzed by heuristic methods in~\cite{tanaka1983detect, zych2012general, hilweg2017gravitationally, Post-Newtonian, cohen1993standard} for the cases where gravitational field and its variation in the interferometer scale is weak, and by Gupta-Bleuler quantization method of optical
fibers in \underline{constant} gravitational potential in~\cite{Mieling}. We consider the following setup Alice and Bob as fundamental observers where Alice generates photons in her rest frame and sends them through a Mach-Zender interferometer to Bob. Bob uses her detectors to observe the interferometer pattern. As the photon in the Mach-Zender interferometer travels along the superposition of paths and experiences different gravitational fields in each path, it is interesting to study the influence of the gravitational field on quantum measurements of Bob and quantum properties like superposition.


\subsection{Propagation of Photon through Interferometer from Alice to Bob}
Alice in her rest frame can generate the wave function of photons by use of $\hat{a}^{\dag}_{\psi}$. For simplicity, we suppose her photons have positive helicity, therefore the creation operator for this wave function is
\begin{equation}\label{c64}
    \hat{a}^{\dag}_{\psi}= \int \frac{\dd^3{\mathbf{K}}}{K}\, 
\left(\psi(\Omega,\mathbf{K})\hat{a}^{\dag}_{+(\Omega,\hat{\mathbf{K}})}\right)  .
\end{equation}
The normalization condition for $\psi(\Omega,\mathbf{K})$ is, $\int\frac{\dd^{3}\mathbf{K} }{K} \abs{  \psi(\Omega,\mathbf{K})}^{2}=1$. To analyze the Bob measurements, we propagate creation operator $ \hat{a}^{\dag}_{\psi}$ in the interferometer. For simplicity, we strongly confine the wave function profile $\psi(\Omega,\mathbf{K})$ in the direction of propagation, therefore, we can write $\psi(\Omega,\mathbf{K})=\psi(\Omega)$, and ignore integration over $\mathbf{\hat{K}}$, we also omit $+$ sign as they do not have any effect in our final result, therefore,
\begin{equation}\label{c66}
    \hat{a}^{\dag}_{\psi}= \int \frac{\dd{\Omega_{A}}}{\Omega_{A}}\, 
\left(\psi (\Omega_{A})\hat{a}^{\dag}_{(\Omega_{A},\hat{\mathbf{K}}_{1})}\right).
\end{equation}
To analyze the effect of Alice's beam splitters in the setup on the photon we follow the same approach as~\cite{gerry2005introductory} and use relations 
\begin{align}\label{c67}
\hat{a}^{\dag}_{(\Omega_{A},\hat{\mathbf{K}}_{0})}\xrightarrow{BS_{1}} \frac{1}{\sqrt{2}}(i\hat{a}^{\dag}_{(\Omega_{A},\hat{\mathbf{K}}_{3})}+\hat{a}^{\dag}_{(\Omega_{A},\hat{\mathbf{K}}_{2})})\, , \\ \hat{a}^{\dag}_{(\Omega_{A},\hat{\mathbf{K}}_{1})}\xrightarrow{BS_{1}} \frac{1}{\sqrt{2}}(i\hat{a}^{\dag}_{(\Omega_{A},\hat{\mathbf{K}}_{2})}+\hat{a}^{\dag}_{(\Omega_{A},\hat{\mathbf{K}}_{3})})\, .
\end{align}
By use of redshift relation $\Omega_{A}=\chi_{A}\,\omega$ and operation of redshift operators $\hat{a}^{\dag}_{(\omega_{},\hat{\mathbf{k}}_{})}=\hat{U}_{(\chi_{A})} \hat{a}^{\dag}_{(\Omega_{A},\hat{\mathbf{K}}_{})} \hat{U}^{\dag}_{(\chi_{A})}=\hat{a}^{\dag}_{(\chi^{-1}_{A}\Omega_{A},\hat{\mathbf{k}}_{})}$, we write creation operators in isotropic coordinate. As the creation operators propagate in spacetime, they get phase and become 
\begin{equation}\label{c68}
 \rightarrow{}  e^{-i\chi^{-1}_{A}\Omega_{A} \,t+i\chi^{-1}_{A}\Omega_{A}\int_{s_{{A}}}^{s_{{B}}}\varkappa_{(\mathbf{x}(s))}\dd s}\hat{a}^{\dag}_{(\chi^{-1}_{A}\Omega_{A},\hat{\mathbf{k}}_{})} \, ,
\end{equation}
where the integration is over path $I$ or $II$ and $s$ is a parametrization of the path, change in $\hat{\mathbf{k}}$ as it propagates in spacetime can be by mirrors or due to confinement in the optical fiber. At Bob's location to obtain what are these creations in Bob Lab, we use redshift relation $\hat{a}^{\dag}_{\psi_{A}}$, and operate redshift operators $\hat{a}^{\dag}_{(\Omega_{B},\hat{\mathbf{K}}_{1})}=\hat{U}^{\dag}_{(\chi_{B})} \hat{a}^{\dag}_{(\omega_{},\hat{\mathbf{k}}_{1})} \hat{U}_{(\chi_{B})}=\hat{a}^{\dag}_{(\chi_{B}\omega_{},\hat{\mathbf{k}}_{1})}$, we obtain  Alice creation operators in Bob frame as
\begin{equation}\label{c69}
 \rightarrow{}  e^{-i\beta\,\Omega_{A} \,\tau_{B}\,+\,i\,\beta\,\Omega_{A}\int_{s_{{A}}}^{s_{{B}}}\varkappa_{(\mathbf{x}(s))}\dd s}\hat{a}^{\dag}_{(\beta\,\Omega_{A},\hat{\mathbf{K}})},
\end{equation}
where $\beta=\chi_{B}\,\chi^{-1}_{A}$ is the total redshift factor from Alice to Bob. The effect of Bob's beam splitter in the creation operators are
\begin{align}\label{c70}
\hat{a}^{\dag}_{(\Omega_{B},\hat{\mathbf{K}}_{4})}\xrightarrow{BS_{2}} \frac{1}{\sqrt{2}}(i\hat{a}^{\dag}_{(\Omega_{B},\hat{\mathbf{K}}_{7})}+\hat{a}^{\dag}_{(\Omega_{B},\hat{\mathbf{K}}_{6})})\, ,\\ \hat{a}^{\dag}_{(\Omega_{B},\hat{\mathbf{K}}_{5})}\xrightarrow{BS_{2}} \frac{1}{\sqrt{2}}(i\hat{a}^{\dag}_{(\Omega_{B},\hat{\mathbf{K}}_{6})}+\hat{a}^{\dag}_{(\Omega_{B},\hat{\mathbf{K}}_{7})})\, .
\end{align}
Putting all these things together, Alice's creation operators transform to
\begin{align}\label{c71}
\hat{a}^{\dag}_{(\Omega_{A},\hat{\mathbf{K}}_{0})}\rightarrow \frac{1}{2} \bigl[  e^{-i\Phi_{I}}(e^{-i\Phi_{II}+i\Phi_{I}}-1)\hat{a}^{\dag}_{(\beta\Omega_{A},\hat{\mathbf{K}}_{6})} +i\,e^{-i\Phi_{I}}(e^{-i\Phi_{II}+i\Phi_{I}}+1)\hat{a}^{\dag}_{(\beta\Omega_{A},\hat{\mathbf{K}}_{7})} \bigl] e^{-i\beta\Omega_{A}\tau_{B}},
\end{align}
\begin{align}\label{c72}
\hat{a}^{\dag}_{(\Omega_{A},\hat{\mathbf{K}}_{1})}\rightarrow\frac{1}{2} \bigl[e^{i\Phi_{I}}(1-e^{i\Phi_{II}-i\Phi_{I}})\hat{a}^{\dag}_{(\beta\Omega_{A},\hat{\mathbf{K}}_{7})} +ie^{i\Phi_{I}}(1+e^{i\Phi_{II}-i\Phi_{I}})\hat{a}^{\dag}_{(\beta\Omega_{A},\hat{\mathbf{K}}_{6})} \bigl] e^{-i\beta\Omega_{A}\tau_{B}},
\end{align}
at Bob's frame. Alice's photon creation operator for $\mathbf{K}_{1}$ input of the interferometer at the location of detectors is
\begin{align}\label{c73}
    \hat{a}^{\dag}_{\psi_{A,B}}=\int \frac{\dd{\Omega_{A}}}{\Omega_{A}}\, 
\bigl[\psi (\Omega_{A})\frac{e^{i\Phi_{I}}}{2} \bigl( (1-e^{i\Phi_{II}-i\Phi_{I}})\hat{a}^{\dag}_{(\beta\Omega_{A},\hat{\mathbf{K}}_{7})}  +i(1+e^{i\Phi_{II}-i\Phi_{I}})\hat{a}^{\dag}_{(\beta\Omega_{A},\hat{\mathbf{K}}_{6})} \bigl)\bigl] e^{-i\beta\Omega_{A}\tau_{B}} ,
\end{align}
where $\Phi_{X}=\beta\int_{s_{{A}}}^{s_{{B}}}\varkappa_{(\mathbf{x}(s))}\dd s,$ is the phase for the photon in path $X\in\{I,II\}$ times redshift factors. By propagating Alice's photon through the interferometer to Bob, $\ket{\Psi_{i}}=  \hat{a}^{\dag}_{\psi_{A}}\ket{0}\rightarrow  \ket{\Psi_{f}}=\hat{a}^{\dag}_{\psi_{A,B}}\ket{0}$, The final state is given by
\begin{align}\label{c76}
\ket{\Psi_{f}}&=\hat{a}^{\dag}_{\psi_{A,B}}\ket{0}\notag\\&=\int \frac{\dd{\Omega_{A}}}{\Omega_{A}}\, 
\bigl[\psi (\Omega_{A})\frac{e^{i\Phi_{I}}}{2} \bigl( (1-e^{i\Phi_{II}-i\Phi_{I}})\hat{a}^{\dag}_{(\beta\Omega_{A},\hat{\mathbf{K}}_{7})}+i(1+e^{i\Phi_{II}-i\Phi_{I}})\hat{a}^{\dag}_{(\beta\Omega_{A},\hat{\mathbf{K}}_{6})} \bigl)\bigl] e^{-i\beta\Omega_{A}\tau_{B}} \ket{0}.
\end{align}


\subsection{Bob's Interferometric Measurements}
 To obtain Bob's detector measurement results, we consider the photon number operators~\cite{bialynicki2013role} as follows
\begin{align}\label{c77}
    \hat{N}_{D_1}=\int_{0}^{+\infty}\frac{\dd\Omega_{B}}{\Omega_{B}}\hat{a}^{\dag}_{(\Omega_{B},\hat{\mathbf{K}}_{6})}\hat{a}_{(\Omega_{B},\hat{\mathbf{K}}_{6})} ,\notag \\  \hat{N}_{D_2}=\int_{0}^{+\infty}\frac{\dd\Omega_{B}}{\Omega_{B}}\hat{a}^{\dag}_{(\Omega_{B},\hat{\mathbf{K}}_{7})}\hat{a}_{(\Omega_{B},\hat{\mathbf{K}}_{7})}.
\end{align}
which count the photons received by detectors $D_{1}$ and $D_{2}$, and $\hat{a}_{(\Omega_{B},\hat{\mathbf{K}}_{})} $ are annihilation operators for Bob detectors. By calculation of the number operators expectation value for the state $ \ket{\Psi_{f}}$,
\begin{align}\label{c78}
P_{D_1/D_2}=\bra{\Psi_{f}}\hat{N}_{D_1/D_2}\ket{\Psi_{f}}=\int_{{0}}^{+\infty}\,\frac{\dd \Omega_{}}{\Omega_{}}\,\abs{\psi(\Omega_{})}^{2}\,\frac{1}{4} \abs{1\mp e^{i\Delta\Phi}},
\end{align}
where $\Delta\Phi=\Phi_{II}-\Phi_{I}$ is the phase difference between the upper path and lower paths of the interferometer times redshift factors.
By use of trigonometry identities and the normalization condition, we can write
\begin{equation}\label{c79}
P_{D_1/D_2}=\frac{1}{2} \left(1 \pm\int_{{0}}^{+\infty}\,\frac{\dd \Omega}{\Omega}\, \abs{\psi(\Omega)}^{2}\,\cos{(\Delta\Phi)}\right)\,,
\end{equation}
where $\Delta\Phi=\beta\,\Omega\oint n(\mathbf{x(s)})\dd s $. It also can be rewritten based on the time difference (Appendix.~\ref{PT}) of interferometer paths as \begin{equation}\label{c81}
P_{D_1/D_2}=\frac{1}{2} \left(1 \pm\int_{{0}}^{+\infty}\,\frac{\dd \Omega}{\Omega}\, \abs{\psi(\Omega)}^{2}\,\cos{(\beta\,\Omega\,\Delta t)}\right),
\end{equation}
where $\Delta t=\oint n(\mathbf{x(s)})\dd s$.
It is convenient to define visibility as
\begin{equation}\label{c82}
    \mathcal{V}=\frac{Max_{_{\psi}}\{P_{D_1/D_2}\}-Min_{_{\psi}}\{P_{D_1/D_2}\}}{Max_{_{\psi}}\{P_{D_1/D_2}\}+Min_{_{\psi}}\{P_{D_1/D_2}\}}.
\end{equation}
It is possible to quantify the which-way information by use of $\mathcal{D}$ which is a measure of distinguishability of the ways and satisfies the complementarity principle (wave-particle duality) relation~\cite{zych2012general,PhysRevLett.77.2154}:
\begin{equation}\label{c83}
    \mathcal{D}^2+\mathcal{V}^2=1\,.
\end{equation}
Based on this relation as the visibility decreases due to gravity, the distinguishability of the paths increases. Therefore, the superposition of the photon in two paths decreases, and it will be erased whenever $ \mathcal{V}=0$, in this case, the photon wave function collapses to one of the two paths.


\subsection{Interferometry with Gaussian-like photon}
To make the analysis more precise, we consider a photon with Gaussian-like waveform,
$\psi_{G(\Omega)}= \frac{1}{\sqrt{\sqrt{2\pi}\sigma}} \sqrt{\Omega} \exp{-\frac{(\Omega-\Omega_{0})^{2}}{4\sigma^{2}}}$. We suppose $\Omega_{0}\gg \sigma$, therefore, we can replace $\int_{0}^{\infty}$ by $\int_{-\infty}^{+\infty}$ in integral and compute $P_{D_1/D_2}$ 
\begin{equation} \label{c85}
P_{D_1/D_2}=\frac{1}{2} \left[1 \pm\, \mathcal{V} \, 
\cos\left({\Omega_{0}\,\beta\,\oint \varkappa_{(s)}\dd s}\right)\right],
\end{equation}
where, $\mathcal{V}=e^{-\frac{1}{2}(\sigma\,\beta\,\oint \varkappa_{(s)}\dd s)^{2}} ,$ 
which shows that the gravitational field causes decoherence in interferometric patterns of photons. We rewrite this equation based on the time difference,
\begin{equation}\label{c87}
P_{D_1/D_2}=\frac{1}{2} \left[1 \pm\, \mathcal{V} \,  \, 
\cos\left(\Omega_{0}\,\beta\,\Delta t \right)\right],
\end{equation}
where, $\mathcal{V}=e^{-\frac{1}{2}(\sigma\,\beta\,\Delta t )^{2}}$. For the case, $\,\beta\Delta t\gg \sigma^{-1} $ the visibility becomes zero and the interference pattern is lost. In this case, due to the complementarity principle relation~\eqref{c83}, $ \mathcal{D}=1$, and the path of the photon became specified by the time it received to the detectors~\cite{Zych2016}. In this case, the photon which is in the superposition of two paths decoheres due to the influence of gravitational field.

\subsection{Reduction of Visibility and Gravitational decoherence}
To investigate the visibility and decoherence more precisely, we consider two setups for the interferometry. The first setup is similar to Tanaka's proposal~\cite{tanaka1983detect} in which the gravitational field is the same in Alice and Bob's frames and the second setup is similar to Zych's proposal~\cite{zych2012general} in which Alice and Bob are at different gravitational fields in all other aspects, the interferometers are equivalent in position and configuration relative to the gravitational field source.
Therefore, $\Delta t$ is the same for both cases, while $\beta=\chi_{B}\chi_{A}^{-1}=1$ in Tanaka's case. For a Gaussian-like photon, we have,
\begin{align} \label{c91}
P_{D_1/D_2}^{Tanaka}&=\frac{1}{2} \left[1 \pm\, e^{-\frac{1}{2}(\sigma\,\Delta t )^{2}} \,  \, 
\cos\left(\Omega_{0}\,\Delta t \right)\right]\, ,
\\ \label{c92}
P_{D_1/D_2}^{Zych}&=\frac{1}{2} \left[1 \pm\, e^{-\frac{1}{2}(\sigma\,\beta\,\Delta t )^{2}} \,  \, 
\cos\left(\Omega_{0}\,\beta\,\Delta t \right)\right].
\end{align}
Apart from the difference in the interferometric pattern due to the absence and presence of redshift factor. In Tanaka's case, decoherence happens only due to the photon-gravity coupling, while, in Zych's setup, the gravitational redshift factor also appears in visibility relation, therefore, it influences decoherence. These motivate us to define two types of decoherence due to the gravitational field:\\
\\
\textit{\textbf{Active Gravitational Decoherence (Phase Decoherence):} Decoherence due to photon-gravity coupling as it propagates in spacetime.}
\\
\\
\textit{\textbf{Passive Gravitational Decoherence (Shift Decoherence):} Decoherence due to the distinct effects of the gravitational field on the photons sources and photons detectors.}
\\

The main conclusion is that the total gravitational decoherence of photon arises from a \textit{nonlinear} combination of active and passive decoherence. Furthermore, in general case \textit{distinct effects of the gravitational field on source and detectors} influences the interference pattern (phase difference) and visibility:
\begin{itemize}
    \item In the case of blueshift from Alice to Bob ($\beta=\chi_{B}\chi^{-1}_{A}>1$), it increases the phase difference and reduces visibility. 

    \item In the case of redshift from Alice to Bob ($\beta=\chi_{B}\chi^{-1}_{A}<1$), it decreases phase difference and increases visibility. 
\end{itemize}
Therefore, it causes asymmetry between interference patterns and visibilities when Alice and Bob change their roles as sender and receiver while their position is prefixed in a Zych's like setup.

The key result is that phase difference, visibility, and distinguishability are observer-dependent concepts. It is crucial to note that the observer dependence of visibility implies the observer dependence of decoherence.


\subsection{Photon Interferometry in Weak Uniform Gravitational Field} \label{IWF}
Let us consider Zych's setup~\cite{zych2012general} in the gravitational field of the Earth, such that the variation of the gravitational acceleration in the interferometer scale is negligible. Vertical paths are equal to $h$ and the horizontal paths are equal to $l$ in the coordinate length. We consider $\mathbf{r}_0$ as the position vector of the center of the interferometer, and $\mathbf{X}$ is the position vector of the interferometer paths, therefore, it is possible to expand $\varkappa_{(\mathbf{x})}$ around $\mathbf{r}_0$ based on $\mathbf{X}$. By asymptotic expansion for large $r_0$ ($r_0\gg \frac{MG}{2}$) and taking into account the value of acceleration in the asymptotic region [Appendix~\ref{Acc}, Equation~\eqref{acc8}], we get:
\begin{equation}
 \varkappa_{(|\mathbf{r}_0 +\mathbf{X}|)}=1+\frac{2MG}{r_0}-2 g_{(r_{0})}\hat{r}_{0}\cdot\mathbf{X} +\cdots \, .
\end{equation} 
It is possible to choose a coordinate in such a form that $\hat{z}=\hat{r}_{0}$, consequently,
\begin{equation}
 \varkappa_{(|\mathbf{r}_0 +\mathbf{X}|)}=1+\frac{2MG}{r_0}-2 g_{(r_{0})} z +\cdots \, .
\end{equation}
A similar expansion procedure for redshift factor $\chi_{(\mathbf{x})}$ in the same conditions is possible and it will result in:
\begin{align}
    \chi_{(\mathbf{x})}=1+\frac{MG}{r_0}- g_{(r_{0})} z +\cdots \, ,\\  \chi_{(\mathbf{x})}^{-1}=1-\frac{MG}{r_0} + g_{(r_{0})} z +\cdots \, .
\end{align}
By calculating $ \Delta t=\oint \varkappa_{(\mathbf{x})}\,\dd s$, $\Delta \Phi=\Omega_0 \Delta t$ and $\beta=\chi^{-1}_{A}\chi_{B}$ for the setup, we get
\begin{align}\label{qq}
    &\Delta \Phi_{E}=-\frac{2\Omega_0 glh}{c^3}\sin{\alpha}\, ,\\\label{qq1}  &\Delta t=-\frac{2glh}{c^3}\sin{\alpha}\, ,\\\label{qq2} &\beta=\chi^{-1}_{A}\chi_{B}= 1-\frac{gh}{c^2}\sin{\alpha},
\end{align}
where $\alpha$ is the angle that the plane of the interferometer forms with the horizontal plane. Here we have added index $E$ to $\Delta\Phi$ to refer to the fact that it is computed in Einstein's theory of gravity. The same value for $\Delta\Phi_{E}$ for classical electromagnetic radiation was obtained earlier by Cohen and Mashhoon~\cite{cohen1993standard} by analyzing Maxwell's equation in Schwarzschild spacetime and deriving the index of refraction for the gravitational field. It is enlightening to mention as Cohen and Mashhoon have pointed out in~\cite{cohen1993standard}, their result is a classical, general relativistic effect in contrast with neutron interferometry relative phase shift in gravitational field (COW experiment) which is purely quantum mechanical. As our result for $\Delta \Phi_{E}$ is based on QED in curved spacetime it is the quantum-general relativistic analog of Cohen and Mashhoon's result. It is important to note the value of $\Delta\Phi_{E}$ is comparable to the value obtained by approaching $v\rightarrow c$ in the non-relativistic results of the well-known COW neutron interferometry in gravitational field experiment~\cite{colella1975observation, rauch2015neutron}:
\begin{equation}\label{qqq}
    \Delta \Phi_{\text{COW}}=-\frac{\Omega_0 glh}{v c^2}\sin{\alpha}\xrightarrow[]{v\rightarrow c}-\frac{\Omega_0 glh}{ c^3}\sin{\alpha}.
\end{equation}
The difference of a factor of two has the same origin as the differences in the bending of rays and particle trajectories in the exterior gravitational field of a massive body~\cite{cohen1993standard}. It is reminiscent of the fact that $\Delta \Phi_{\text{COW}}$ is the leading non-relativistic term of the relativistic relative phase shift $\Delta\Phi_{\text{neutron}}$,
\begin{equation}
   \Delta\Phi_{\text{COW}}=\lim_{\frac{v}{c} \to 0} \Delta \Phi_{\text{neutron}} \, ,
\end{equation}
where $\Delta\Phi_{\text{neutron}}$ is obtainable by considering Dirac equations in gravitational field~\cite{borde2001relativistic}. Therefore, equation~\eqref{qqq} will not give the correct result for the massless case due to,
\begin{align}
    \lim_{v\to c} \Delta \Phi_{\text{COW}}=\lim_{v\to c} \left(\lim_{\frac{v}{c} \to 0} \Delta \Phi_{\text{neutron}}\right)\neq \lim_{\substack{v\to c\\ \text{or}\\ m\to 0}} \Delta \Phi_{\text{neutron}} =\Delta \Phi_{E} \, .
\end{align}
After this digression regarding the origin of relative phase shift, we return to interferometry. For a photon in the optical regime $\Omega_{0}=10^{15}\, \text{s}^{-1}$ with bandwidth $\sigma=10^3\, \text{s}^{-1}$ and an interferometer with $h=1\,\text{m} $, $l=10^{5}\,\text{m}$, in the gravitational field of the Earth $g=9.8\, \text{m s}^{-2}$, the redshift influence (distinct effects of the gravitational field on sources and detectors) on the interferometry pattern is negligible ($\beta\simeq1-10^{-15}\approx 1$). Rewriting Equation~\eqref{c92} based on~\eqref{qq}--\eqref{qq2} and the above quantities, gives
\begin{align} \label{}
P_{D_1/D_2}^{Zych}\simeq\frac{1}{2} \bigl[1 \pm\, e^{-\frac{1}{2}\left(\frac{2\sigma glh}{c^3}\sin{\alpha}\right)^{2}} \cos\bigl(\frac{2\Omega_{0}glh}{c^3}\,\sin{\alpha}\bigl)\bigl].
\end{align}
In~\cite{zych2012general} Zych et al. analyzed the photon interferometry in the weak gravitational field as a photonic version of the "quantum interferometry of clocks" proposal~\cite{zych2011quantum} and some of its experimental outcomes are predicted. It is pointed out, that an equation similar to Equation~\eqref{c81} which is obtained by considering Klein-Gordon equation, has two limits that correspond to tests of two different effects (the relative phase shift that is predicted in their work is equal to $\lim_{v\to c}\Delta\Phi_{\text{COW}}$ and half the one we calculated here. In addition, the factor $\frac{1}{\Omega}$ in measure of integral and factor $\beta=\chi_{B}\chi^{-1}_{A}$ are missed in~\cite{zych2012general} results): 
\begin{itemize}
    \item Case I: When $\Delta t$ is smaller than the frequency-width of the photon wave function, the visibility is maximal. Due to the absence of time dilation in the Newtonian regime, this relative phase shift is explained based on the interaction of photon energy and gravitational field. Because of its similarity to gravitational redshift (Pound-Rebka wrong interpretation), its observation is considered as an experimental test of mass-energy equivalence, not as a test of the time dilation (Shapiro time delay) or as a general relativistic effect~\cite{zych2012general}.
    \\
    \item Case II: When $\Delta t$ is greater than the frequency-width of the photon wave function, due to the oscillating nature of the cosine function the integrated becomes zero, therefore the visibility becomes zero (interference pattern disappears). In this regime, the gravitational field decoheres the photon in a spatial superposition of two paths, and the interference pattern disappears. This is explainable based on gravitational time dilation for a single photon, and its observation is considered as a test of time dilation (Shapiro time delay) and a general relativistic effect~\cite{zych2012general}.
\end{itemize}
Based on our analysis, the relative phase shift is $ \Delta \Phi_{E}$ in Equation~\eqref{qq} and NOT $\lim_{v\to c}\Delta\Phi_{\text{COW}}$ as mentioned in~\cite{tanaka1983detect, zych2012general, hilweg2017gravitationally, Post-Newtonian}. Furthermore, gravitational active decoherence originates from the photon-gravity coupling, and their observations must be considered as a test of the coupling of the photon wave vector and the gravitational field and as a completely quantum-general relativistic effect. It is important to notice gravitational time delay (Shapiro time delay) of photon also originates from the photon-gravity coupling (see Appendix~\ref{PT}).

To make this statement about the quantum-general relativistic nature of photon-gravity coupling more well-established, we analyze the photon interferometry in Newtonian gravity in the next subsection in detail.


\subsection{Photon Interferometry in Newtonian Gravity}\label{ING}
In this section, we suppose gravity is Newtonian and try to couple it to photons. This premise conflicts with observations, especially light bending by the sun (see Appendix.~\ref{Def} for the calculations of light deflection in Einstein's gravity and Newtonian gravity). In light bending by the sun, photon rays follow a path in a local Newtonian gravitational field (see Appendix.~\ref{Acc} for the definition of the local Newtonian gravitational field), but its deflection angle is properly predicted by Einstein's theory of gravity, therefore the gravitational field is not Newtonian. However, this investigation can provide valuable information about the quantum-gravity interface. Especially, the status of Newtonian gravitation theory, and the Einstein equivalence principle in quantum regimes.

 Electromagnetism in contrast to Newtonian gravity is a relativistic theory (here by relativistic we mean special relativity), therefore it is not possible to couple electromagnetic field to Newtonian gravity in a consistent form (one of the main motivations of Einstein to develop the general theory of relativity was to construct a relativistic gravitation theory~\cite{einstein2003meaning, MTW}). To resolve this conundrum, we develop two schemes to couple electromagnetism with Newtonian gravity. The first scheme is based on the Newtonian limit of Einstein's gravity, and the second one is based on the Einstein equivalence principle.
 The important observable effect in photon interferometry is the relative phase shift. As we pointed out in~\ref{IWF} subsection, the gravitational redshift does not have any significant influence on the results of proposed experiments~\cite{hilweg2017gravitationally,zych2012general}, so we ignore it.
 \begin{itemize}
     \item Scheme I: Einstein's theory of gravity reduces to Newtonian gravity only if, in Newtonian coordinates, $g_{00}=-\bigl(1+2V(r)\bigl)$ where $V(r)=-\frac{MG}{r}$ is Newtonian potential~\cite{MTW}. Therefore, the Newtonian metric is:
\begin{equation}\label{n1}
    \dd s^{2}=-(1-\frac{2MG}{r})\dd t^{2} +(\dd x^{2}+\dd y^{2}+\dd z^{2})\, ,
\end{equation}
where $t$ is Newtonian time and $(x, y, z)$ are Newtonian coordinates~\cite{MTW}. It is tempting to repeat what we have done in previous sections for this metric and analyze the interferometry setup in the uniform gravitational field limit, which will result in
\begin{align}\label{n2}
    &\Delta \Phi_{N_I}=-\frac{\Omega_0 glh}{c^3}\sin{\alpha}\, , \\  &\Delta t_{N_I}=-\frac{glh}{c^3}\sin{\alpha}\, .
\end{align}
These calculations may be doubted due to the noticeability of relativistic effects for light, which is in contrast with one of the premises for derivation of~\eqref{n1} metric, which is "the special relativistic effects must be negligible"~\cite{MTW}. This scheme is the best method to couple Newtonian gravity in its full form with electromagnetism.

\item Scheme II: As the setup we discussed in subsection~\ref{IWF} is practically in a uniform gravitational field, by use of the Einstein equivalence principle it can be replaced with a congruence of uniformly accelerated frames in Minkowski spacetime with acceleration opposite and equal to gravitational acceleration. Therefore, the problem of the electromagnetic field in a uniform locally Newtonian gravitational field and its coupling to gravity by use of the Einstein equivalence principle are replaced by the problem of electromagnetic in Minkowski spacetime from the point of view of the uniform accelerated observers.  

It is possible to construct the Fermi coordinate $(T,X,Y,Z)$ for this congruence of observers~\cite{mashhoon2017nonlocal}, where $T$ is the proper time of one of the accelerated frames which is called reference observer and $(X,Y,Z)$ are geodesic coordinate constructed along reference observer world line. The background Minkowski metric in terms of this Fermi coordinate takes the following form~\cite{mashhoon2017nonlocal}:
\begin{equation}\label{n3}
    \dd s^{2}=-(1+\mathbf{g}_{(r_0)}\cdot\mathbf{X})^2\,\dd T^{2} +(\dd X^{2}+\dd Y^{2}+\dd Z^{2})
\end{equation}
It is important to note the run time of the experiment relative to the characteristic timescale of the Earth acceleration, $T_0=\frac{c}{g}=1$ Year, and the scale of the interferometer relative to the characteristic length scale of the Earth acceleration, $L_0=\frac{c^2}{g}=1$ Light year, are not noticeable. Now we can formulate electromagnetism based on observers' metric. As this metric is isotropic, we can repeat a similar step we did in previous sections and follow the same quantization procedure and all other analyses. For interferometry experiment which is performed in this uniform acceleration, it is found,
\begin{align}\label{n5}
    &\Delta \Phi_{{N_{II}}}=-\frac{\Omega_0 glh}{c^3}\sin{\alpha}\, , \\  &\Delta T_{{N_{II}}}=-\frac{glh}{c^3}\sin{\alpha}\, .
\end{align}
 \end{itemize}
 The values obtained by these two schemes,~\eqref{n2} and~\eqref{n5} are equal. Consequently, it seems reasonable for the relative interferometric phase shift of photon in Newtonian gravity to accept the value, 
\begin{equation}\label{n7}
   \Delta\Phi_{N}= -\frac{\Omega_0 glh}{c^3}\sin{\alpha}\,.
\end{equation}
It is interesting to note this value is the value obtained by taking the relativistic limit of $\Delta\Phi_{\text{COW}}$. There is the following relation between what we have calculated in subsection~\ref{IWF}, Equation~\eqref{qq}, based on QED in Einstein's theory of gravity and QED in Newtonian gravity
\begin{equation}\label{n8}
   \Delta\Phi_{E}= 2\Delta\Phi_{N}\,.
\end{equation}


\subsection{Photon Gravitational Interferometry and the Status of Newtonian Gravity and Einstein Equivalence Principle in Quantum Regime}

Single photon interferometry in the weak gravitational field is feasible with current-day technology~\cite{zych2012general,hilweg2017gravitationally} and it is going to be performed in the near future. This experiment will be the first experiment that probes the interface of general relativity and quantum and the second one that investigates the quantum-gravity interface (the first experiment was COW which investigated the influence of gravity on Neutron interferometric phase shift~\cite{colella1975observation}, this experiment is explainable either by Einstein equivalence principle or Newtonian gravity). It is interesting to discuss the physical implications of this experiment's possible results based on what we have done in this paper. 

Based on our investigation, QED in CST predicts $\Delta\Phi_{E}$ for the value of the interferometric phase shift for a photon in the Earth's gravitational field. On the other hand, coupling photon to Newton's gravity by Newtonian metric~\eqref{n2} or by the Einstein equivalence principle~\eqref{n5} results in $\Delta\Phi_{N}=\frac{1}{2} \Delta\Phi_{E}$.

Falsification of the prediction $\Delta\Phi_{N}$ will shield light in the quantum-gravity interface, especially, it will doubt the validity of Newtonian gravity in the quantum regime (even in the weak uniform gravitational field of the Earth) and doubt the validity of extension of the Einstein equivalence principle beyond a single point in the quantum regime (even for a uniform weak gravitational field). See Appendix.~\ref{EEP} for more discussion about the Einstein equivalence principle.

It is enlightening to mention that the observation of gravitational light bending by the sun was a crucial classical test of Einstein's theory of gravity on the astronomical scale and the classical regime falsification of Newtonian gravity in a region where the gravitational field is locally Newtonian, similarly, the observation of $\Delta\Phi_{E}$ in interferometric phase shift for a photon in the Earth's gravitational field is a crucial test of QED in CST in the Lab scale and the quantum regime falsification of Newtonian gravity in a region where the gravitational field is locally Newtonian and almost uniform.


\section{Discussion}\label{Disc}
Investigating the influence of the gravitational field on photons is an important subject matter in view of the fundamental nature of the quantum-gravity interface. In this paper, we have developed a consistent formulation of QED in CST. We have investigated in detail the influence of the Schwarzschild gravitational field on photons. We have shown the photon wave vector couples to the gravitational field (\textit{photon-gravity coupling}) and there is no coupling between the energy of the propagating photon and the static gravitational field. Furthermore, the claim that the gravitational redshift is a shift in the sharp frequencies of the photons for all frequencies of the spectrum is proved. Our analysis shows that observable effects can be considered as the \text{nonlinear} combination of active gravitational effects (the influence of the gravitational field on photons due to photon-gravity coupling) and passive gravitational effects (the influence of the gravitational field on sources and detectors of photons). 
We also calculated the proper value of the photon gravitational interferometric relative phase shift and demonstrated it is genuinely a quantum-general relativistic effect, and its observation even in a uniform locally Newtonian gravitational field is a test of QED in CST. The physical implications of the observation of photon gravitational interferometric relative phase shift for the structure of spacetime and the status of Newtonian gravitation theory and the Einstein equivalence principle were discussed in detail. The gravitational decoherence has been derived as a result of photon-gravity coupling, and it is shown it is observer-dependent due to passive gravitational effects. The derivation of the \textit{gravitational decoherence of quantum superpositions} for photons in the framework of quantum field theory in curved spacetime can be considered to be the proof of concept. Therefore, it is nearly impossible to argue that gravitational decoherence is not a fundamental property of the quantum-gravity interface, and it will be found as soon as the necessary experimental accuracy is achieved. See ~\cite{Bassi, question, replyto} for the ongoing debate about the interpretation and testability of similar types of phenomena, which were discussed in~\cite{pikovski2015universal, zych2011quantum}.

In this paper, we restricted the application of formalism to investigate single photons in Schwarzschild spacetime. The generalization to multi-photon is straightforward~\cite{Smith,Smith_2007}, however, an extension of Glauber’s measurement
theory in curved spacetime is required to analyze multi-photon measurements. photon-gravity coupling in Kerr spacetime is discussed in Appendix \ref{Kerr}.


\section*{Acknowledgments}

I am deeply appreciative of the discussions I had with Bahram Mashhoon and would like to express my gratitude to Shant Baghram and David Edward Bruschi for their meticulous review and insightful feedback on the manuscript. I am also grateful to Laleh Memarzadeh, Armin Ghazi, Arad Nasiri, and Sina Soltani for their insightful comments.


\appendix

\section{Generalized Poisson Brackets}\label{GPB}
The generalized Poisson brackets for~\eqref{b12} Hamiltonian system like~\cite{QED} are defined as
\begin{align}\label{b13}
    \{\mathcal{F},\mathcal{G}\}=\int \dd x^{3} \biggl[\fdv{\mathcal{F}}{\mathbf{D}(\mathbf{x})}\cdot \nabla\times\fdv{\mathcal{G}}{\mathbf{B}(\mathbf{x})}-\fdv{\mathcal{G}}{\mathbf{D}(\mathbf{x})}\cdot \nabla\times\fdv{\mathcal{F}}{\mathbf{B}(\mathbf{x})} \biggl]\, ,
\end{align}
where $\mathcal{F}$ and $\mathcal{G}$ are analytical functional in $\mathbf{D}(\mathbf{x})$ and $\mathbf{B}(\mathbf{x})$. Integrating by parts results in
\begin{align}\label{b14}
    \{\mathcal{F},\mathcal{G}\}=\int \dd x^{3} \biggl[ \fdv{\mathcal{G}}{\mathbf{B}(\mathbf{x})}\cdot \nabla\times\fdv{\mathcal{F}}{\mathbf{D}(\mathbf{x})}-\fdv{\mathcal{F}}{\mathbf{B}(\mathbf{x})}\cdot \nabla\times\fdv{\mathcal{G}}{\mathbf{D}(\mathbf{x})}\biggl]\, .
\end{align}
Generalized Poisson brackets satisfy~\cite{QED}
\begin{align}\label{b15}
    &\{\mathcal{F},\mathcal{G}\} =-\{\mathcal{G},\mathcal{F}\},\\
&\{\lambda_{1}\mathcal{F}_{1}+\lambda_{2}\mathcal{F}_{2},\mathcal{G}\}=\lambda_{1}\{\mathcal{F}_{1},\mathcal{G}\}+\lambda_{2}\{\mathcal{F}_{2},\mathcal{G}\},\\
    &\{\mathcal{F}_{1},\{\mathcal{F}_{2},\mathcal{F}_{3}\}\}+\{\mathcal{F}_{3},\{\mathcal{F}_{1},\mathcal{F}_{2}\}\}+\{\mathcal{F}_{2},\{\mathcal{F}_{3},\mathcal{F}_{1}\}\}=0, \\
    &\{\mathcal{F}_{1}\mathcal{F}_{2},\mathcal{G}\}=\mathcal{F}_{1}\{\mathcal{F}_{2},\mathcal{G}\}+\mathcal{F}_{2}\{\mathcal{F}_{1},\mathcal{G}\}\, .
\end{align}
whenever $\mathcal{F}$ and $\mathcal{G}$ have the following form
\begin{align}\label{b17}
    \mathcal{F}(\mathbf{x})=\int \dd z^{3} \mathbf{F}\bigl(\mathbf{D}(\mathbf{z}),\mathbf{B}(\mathbf{z})\bigl)\delta(\mathbf{z}-\mathbf{x}) \, ,\notag \\ \mathcal{G}(\mathbf{y})=\int \dd z^{3} \mathbf{G}\bigl(\mathbf{D}(\mathbf{z}),\mathbf{B}(\mathbf{z})\bigl)\delta(\mathbf{z}-\mathbf{y})\, ,
\end{align}
it is not hard to show
\begin{align}\label{b18}
    \{\mathbf{F}\bigl(\mathbf{D}(\mathbf{x}),\mathbf{B}(\mathbf{x})\bigl),\mathbf{G}\bigl(\mathbf{D}(\mathbf{y}),\mathbf{B}(\mathbf{y})\bigl)\}
    =\left[\pdv{\mathbf{F}(\mathbf{x})}{D_{i}(\mathbf{x})}\pdv{\mathbf{G}(\mathbf{y})}{B_{j}(\mathbf{y})}-\pdv{\mathbf{F}(\mathbf{x})}{B_{i}(\mathbf{x})}\pdv{\mathbf{G}(\mathbf{y})}{D_{j}(\mathbf{y})}\right]\epsilon_{ijk}\partial_{k}\delta(\mathbf{x}-\mathbf{y})\,.
\end{align}
\section{Polarization Basis}\label{Pol}
To fix the notation for the polarization of electromagnetic waves, consider a plane monochromatic electromagnetic wave of frequency $\omega$ propagating along the $\hat{k}$ direction. We choose $\hat{\ell}_{1}$ and $\hat{\ell}_{2}$ such that form a right-oriented orthonormal triad
\begin{equation}\label{P1}
\hat{\ell}_1 \times \hat{\ell}_2 =  \hat{\mathbf{k}}\, .
\end{equation}
An orthonormal basis is formed by $(\hat{\ell}_{1},\hat{\ell}_{2})$, for linear polarization of the electromagnetic wave propagation in $\hat{\mathbf{k}}$ direction. We can use them and construct a basis $( \hat{\ell}_{+}, \hat{\ell}_{-})$ for circular polarization, 
\begin{equation}\label{P2}
 \hat{\ell}_{\pm} = \hat{\ell}_1\pm i \hat{\ell}_2\,.
\end{equation}
It is important to note $( \hat{\ell}_{+}, \hat{\ell}_{-})$ is orthogonal basis under inner product $\hat{a}_{}\cdot\bar{\hat{b}}_{}$, where bar is complex conjugation. As $\hat{\ell}_{\pm}$ are complex conjugate of each other for convenience, we write $\hat{\ell}_{}=\hat{\ell}_{+}$.
There are the following relations between $\hat{\ell}(\hat{\mathbf{k}})$, $\hat{\mathbf{k}}$, and $\mathbf{k}$;
\begin{align}
    \mathbf{k}=\abs{k}\,\hat{\mathbf{k}}=\omega\, \hat{\mathbf{k}}\, ,\\  \hat{\ell}(\hat{\mathbf{k}})\cdot \bar{\hat{\ell}}(\hat{\mathbf{k}})=2\, ,\\ \hat{\ell}(\hat{\mathbf{k}})\cdot {\hat{\ell}}(\hat{\mathbf{k}})=0\, , \\ \hat{\ell}(\hat{\mathbf{k}})\cdot\hat{\mathbf{k}}=0 \, ,
\end{align}
\begin{equation}
    i\,\mathbf{k}\times\hat{\ell}(\hat{\mathbf{k}})=\abs{k}\, \hat{\ell}(\hat{\mathbf{k}})=\omega\, \hat{\ell}(\hat{\mathbf{k}}) \, .
\end{equation}
\section{Local Acceleration}\label{Acc}
Acceleration of observer with four-velocity, $u^{\alpha}$, is~\cite{MTW},
\begin{equation}\label{acc1}
    a^{\mu}=u^{\alpha}\nabla_{\alpha}u^{\mu}.
\end{equation}
By considering the Fundamental observer four-velocity, $u^{\nu}=e_{\hat{0}}  ^{\nu}$, the acceleration of fundamental frames in Schwarzschild metric is
\begin{equation}\label{acc2}
    a^{i}=\bigl(\frac{MG x^{i}}{r^{3}}\bigl)\bigl(1+\frac{MG}{2r}\bigl)^{-5}\bigl(1-\frac{MG}{2r}\bigl)^{-1}\,\, .
\end{equation}
Projecting the acceleration to the fundamental frame gives
 \begin{equation}\label{acc3}
    a^{\hat{i}}=\bigl(\frac{MG x^{i}}{r^{3}}\bigl)\bigl(1+\frac{MG}{2r}\bigl)^{-3}\bigl(1-\frac{MG}{2r}\bigl)^{-1}\,\,.
\end{equation}
The norm of this vector is 
\begin{equation}\label{acc4}
    g_{(r)}=\bigl(\frac{MG }{r^{2}}\bigl)\bigl(1+\frac{MG}{2r}\bigl)^{-3}\bigl(1-\frac{MG}{2r}\bigl)^{-1}\,\, .
\end{equation}
It is the gravitational acceleration that fundamental observers measure by their accelerometers.
 Asymptotic expansion for large $r$ gives
\begin{equation}\label{acc7}
     g_{(r)}=\frac{MG}{r^2}-\frac{(MG)^{2}}{r^3}+\frac{(MG)^{3}}{r^{4}}-\frac{3(MG)^{4}}{4\,r^{5}}+\cdots\,\, .
\end{equation}
For $r\gg \frac{MG}{2}$(asymptotic region or weak field region), the gravitational acceleration  is 
\begin{equation}\label{acc8}
     \mathbf{g}_{(r)}=-\frac{MG}{r^2}\hat{r}\, .
\end{equation}
Which is equal to its Newtonian gravity counterpart (see Equation~\eqref{acc9}).
This acceleration can be used to assign the following time and length scale to spacetime~\cite{mashhoon2017nonlocal}:
\begin{equation}\label{acc5}
L_{0}=\frac{c^{2}}{g}\, ,  \qquad   \qquad T_{0}=\frac{c}{g}\, .
\end{equation}
For static observers on the Earth's surface, $L_0\simeq1$ Light year and $t_0\simeq1$ year. We use these scales and assign the following threshold frequency and wavelength to
 a physical field in curved spacetime,
\begin{equation}\label{acc6}
\lambda_{0}=\frac{c^{2}}{g}\, ,  \qquad   \qquad \omega_{0}=\frac{2\pi\, g}{c}\, .
\end{equation}
A similar analysis for fundamental frames in Newtonian metric~\ref{ING} or just by the use of Newton's second law in the Newtonian gravitational field gives:
\begin{equation}\label{acc9}
     \mathbf{g}_{N(r)}=-\frac{MG}{r^2}\hat{r}\,\, .
\end{equation}
Comparison the value obtained in Equation~\eqref{acc8} for the value of acceleration in the weak field limit of Einstein's theory and the value obtained in Equation~\eqref{acc9} for the value of acceleration in Newtonian gravity, shows that in the weak field limit the Einstein gravity is locally Newtonian in the sense that by use of accelerometers which are static relative to the source in radius $r\gg \frac{MG}{2}$, we can not distinguish whether gravity is Einsteinian or Newtonian. 


\section{Field Operators and Ladder Operators Consistency Check }\label{consis}

If we consider field operators~\eqref{c11} and impose the following commutation relations for creation and annihilation operators 
\begin{align}\label{c13-1}
    [\hat{a}_{\lambda,(\omega,\hat{\mathbf{k}})}, \hat{a}^{\dag}_{\lambda',(\omega,\hat{\mathbf{k'}})}]=\hbar^{-1}\omega^{-1} \delta_{\lambda\lambda'}\delta(\omega-\omega')\delta^{2}(\hat{\mathbf{k}}-\hat{\mathbf{k'}}) =\hbar\omega \delta_{\lambda\lambda'}\delta^{3}({\mathbf{k}}-{\mathbf{k'}}) \, ,
\end{align}
    \begin{equation}\label{c14-1}
    [\hat{a}_{\lambda,(\omega,\hat{\mathbf{k}})}, \hat{a}_{\lambda',(\omega,\hat{\mathbf{k'}})}]=0\, ,
\end{equation}
and compute $[ \hat{\mathbf{S}}_{i}(t,\mathbf{x}),\hat{\mathbf{S}}^{\dag}_{j}(t,\mathbf{y})]$, we get
\begin{equation}\label{c15}
    [ \hat{\mathbf{S}}_{i}(t,\mathbf{x}),\hat{\mathbf{S}}_{j}^{\dag}(t,\mathbf{y})]=-2\hbar\epsilon_{ijk}\pdv{}{x^k}I(\mathbf{x},\mathbf{y})\, ,
\end{equation}
where
\begin{align}\label{c16}
I(\mathbf{x},\mathbf{y})=\int \frac{\dd^3 {\mathbf{k}}}{(2\pi)^{3/2}}  \varkappa_{(\mathbf{y})} \bigl( e^{i\omega \int^{\mathbf{x}} \varkappa_{(\mathbf{x}')} \hat{\mathbf{k}} \cdot \dd x'-i\omega \int^{\mathbf{y}}  \varkappa_{(\mathbf{y}')} \hat{\mathbf{k}} \cdot \dd y'}+e^{-i\omega \int^{\mathbf{x}} \varkappa_{(\mathbf{x}')} \hat{\mathbf{k}} \cdot \dd x'+i\omega \int^{\mathbf{y}} \varkappa_{(\mathbf{y}')} \hat{\mathbf{k}} \cdot \dd y'}\bigl).
\end{align}
To show that   $I(\mathbf{x},\mathbf{y}) \rightarrow \delta(\mathbf{x}-\mathbf{y})$, we rewrite this integral by use of elementary calculus identity and $\mathbf{k}=\omega\hat{\mathbf{k}}$, as 
\begin{align}\label{c17}
I(\mathbf{x},\mathbf{y})=\int\frac{\dd^{3}\mathbf{k}}{(2\pi)^{3/2}} \varkappa_{(\mathbf{y})}\bigl(e^{i\int_{\mathbf{y}}^{\mathbf{x}} \varkappa_{(\mathbf{x}')} \mathbf{k} \cdot \dd x'}+e^{-i \int_{\mathbf{y}}^{\mathbf{x}} \varkappa_{(\mathbf{x}')} \mathbf{k} \cdot \dd x'} \bigl).
\end{align}
For the cases $\mathbf{x}\neq \mathbf{y}$, the integral is equal to zero due to anti-symmetry under $\mathbf{k}\rightarrow -\mathbf{k} $.
For the cases $\mathbf{x}\simeq \mathbf{y}$, we consider $\mathbf{x}=\mathbf{y}+\mathbf{\Delta y}$,
\begin{align}\label{c18}
I(\mathbf{y} + \mathbf{\Delta y}, \mathbf{y})=\int\frac{\dd^{3}\mathbf{k}}{(2\pi)^{3/2}} \varkappa_{(\mathbf{y})} \bigl( e^{i\, \varkappa_{(\mathbf{y})} \mathbf{k} \cdot \mathbf{\Delta \mathbf{y}}}+e^{-i\,\varkappa_{(\mathbf{y})} \mathbf{k} \cdot \mathbf{\Delta \mathbf{y}}} \bigl)=\delta^3(\mathbf{\Delta \mathbf{y}}) ,
\end{align}
So,
\begin{equation}\label{c19}
    I(\mathbf{x},\mathbf{y})=\delta(\mathbf{x}-\mathbf{y})\, .
\end{equation}
Therefore, the second quantization procedure is consistent and the field operators' commutation relations and creation-annihilation operators' commutation relations are equivalent.

\section{Phase Difference and Time Delay} \label{PT}
The covariant field tensor of self-dual and anti-self-dual of the solutions~\eqref{c7} and~\eqref{c8} are
    \begin{equation}
        {{^{\pm}}f_{\mu\nu}}= A_{\pm}\ell^{\pm}_{\mu\nu}(\mathbf{x}) \,e^{-i\omega\,t + i\,\omega \int^{\mathbf{x}} \varkappa_{(\mathbf{x}')} \hat{\mathbf{k}} \cdot \dd x'}\, \, ,
    \end{equation}
where,
\begin{equation}
    \ell^{\pm}_{\mu\nu}=\left(
    \begin{array}{cccc}
     0  -\hat{\ell}^{\pm}_{1}  -\hat{\ell}^{\pm}_{2}  -\hat{\ell}^{\pm}_{3} \\
      \hat{\ell}^{\pm}_{1}  0  \pm i \varkappa_{(\mathbf{x})} \hat{\ell}^{\pm}_{3}  \mp i\varkappa_{(\mathbf{x})}\hat{\ell}^{\pm}_{2}\\
      \hat{\ell}^{\pm}_{2}  \mp i\varkappa_{(\mathbf{x})}\hat{\ell}^{\pm}_{3}    0    \pm i \varkappa_{(\mathbf{x})}\hat{\ell}^{\pm}_{1}\\
     \hat{\ell}^{\pm}_{3}    \pm i \varkappa_{(\mathbf{x})}\hat{\ell}^{\pm}_{2}  \mp i \varkappa_{(\mathbf{x})}\hat{\ell}^{\pm}_{1}  0
    \end{array}\right)\, .
\end{equation}
We ignore amplitudes and consider the action as:
\begin{equation}\label{P14}
   \mathcal{S}^{}(t_{i},\mathbf{x}_{i};t_{f},\mathbf{x}_{f})= -\omega\int_{t_{i}}^{t_{f}}\,\dd t + \,\omega \int_{{\mathbf{x}_{i}}}^{\mathbf{x}_{f}} \varkappa_{(\mathbf{x}')} \,\hat{\mathbf{k}} \cdot dx' \, \, ,
\end{equation}
By using this action and putting it in path integral formalism~\cite{Sakurai} and computing the quantum phase difference for different paths. Using the path integral formalism, the geometric phase difference between two arbitrary paths is:
\begin{align}\label{P15}
    \Delta\phi=\phi^{}_{II}-\phi^{}_{I} =[\mathcal{S}^{}(t_{i},\mathbf{x}_{i};t_{f},\mathbf{x}_{f})]_{\text{path II}}-[\mathcal{S}^{}(t_{i},\mathbf{x}_{i};t_{f},\mathbf{x}_{f}) ]_{\text{path I}}=\omega\oint \varkappa_{(\mathbf{x}(s))}
    \,\dd s  \, \, ,
\end{align}
where we have used $dx'=\hat{\mathbf{k}}\dd s$ where $s$ is a parameter that we use to parametrize the paths.
From the principle of Fermat (principle of least time), we can compute the gravitational time delay,
\begin{equation}
    \delta t= \int_{{\mathbf{x}_{i}}}^{\mathbf{x}_{f}} \varkappa_{(\mathbf{x}')} \hat{\mathbf{k}} \cdot \dd x' =\int_{{s_{i}}}^{s_{f}} \varkappa_{(s)} \,\dd s\, \, ,
\end{equation}
and then compute the time difference of photons in different paths of the interferometer in gravitational field, 
\begin{equation}\label{P19}
    \Delta t=\delta t_{II}-\delta t_{I}  =\oint \varkappa_{(s)}\,\dd s \, \, .
\end{equation}


\section{Coherent Photons}\label{coh}

Coherent states which describe, in a good approximation, light emitted by
lasers are superpositions of the states with different numbers of identical photons based on~\cite{bialynicki2019photons} described by the wave function $\psi(\mathbf{k},\lambda)$ are
\begin{align}\label{c56}
    \ket{\psi_{coh}}=e^{\frac{\overline{N}}{2}}\left( \ket{0}+\frac{\overline{N}^{1/2}}{1!}\hat{a}^{\dag}_{\psi}\ket{0}+\frac{\overline{N}}{2!}(\hat{a}^{\dag}_{\psi})^{2}\ket{0}+\cdots\right)=e^{\frac{\overline{N}}{2}}\exp{{\overline{N}^{1/2}}\hat{a}^{\dag}_{\psi}}\ket{0},
\end{align}
$\overline{N}$ is the average number of photons in the coherent state. Fidelity of coherent states $\mathcal{F}_{c}$ is defined in~\cite{bialynicki2020three} as,
\begin{equation}\label{c57}
   \mathcal{F}_{c}=\frac{\abs{\bra{\phi_{coh}}\ket{\psi_{coh}}}^{2}}{\bra{\phi_{coh}}\ket{\phi_{coh}}\bra{\psi_{coh}}\ket{\psi_{coh}}}\, \, .
\end{equation}
Since coherent states are defined in terms of the photon wave function in momentum representation, there exists a relation between $\mathcal{F}_{c}$ and $\mathcal{F}_{m}$. This relation depends not only on fidelity but also on the phase
difference $\varphi$ of photon wave functions ($\Delta=\abs{\Delta}\,e^{i\varphi}$)~\cite{bialynicki2020three}
\begin{equation}\label{c59}
    \mathcal{F}_{c}=\exp{-2\overline{N}(1-\cos{\varphi}\sqrt{\mathcal{F}_{m}})}\, .
\end{equation}

\section{Photon Deflection}\label{Def}

To specify the classical path of photons in the gravitational field of the Schwarzschild metric, we use the action~\eqref{P14}, the differential equation of rays has the following form;
\begin{equation}\label{Dfe1}
    \dv{}{\tau}\left( \dv{\mathbf{x}}{\tau}\right)=\grad \varkappa_{(\mathbf{x})}\, .
\end{equation}
The equation of null rays in a Schwarzschild gravitational field in isotropic coordinates may be written in the form of:
\begin{equation}\label{Def2}
    \theta=b\int^{r}\frac{\dd r}{r\sqrt{\varkappa_{(r)}^{2}\,r^{2}-b^{2}}}\, .
\end{equation}
The deflection angle of the light ray with impact factor $b$ and closest approach radius $r_{min}$ is~\cite{roy2019study}
\begin{equation}\label{Def3}
      \delta\theta=2 b\int_{r_{min}}^{+\infty}\frac{\dd r}{r\sqrt{\varkappa_{(r)}^{2}\,r^{2}-b^{2}}}-\pi\, .
\end{equation}
The method we discussed in this section, to obtain photon deflection angle, is semiclassical. Another approach to obtaining the deflection angle of a photon by the Schwarzschild gravitational field is possible based on~\cite{mashhoon1973scattering}, where scattering of photons by Schwarzschild black hole is discussed. The origin of this effect for light is photon-gravity coupling (the coupling of photon wave vector and gravitational field.) This equation for the weak field limit of the Schwarzschild spacetime gives
\begin{equation}\label{Def4}
      \delta\theta_{E}= -\frac{4GM}{c^2\,r_{min}}\, .
\end{equation}
Here we have added index $E$ to refer to the fact that it is computed in Einstein’s theory of gravity.
A similar analysis, for Newtonian metric~\eqref{n1}, results in
\begin{equation}\label{Def5}
      \delta\theta_{N}= -\frac{2GM}{c^2\,r_{min}}\, ,
\end{equation}
for the deflection angle. Which is the half value obtained by Einstein's theory of gravity.

It is important to note the values $\delta\theta_{E}$ and $\delta\theta_{N}$ are obtained in a local Newtonian gravitational field in a sense that throughout the classical path of the photon, the gravitational acceleration is $g_{(r)}=\frac{MG}{r^2}$, therefore, it is not possible to distinguish whether gravity is Einsteinian or Newtonian by use of accelerometers in fundamental frames.

\section{Einstein Equivalence Principle}\label{EEP}

Einstein's Equivalence principle is usually stated as "\textit{all effects of a uniform gravitational field are identical to the effects of a uniform acceleration of the coordinate system}"\cite{MTW}. 
This principle had historical importance for the discovery of Einstein gravitation theory and was the key point for Einstein to formulate the geometric theory of gravitation~\cite{synge1960relativity}. It is important to mention that as it is emphasized in \cite{mashhoon2017nonlocal} this principle is extremely local, and it is valid only point-wise. Therefore, the falsification of its extension beyond a single pint is not in contrast with the foundation of Einstein's gravitation theory. For more arguments against the validity of the equivalence principle in quantum-gravity interface, see~\cite{Singleton, lebed2022breakdown}. 

\section{Photon-Gravity Coupling in Kerr Spacetime}\label{Kerr}

In this Appendix, photon-gravity coupling in Kerr spacetime which is a more realistic model of astronomical objects, is outlined.
The Kerr spacetime in the Boyer–Lindquist coordinate is given in~\cite{chandra}. We apply the transformation $r\rightarrow\bar{r}$, where $ r=\varrho^{2}\, \bar{r} =\bar{r}\bigl(1+\frac{M+a}{2\bar{r}} \bigl) \bigl(1+\frac{M-a}{2\bar{r}} \bigl)$ \cite{IsoKerr}
and obtain it in quasi-isotropic coordinates, then by expansion based on $a$ and represent $\bar{r}$ by $r$, we get
\begin{align}
    \dd s^{2}=-\biggl(\frac{\mathcal{B}_{-}(r)}{\mathcal{B}_{+}(r)}\biggl)^2\dd t^{2}-\,4\mathcal{A}(r)\cdot d{x} \dd t
    +\mathcal{B}^{\,4}_{+}(r)\delta_{ij}dx^idx^j+\mathcal{O}(a^{2}).
\end{align}
 Here $\mathcal{B}_{\pm}(r)=1\pm\frac{MG}{2r}$, $\mathcal{A}(r)= \frac{ \mathcal{J}\times x}{\bar{r}^{3}(1+\frac{M}{2\bar{r}})^2}$, where $r$ is the radius $r^{2}=x^{2}+y^{2}+z^{2}$, and $\abs{\mathcal{J}}=M a$. This metric describes Schwarzschild spacetime, which is disturbed by the slow rotation of source mass with angular momentum $\mathcal{J}$. It is possible to show for this metric $\boldsymbol{\kappa}_{li}=\,\varkappa\,\delta^{li}\pm i\epsilon^{jil}\, \mho_{j}\, .$ where $\varkappa=\mathcal{B}_{+}^{3}\mathcal{B}_{-}^{-1}$ and  $\mho_{j}=-2\mathcal{B}_{+}^{2}\mathcal{B}_{-}^{-2}\mathcal{A}_{j}$
It is straightforward to show Maxwell's equation in slowly rotating Kerr spacetime has solutions in the form of
\begin{equation}
\mathbf{F}^{\pm}=A_{\pm}(\omega,\hat{\mathbf{k}})\, \hat{\ell}_{\pm}(\hat{\mathbf{k}}) \,e^{\left[-i\omega\,t + i\,\omega \int^{\mathbf{x}} \varkappa_{(x')} \hat{\mathbf{k}} \cdot d\mathbf{x'} + i\,\omega \int^{\mathbf{x}}\mho_{(x)} \cdot d\mathbf{x'}\right]}
\end{equation} 
at the zero order in $\omega^{-1}$. These solutions in the asymptotic region far from the source where the gravitational field is absent take the form of plane wave solutions of Maxwell equations in Minkowski spacetime.
As it is apparent from these solutions \textit{the wave vector of photon $\vec{\mathbf{k}}$ in the asymptotic region far from the source where the gravitational field is absent is replaced by $\varkappa_{(x)}\,\vec{\mathbf{k}}+\, \vec{\mho}_{(x)}$, in regions the gravitational field is nonzero}. Therefore, in the slowly rotating Kerr, the gravitational field couples to the wave vector through $\varkappa_{(x)}$ and $\vec{\mho}_{(x)}$. 
It is not possible to construct an exact second quantized theory in Kerr. This originates from the absence of a hypersurface orthogonal timelike Killing vector. However, as the slowly rotating Kerr can be considered as a perturbed Schwarzshild metric, we can use these solutions which contain corrections due to rotation, and develop an approximate picture to analyze photons in slowly rotating Kerr.


\begin{thebibliography}{67}%
\makeatletter
\providecommand \@ifxundefined [1]{%
 \@ifx{#1\undefined}
}%
\providecommand \@ifnum [1]{%
 \ifnum #1\expandafter \@firstoftwo
 \else \expandafter \@secondoftwo
 \fi
}%
\providecommand \@ifx [1]{%
 \ifx #1\expandafter \@firstoftwo
 \else \expandafter \@secondoftwo
 \fi
}%
\providecommand \natexlab [1]{#1}%
\providecommand \enquote  [1]{``#1''}%
\providecommand \bibnamefont  [1]{#1}%
\providecommand \bibfnamefont [1]{#1}%
\providecommand \citenamefont [1]{#1}%
\providecommand \href@noop [0]{\@secondoftwo}%
\providecommand \href [0]{\begingroup \@sanitize@url \@href}%
\providecommand \@href[1]{\@@startlink{#1}\@@href}%
\providecommand \@@href[1]{\endgroup#1\@@endlink}%
\providecommand \@sanitize@url [0]{\catcode `\\12\catcode `\$12\catcode `\&12\catcode `\#12\catcode `\^12\catcode `\_12\catcode `\%12\relax}%
\providecommand \@@startlink[1]{}%
\providecommand \@@endlink[0]{}%
\providecommand \url  [0]{\begingroup\@sanitize@url \@url }%
\providecommand \@url [1]{\endgroup\@href {#1}{\urlprefix }}%
\providecommand \urlprefix  [0]{URL }%
\providecommand \Eprint [0]{\href }%
\providecommand \doibase [0]{https://doi.org/}%
\providecommand \selectlanguage [0]{\@gobble}%
\providecommand \bibinfo  [0]{\@secondoftwo}%
\providecommand \bibfield  [0]{\@secondoftwo}%
\providecommand \translation [1]{[#1]}%
\providecommand \BibitemOpen [0]{}%
\providecommand \bibitemStop [0]{}%
\providecommand \bibitemNoStop [0]{.\EOS\space}%
\providecommand \EOS [0]{\spacefactor3000\relax}%
\providecommand \BibitemShut  [1]{\csname bibitem#1\endcsname}%
\let\auto@bib@innerbib\@empty
\bibitem [{\citenamefont {Chandrasekhar}(1985)}]{chandra}%
  \BibitemOpen
  \bibfield  {author} {\bibinfo {author} {\bibfnamefont {S.}~\bibnamefont {Chandrasekhar}},\ }\href@noop {} {\emph {\bibinfo {title} {{The mathematical theory of black holes}}}}\ (\bibinfo {year} {1985})\BibitemShut {NoStop}%
\bibitem [{\citenamefont {Weinberg}(1972)}]{weinberg1972gravitation}%
  \BibitemOpen
  \bibfield  {author} {\bibinfo {author} {\bibfnamefont {S.}~\bibnamefont {Weinberg}},\ }\href@noop {} {\emph {\bibinfo {title} {{Gravitation and Cosmology}: {Principles and Applications of the General Theory of Relativity}}}}\ (\bibinfo  {publisher} {John Wiley and Sons},\ \bibinfo {address} {New York},\ \bibinfo {year} {1972})\BibitemShut {NoStop}%
\bibitem [{\citenamefont {Einstein}(1911)}]{einstein1911influence}%
  \BibitemOpen
  \bibfield  {author} {\bibinfo {author} {\bibfnamefont {A.}~\bibnamefont {Einstein}},\ }\bibfield  {title} {\bibinfo {title} {{On The influence of gravitation on the propagation of light}},\ }\href {https://doi.org/10.1002/andp.200590033} {\bibfield  {journal} {\bibinfo  {journal} {Annalen Phys.}\ }\textbf {\bibinfo {volume} {35}},\ \bibinfo {pages} {898} (\bibinfo {year} {1911})}\BibitemShut {NoStop}%
\bibitem [{\citenamefont {Misner}\ \emph {et~al.}(1973)\citenamefont {Misner}, \citenamefont {Thorne},\ and\ \citenamefont {Wheeler}}]{MTW}%
  \BibitemOpen
  \bibfield  {author} {\bibinfo {author} {\bibfnamefont {C.~W.}\ \bibnamefont {Misner}}, \bibinfo {author} {\bibfnamefont {K.~S.}\ \bibnamefont {Thorne}},\ and\ \bibinfo {author} {\bibfnamefont {J.~A.}\ \bibnamefont {Wheeler}},\ }\href@noop {} {\emph {\bibinfo {title} {{Gravitation}}}}\ (\bibinfo  {publisher} {W. H. Freeman},\ \bibinfo {address} {San Francisco},\ \bibinfo {year} {1973})\BibitemShut {NoStop}%
\bibitem [{\citenamefont {Pound}\ and\ \citenamefont {Rebka}(1960)}]{pound1960apparent}%
  \BibitemOpen
  \bibfield  {author} {\bibinfo {author} {\bibfnamefont {R.~V.}\ \bibnamefont {Pound}}\ and\ \bibinfo {author} {\bibfnamefont {G.~A.}\ \bibnamefont {Rebka}},\ }\bibfield  {title} {\bibinfo {title} {Apparent weight of photons},\ }\href {https://doi.org/10.1103/PhysRevLett.4.337} {\bibfield  {journal} {\bibinfo  {journal} {Phys. Rev. Lett.}\ }\textbf {\bibinfo {volume} {4}},\ \bibinfo {pages} {337} (\bibinfo {year} {1960})}\BibitemShut {NoStop}%
\bibitem [{\citenamefont {Zych}\ \emph {et~al.}(2012)\citenamefont {Zych}, \citenamefont {Costa}, \citenamefont {Pikovski}, \citenamefont {Ralph},\ and\ \citenamefont {Brukner}}]{zych2012general}%
  \BibitemOpen
  \bibfield  {author} {\bibinfo {author} {\bibfnamefont {M.}~\bibnamefont {Zych}}, \bibinfo {author} {\bibfnamefont {F.}~\bibnamefont {Costa}}, \bibinfo {author} {\bibfnamefont {I.}~\bibnamefont {Pikovski}}, \bibinfo {author} {\bibfnamefont {T.~C.}\ \bibnamefont {Ralph}},\ and\ \bibinfo {author} {\bibfnamefont {C.}~\bibnamefont {Brukner}},\ }\bibfield  {title} {\bibinfo {title} {{General relativistic effects in quantum interference of photons}},\ }\href {https://doi.org/10.1088/0264-9381/29/22/224010} {\bibfield  {journal} {\bibinfo  {journal} {Class. Quant. Grav.}\ }\textbf {\bibinfo {volume} {29}},\ \bibinfo {pages} {224010} (\bibinfo {year} {2012})},\ \Eprint {https://arxiv.org/abs/1206.0965} {arXiv:1206.0965 [quant-ph]} \BibitemShut {NoStop}%
\bibitem [{\citenamefont {Hilweg}\ \emph {et~al.}(2017)\citenamefont {Hilweg}, \citenamefont {Massa}, \citenamefont {Martynov}, \citenamefont {Mavalvala}, \citenamefont {Chru\'sciel},\ and\ \citenamefont {Walther}}]{hilweg2017gravitationally}%
  \BibitemOpen
  \bibfield  {author} {\bibinfo {author} {\bibfnamefont {C.}~\bibnamefont {Hilweg}}, \bibinfo {author} {\bibfnamefont {F.}~\bibnamefont {Massa}}, \bibinfo {author} {\bibfnamefont {D.}~\bibnamefont {Martynov}}, \bibinfo {author} {\bibfnamefont {N.}~\bibnamefont {Mavalvala}}, \bibinfo {author} {\bibfnamefont {P.~T.}\ \bibnamefont {Chru\'sciel}},\ and\ \bibinfo {author} {\bibfnamefont {P.}~\bibnamefont {Walther}},\ }\bibfield  {title} {\bibinfo {title} {{Gravitationally induced phase shift on a single photon}},\ }\href {https://doi.org/10.1088/1367-2630/aa638f} {\bibfield  {journal} {\bibinfo  {journal} {New J. Phys.}\ }\textbf {\bibinfo {volume} {19}},\ \bibinfo {pages} {033028} (\bibinfo {year} {2017})},\ \Eprint {https://arxiv.org/abs/1612.03612} {arXiv:1612.03612 [quant-ph]} \BibitemShut {NoStop}%
\bibitem [{\citenamefont {Exirifard}\ \emph {et~al.}(2021)\citenamefont {Exirifard}, \citenamefont {Culf},\ and\ \citenamefont {Karimi}}]{exirifard2021towards}%
  \BibitemOpen
  \bibfield  {author} {\bibinfo {author} {\bibfnamefont {Q.}~\bibnamefont {Exirifard}}, \bibinfo {author} {\bibfnamefont {E.}~\bibnamefont {Culf}},\ and\ \bibinfo {author} {\bibfnamefont {E.}~\bibnamefont {Karimi}},\ }\bibfield  {title} {\bibinfo {title} {{Towards Communication in a Curved Spacetime Geometry}},\ }\href {https://doi.org/10.1038/s42005-021-00671-8} {\bibfield  {journal} {\bibinfo  {journal} {Commun. Phys.}\ }\textbf {\bibinfo {volume} {4}},\ \bibinfo {pages} {171} (\bibinfo {year} {2021})},\ \Eprint {https://arxiv.org/abs/2009.04217} {arXiv:2009.04217 [gr-qc]} \BibitemShut {NoStop}%
\bibitem [{\citenamefont {Rodríguez}\ \emph {et~al.}(2023)\citenamefont {Rodríguez}, \citenamefont {Schell},\ and\ \citenamefont {Bruschi}}]{rodriguez2023introduction}%
  \BibitemOpen
  \bibfield  {author} {\bibinfo {author} {\bibfnamefont {L.~A.~A.}\ \bibnamefont {Rodríguez}}, \bibinfo {author} {\bibfnamefont {A.~W.}\ \bibnamefont {Schell}},\ and\ \bibinfo {author} {\bibfnamefont {D.~E.}\ \bibnamefont {Bruschi}},\ }\bibfield  {title} {\bibinfo {title} {Introduction to gravitational redshift of quantum photons propagating in curved spacetime}\ }(\bibinfo  {publisher} {IOP Publishing},\ \bibinfo {year} {2023})\ p.\ \bibinfo {pages} {012016}\BibitemShut {NoStop}%
\bibitem [{\citenamefont {Bruschi}\ \emph {et~al.}(2021)\citenamefont {Bruschi}, \citenamefont {Chatzinotas}, \citenamefont {Wilhelm},\ and\ \citenamefont {Schell}}]{bruschi2021spacetime}%
  \BibitemOpen
  \bibfield  {author} {\bibinfo {author} {\bibfnamefont {D.~E.}\ \bibnamefont {Bruschi}}, \bibinfo {author} {\bibfnamefont {S.}~\bibnamefont {Chatzinotas}}, \bibinfo {author} {\bibfnamefont {F.~K.}\ \bibnamefont {Wilhelm}},\ and\ \bibinfo {author} {\bibfnamefont {A.~W.}\ \bibnamefont {Schell}},\ }\bibfield  {title} {\bibinfo {title} {Spacetime effects on wavepackets of coherent light},\ }\href {https://doi.org/10.1103/PhysRevD.104.085015} {\bibfield  {journal} {\bibinfo  {journal} {Phys. Rev. D}\ }\textbf {\bibinfo {volume} {104}},\ \bibinfo {pages} {085015} (\bibinfo {year} {2021})}\BibitemShut {NoStop}%
\bibitem [{\citenamefont {Bruschi}\ and\ \citenamefont {Schell}(2022)}]{bruschi2023gravitational}%
  \BibitemOpen
  \bibfield  {author} {\bibinfo {author} {\bibfnamefont {D.~E.}\ \bibnamefont {Bruschi}}\ and\ \bibinfo {author} {\bibfnamefont {A.~W.}\ \bibnamefont {Schell}},\ }\bibfield  {title} {\bibinfo {title} {{Gravitational Redshift Induces Quantum Interference}},\ }\href {https://doi.org/10.1002/andp.202200468} {\bibfield  {journal} {\bibinfo  {journal} {Annals Phys.}\ }\textbf {\bibinfo {volume} {535}},\ \bibinfo {pages} {2200468} (\bibinfo {year} {2022})},\ \Eprint {https://arxiv.org/abs/2109.00728} {arXiv:2109.00728 [quant-ph]} \BibitemShut {NoStop}%
\bibitem [{\citenamefont {Bruschi}\ \emph {et~al.}(2014)\citenamefont {Bruschi}, \citenamefont {Ralph}, \citenamefont {Fuentes}, \citenamefont {Jennewein},\ and\ \citenamefont {Razavi}}]{bruschi2014spacetime}%
  \BibitemOpen
  \bibfield  {author} {\bibinfo {author} {\bibfnamefont {D.~E.}\ \bibnamefont {Bruschi}}, \bibinfo {author} {\bibfnamefont {T.~C.}\ \bibnamefont {Ralph}}, \bibinfo {author} {\bibfnamefont {I.}~\bibnamefont {Fuentes}}, \bibinfo {author} {\bibfnamefont {T.}~\bibnamefont {Jennewein}},\ and\ \bibinfo {author} {\bibfnamefont {M.}~\bibnamefont {Razavi}},\ }\bibfield  {title} {\bibinfo {title} {Spacetime effects on satellite-based quantum communications},\ }\href {https://doi.org/10.1103/PhysRevD.90.045041} {\bibfield  {journal} {\bibinfo  {journal} {Phys. Rev. D}\ }\textbf {\bibinfo {volume} {90}},\ \bibinfo {pages} {045041} (\bibinfo {year} {2014})}\BibitemShut {NoStop}%
\bibitem [{\citenamefont {Mieling}\ \emph {et~al.}(2022)\citenamefont {Mieling}, \citenamefont {Hilweg},\ and\ \citenamefont {Walther}}]{mieling2022measuring}%
  \BibitemOpen
  \bibfield  {author} {\bibinfo {author} {\bibfnamefont {T.~B.}\ \bibnamefont {Mieling}}, \bibinfo {author} {\bibfnamefont {C.}~\bibnamefont {Hilweg}},\ and\ \bibinfo {author} {\bibfnamefont {P.}~\bibnamefont {Walther}},\ }\bibfield  {title} {\bibinfo {title} {{Measuring space-time curvature using maximally path-entangled quantum states}},\ }\href {https://doi.org/10.1103/PhysRevA.106.L031701} {\bibfield  {journal} {\bibinfo  {journal} {Phys. Rev. A}\ }\textbf {\bibinfo {volume} {106}},\ \bibinfo {pages} {L031701} (\bibinfo {year} {2022})},\ \Eprint {https://arxiv.org/abs/2202.12562} {arXiv:2202.12562 [gr-qc]} \BibitemShut {NoStop}%
\bibitem [{\citenamefont {Frolov}\ and\ \citenamefont {Shoom}(2011)}]{Frolov}%
  \BibitemOpen
  \bibfield  {author} {\bibinfo {author} {\bibfnamefont {V.~P.}\ \bibnamefont {Frolov}}\ and\ \bibinfo {author} {\bibfnamefont {A.~A.}\ \bibnamefont {Shoom}},\ }\bibfield  {title} {\bibinfo {title} {Spinoptics in a stationary spacetime},\ }\href {https://doi.org/10.1103/PhysRevD.84.044026} {\bibfield  {journal} {\bibinfo  {journal} {Phys. Rev. D}\ }\textbf {\bibinfo {volume} {84}},\ \bibinfo {pages} {044026} (\bibinfo {year} {2011})}\BibitemShut {NoStop}%
\bibitem [{\citenamefont {Lambiase}\ and\ \citenamefont {Papini}(2021)}]{Lambiase}%
  \BibitemOpen
  \bibfield  {author} {\bibinfo {author} {\bibfnamefont {G.}~\bibnamefont {Lambiase}}\ and\ \bibinfo {author} {\bibfnamefont {G.}~\bibnamefont {Papini}},\ }\bibfield  {title} {\bibinfo {title} {{The Interaction of Spin with Gravity in Particle Physics: Low Energy Quantum Gravity}},\ }\href {https://doi.org/10.1007/978-3-030-84771-5} {\bibfield  {journal} {\bibinfo  {journal} {Lect. Notes Phys.}\ }\textbf {\bibinfo {volume} {993}},\ \bibinfo {pages} {pp.} (\bibinfo {year} {2021})}\BibitemShut {NoStop}%
\bibitem [{\citenamefont {Andersson}\ and\ \citenamefont {Oancea}(2023)}]{Andersson}%
  \BibitemOpen
  \bibfield  {author} {\bibinfo {author} {\bibfnamefont {L.}~\bibnamefont {Andersson}}\ and\ \bibinfo {author} {\bibfnamefont {M.~A.}\ \bibnamefont {Oancea}},\ }\bibfield  {title} {\bibinfo {title} {Spin hall effects in the sky},\ }\href {https://doi.org/10.1088/1361-6382/ace021} {\bibfield  {journal} {\bibinfo  {journal} {Classical and Quantum Gravity}\ }\textbf {\bibinfo {volume} {40}},\ \bibinfo {pages} {154002} (\bibinfo {year} {2023})}\BibitemShut {NoStop}%
\bibitem [{\citenamefont {Bini}\ \emph {et~al.}(2018)\citenamefont {Bini}, \citenamefont {Chicone}, \citenamefont {Mashhoon},\ and\ \citenamefont {Rosquist}}]{bini2018spinning}%
  \BibitemOpen
  \bibfield  {author} {\bibinfo {author} {\bibfnamefont {D.}~\bibnamefont {Bini}}, \bibinfo {author} {\bibfnamefont {C.}~\bibnamefont {Chicone}}, \bibinfo {author} {\bibfnamefont {B.}~\bibnamefont {Mashhoon}},\ and\ \bibinfo {author} {\bibfnamefont {K.}~\bibnamefont {Rosquist}},\ }\bibfield  {title} {\bibinfo {title} {{Spinning Particles in Twisted Gravitational Wave Spacetimes}},\ }\href {https://doi.org/10.1103/PhysRevD.98.024043} {\bibfield  {journal} {\bibinfo  {journal} {Phys. Rev. D}\ }\textbf {\bibinfo {volume} {98}},\ \bibinfo {pages} {024043} (\bibinfo {year} {2018})},\ \Eprint {https://arxiv.org/abs/1805.07080} {arXiv:1805.07080 [gr-qc]} \BibitemShut {NoStop}%
\bibitem [{\citenamefont {Bini}\ \emph {et~al.}(2012)\citenamefont {Bini}, \citenamefont {Chicone},\ and\ \citenamefont {Mashhoon}}]{bini2012spacetime}%
  \BibitemOpen
  \bibfield  {author} {\bibinfo {author} {\bibfnamefont {D.}~\bibnamefont {Bini}}, \bibinfo {author} {\bibfnamefont {C.}~\bibnamefont {Chicone}},\ and\ \bibinfo {author} {\bibfnamefont {B.}~\bibnamefont {Mashhoon}},\ }\bibfield  {title} {\bibinfo {title} {{Spacetime Splitting, Admissible Coordinates and Causality}},\ }\href {https://doi.org/10.1103/PhysRevD.85.104020} {\bibfield  {journal} {\bibinfo  {journal} {Phys. Rev. D}\ }\textbf {\bibinfo {volume} {85}},\ \bibinfo {pages} {104020} (\bibinfo {year} {2012})},\ \Eprint {https://arxiv.org/abs/1203.3454} {arXiv:1203.3454 [gr-qc]} \BibitemShut {NoStop}%
\bibitem [{\citenamefont {Mashhoon}(1973{\natexlab{a}})}]{mashhoon1973scattering}%
  \BibitemOpen
  \bibfield  {author} {\bibinfo {author} {\bibfnamefont {B.}~\bibnamefont {Mashhoon}},\ }\bibfield  {title} {\bibinfo {title} {{Scattering of Electromagnetic Radiation from a Black Hole}},\ }\href {https://doi.org/10.1103/PhysRevD.7.2807} {\bibfield  {journal} {\bibinfo  {journal} {Phys. Rev. D}\ }\textbf {\bibinfo {volume} {7}},\ \bibinfo {pages} {2807} (\bibinfo {year} {1973}{\natexlab{a}})}\BibitemShut {NoStop}%
\bibitem [{\citenamefont {Mashhoon}(1974{\natexlab{a}})}]{mashhoon1974electromagnetic}%
  \BibitemOpen
  \bibfield  {author} {\bibinfo {author} {\bibfnamefont {B.}~\bibnamefont {Mashhoon}},\ }\bibfield  {title} {\bibinfo {title} {{Electromagnetic scattering from a black hole and the glory effect}},\ }\href {https://doi.org/10.1103/PhysRevD.10.1059} {\bibfield  {journal} {\bibinfo  {journal} {Phys. Rev. D}\ }\textbf {\bibinfo {volume} {10}},\ \bibinfo {pages} {1059} (\bibinfo {year} {1974}{\natexlab{a}})}\BibitemShut {NoStop}%
\bibitem [{\citenamefont {Mashhoon}(1974{\natexlab{b}})}]{mashhoon1974can}%
  \BibitemOpen
  \bibfield  {author} {\bibinfo {author} {\bibfnamefont {B.}~\bibnamefont {Mashhoon}},\ }\bibfield  {title} {\bibinfo {title} {Can {E}instein's theory of gravitation be tested beyond the geometrical optics limit?},\ }\href {https://doi.org/10.1038/250316a0} {\bibfield  {journal} {\bibinfo  {journal} {Nature}\ }\textbf {\bibinfo {volume} {250}},\ \bibinfo {pages} {316} (\bibinfo {year} {1974}{\natexlab{b}})}\BibitemShut {NoStop}%
\bibitem [{\citenamefont {Mashhoon}(1975)}]{mashhoon1975influence}%
  \BibitemOpen
  \bibfield  {author} {\bibinfo {author} {\bibfnamefont {B.}~\bibnamefont {Mashhoon}},\ }\bibfield  {title} {\bibinfo {title} {{Influence of Gravitation on the Propagation of Electromagnetic Radiation}},\ }\href {https://doi.org/10.1103/PhysRevD.11.2679} {\bibfield  {journal} {\bibinfo  {journal} {Phys. Rev. D}\ }\textbf {\bibinfo {volume} {11}},\ \bibinfo {pages} {2679} (\bibinfo {year} {1975})}\BibitemShut {NoStop}%
\bibitem [{\citenamefont {Kopeikin}\ and\ \citenamefont {Mashhoon}(2002)}]{kopeikin2002gravitomagnetic}%
  \BibitemOpen
  \bibfield  {author} {\bibinfo {author} {\bibfnamefont {S.}~\bibnamefont {Kopeikin}}\ and\ \bibinfo {author} {\bibfnamefont {B.}~\bibnamefont {Mashhoon}},\ }\bibfield  {title} {\bibinfo {title} {{Gravitomagnetic effects in the propagation of electromagnetic waves in variable gravitational fields of arbitrary moving and spinning bodies}},\ }\href {https://doi.org/10.1103/PhysRevD.65.064025} {\bibfield  {journal} {\bibinfo  {journal} {Phys. Rev. D}\ }\textbf {\bibinfo {volume} {65}},\ \bibinfo {pages} {064025} (\bibinfo {year} {2002})},\ \Eprint {https://arxiv.org/abs/gr-qc/0110101} {arXiv:gr-qc/0110101} \BibitemShut {NoStop}%
\bibitem [{\citenamefont {{Skrotskii}}(1957)}]{skrotskii1957influence}%
  \BibitemOpen
  \bibfield  {author} {\bibinfo {author} {\bibfnamefont {G.~V.}\ \bibnamefont {{Skrotskii}}},\ }\bibfield  {title} {\bibinfo {title} {{The Influence of Gravitation on the Propagation of Light}}\ }(\bibinfo {year} {1957})\ p.\ \bibinfo {pages} {226}\BibitemShut {NoStop}%
\bibitem [{\citenamefont {Plebanski}(1960)}]{plebanski1960electromagnetic}%
  \BibitemOpen
  \bibfield  {author} {\bibinfo {author} {\bibfnamefont {J.}~\bibnamefont {Plebanski}},\ }\bibfield  {title} {\bibinfo {title} {Electromagnetic waves in gravitational fields},\ }\href {https://doi.org/10.1103/PhysRev.118.1396} {\bibfield  {journal} {\bibinfo  {journal} {Phys. Rev.}\ }\textbf {\bibinfo {volume} {118}},\ \bibinfo {pages} {1396} (\bibinfo {year} {1960})}\BibitemShut {NoStop}%
\bibitem [{\citenamefont {de~Felice}(1971)}]{de1971gravitational}%
  \BibitemOpen
  \bibfield  {author} {\bibinfo {author} {\bibfnamefont {F.}~\bibnamefont {de~Felice}},\ }\bibfield  {title} {\bibinfo {title} {{On the Gravitational field acting as an optical medium}},\ }\href {https://doi.org/10.1007/BF00758153} {\bibfield  {journal} {\bibinfo  {journal} {Gen. Rel. Grav.}\ }\textbf {\bibinfo {volume} {2}},\ \bibinfo {pages} {347} (\bibinfo {year} {1971})}\BibitemShut {NoStop}%
\bibitem [{\citenamefont {Volkov}\ \emph {et~al.}(1971)\citenamefont {Volkov}, \citenamefont {Izmest'ev},\ and\ \citenamefont {Skrotskii}}]{volkov1971propagation}%
  \BibitemOpen
  \bibfield  {author} {\bibinfo {author} {\bibfnamefont {A.}~\bibnamefont {Volkov}}, \bibinfo {author} {\bibfnamefont {A.}~\bibnamefont {Izmest'ev}},\ and\ \bibinfo {author} {\bibfnamefont {G.}~\bibnamefont {Skrotskii}},\ }\bibfield  {title} {\bibinfo {title} {The propagation of electromagnetic waves in a riemannian space},\ }\href@noop {} {\bibfield  {journal} {\bibinfo  {journal} {Sov. Phys. JETP}\ }\textbf {\bibinfo {volume} {32}},\ \bibinfo {pages} {686} (\bibinfo {year} {1971})}\BibitemShut {NoStop}%
\bibitem [{\citenamefont {Bialynicki-Birula}\ and\ \citenamefont {Bialynicka-Birula}(2013)}]{bialynicki2013role}%
  \BibitemOpen
  \bibfield  {author} {\bibinfo {author} {\bibfnamefont {I.}~\bibnamefont {Bialynicki-Birula}}\ and\ \bibinfo {author} {\bibfnamefont {Z.}~\bibnamefont {Bialynicka-Birula}},\ }\bibfield  {title} {\bibinfo {title} {The role of the {R}iemann--{S}ilberstein vector in classical and quantum theories of electromagnetism},\ }\href {https://doi.org/10.1088/1751-8113/46/5/053001} {\bibfield  {journal} {\bibinfo  {journal} {Journal of Physics A: Mathematical and Theoretical}\ }\textbf {\bibinfo {volume} {46}},\ \bibinfo {pages} {053001} (\bibinfo {year} {2013})}\BibitemShut {NoStop}%
\bibitem [{\citenamefont {Bialynicki-Birula}\ and\ \citenamefont {Bialynicka-Birula}(2012)}]{Reply}%
  \BibitemOpen
  \bibfield  {author} {\bibinfo {author} {\bibfnamefont {I.}~\bibnamefont {Bialynicki-Birula}}\ and\ \bibinfo {author} {\bibfnamefont {Z.}~\bibnamefont {Bialynicka-Birula}},\ }\bibfield  {title} {\bibinfo {title} {Bialynicki-birula and bialynicka-birula reply:},\ }\href {https://doi.org/10.1103/PhysRevLett.109.188902} {\bibfield  {journal} {\bibinfo  {journal} {Phys. Rev. Lett.}\ }\textbf {\bibinfo {volume} {109}},\ \bibinfo {pages} {188902} (\bibinfo {year} {2012})}\BibitemShut {NoStop}%
\bibitem [{\citenamefont {Mashhoon}(1973{\natexlab{b}})}]{mashhoonEMW}%
  \BibitemOpen
  \bibfield  {author} {\bibinfo {author} {\bibfnamefont {B.}~\bibnamefont {Mashhoon}},\ }\bibfield  {title} {\bibinfo {title} {Electromagnetic waves in an expanding universe},\ }\href {https://doi.org/10.1103/PhysRevD.8.4297} {\bibfield  {journal} {\bibinfo  {journal} {Phys. Rev. D}\ }\textbf {\bibinfo {volume} {8}},\ \bibinfo {pages} {4297} (\bibinfo {year} {1973}{\natexlab{b}})}\BibitemShut {NoStop}%
\bibitem [{\citenamefont {Bialynicki-Birula}\ and\ \citenamefont {Bialynicka-Birula}(1975)}]{QED}%
  \BibitemOpen
  \bibfield  {author} {\bibinfo {author} {\bibfnamefont {I.}~\bibnamefont {Bialynicki-Birula}}\ and\ \bibinfo {author} {\bibfnamefont {Z.}~\bibnamefont {Bialynicka-Birula}},\ }\href@noop {} {\emph {\bibinfo {title} {Quantum Electrodynamics}}},\ \bibinfo {series} {International series of monographs in natural philosophy}, Vol.~\bibinfo {volume} {70}\ (\bibinfo  {publisher} {Oxford university press},\ \bibinfo {year} {1975})\BibitemShut {NoStop}%
\bibitem [{\citenamefont {Mashhoon}(2017)}]{mashhoon2017nonlocal}%
  \BibitemOpen
  \bibfield  {author} {\bibinfo {author} {\bibfnamefont {B.}~\bibnamefont {Mashhoon}},\ }\href@noop {} {\emph {\bibinfo {title} {{Nonlocal Gravity}}}},\ International Series of Monographs on Physics\ (\bibinfo  {publisher} {Oxford University Press},\ \bibinfo {year} {2017})\BibitemShut {NoStop}%
\bibitem [{\citenamefont {Bialynicki-Birula}\ and\ \citenamefont {Bialynicka-Birula}(2020)}]{bialynicki2020three}%
  \BibitemOpen
  \bibfield  {author} {\bibinfo {author} {\bibfnamefont {I.}~\bibnamefont {Bialynicki-Birula}}\ and\ \bibinfo {author} {\bibfnamefont {Z.}~\bibnamefont {Bialynicka-Birula}},\ }\bibfield  {title} {\bibinfo {title} {Three measures of fidelity for photon states},\ }\href {https://doi.org/10.1103/PhysRevA.102.042201} {\bibfield  {journal} {\bibinfo  {journal} {Phys. Rev. A}\ }\textbf {\bibinfo {volume} {102}},\ \bibinfo {pages} {042201} (\bibinfo {year} {2020})}\BibitemShut {NoStop}%
\bibitem [{\citenamefont {Earman}\ and\ \citenamefont {Glymour}(1980)}]{earman1980gravitational}%
  \BibitemOpen
  \bibfield  {author} {\bibinfo {author} {\bibfnamefont {J.}~\bibnamefont {Earman}}\ and\ \bibinfo {author} {\bibfnamefont {C.}~\bibnamefont {Glymour}},\ }\bibfield  {title} {\bibinfo {title} {The gravitational red shift as a test of general relativity: History and analysis},\ }\href {https://doi.org/https://doi.org/10.1016/0039-3681(80)90025-4} {\bibfield  {journal} {\bibinfo  {journal} {Studies in History and Philosophy of Science Part A}\ }\textbf {\bibinfo {volume} {11}},\ \bibinfo {pages} {175} (\bibinfo {year} {1980})}\BibitemShut {NoStop}%
\bibitem [{\citenamefont {Vessot}\ \emph {et~al.}(1980)\citenamefont {Vessot} \emph {et~al.}}]{vessot1980test}%
  \BibitemOpen
  \bibfield  {author} {\bibinfo {author} {\bibfnamefont {R.~F.~C.}\ \bibnamefont {Vessot}} \emph {et~al.},\ }\bibfield  {title} {\bibinfo {title} {{Test of Relativistic Gravitation with a Space-Borne Hydrogen Maser}},\ }\href {https://doi.org/10.1103/PhysRevLett.45.2081} {\bibfield  {journal} {\bibinfo  {journal} {Phys. Rev. Lett.}\ }\textbf {\bibinfo {volume} {45}},\ \bibinfo {pages} {2081} (\bibinfo {year} {1980})}\BibitemShut {NoStop}%
\bibitem [{\citenamefont {{Greenstein}}\ \emph {et~al.}(1971)\citenamefont {{Greenstein}}, \citenamefont {{Oke}},\ and\ \citenamefont {{Shipman}}}]{greenstein1971effective}%
  \BibitemOpen
  \bibfield  {author} {\bibinfo {author} {\bibfnamefont {J.~L.}\ \bibnamefont {{Greenstein}}}, \bibinfo {author} {\bibfnamefont {J.~B.}\ \bibnamefont {{Oke}}},\ and\ \bibinfo {author} {\bibfnamefont {H.~L.}\ \bibnamefont {{Shipman}}},\ }\bibfield  {title} {\bibinfo {title} {{Effective Temperature, Radius, and Gravitational Redshift of Sirius B}},\ }\href {https://doi.org/10.1086/151174} {\bibfield  {journal} {\bibinfo  {journal} {\apj}\ }\textbf {\bibinfo {volume} {169}},\ \bibinfo {pages} {563} (\bibinfo {year} {1971})}\BibitemShut {NoStop}%
\bibitem [{\citenamefont {Okun}\ \emph {et~al.}(2000)\citenamefont {Okun}, \citenamefont {Selivanov},\ and\ \citenamefont {Telegdi}}]{okun2000interpretation}%
  \BibitemOpen
  \bibfield  {author} {\bibinfo {author} {\bibfnamefont {L.~B.}\ \bibnamefont {Okun}}, \bibinfo {author} {\bibfnamefont {K.~G.}\ \bibnamefont {Selivanov}},\ and\ \bibinfo {author} {\bibfnamefont {V.~L.}\ \bibnamefont {Telegdi}},\ }\bibfield  {title} {\bibinfo {title} {{On the interpretation of the redshift in a static gravitational field}},\ }\href {https://doi.org/10.1119/1.19382} {\bibfield  {journal} {\bibinfo  {journal} {Am. J. Phys.}\ }\textbf {\bibinfo {volume} {68}},\ \bibinfo {pages} {115} (\bibinfo {year} {2000})},\ \Eprint {https://arxiv.org/abs/physics/9907017} {arXiv:physics/9907017} \BibitemShut {NoStop}%
\bibitem [{\citenamefont {Okun}(2000)}]{okun2000photons}%
  \BibitemOpen
  \bibfield  {author} {\bibinfo {author} {\bibfnamefont {L.~B.}\ \bibnamefont {Okun}},\ }\bibfield  {title} {\bibinfo {title} {{Photons and static gravity}},\ }\href {https://doi.org/10.1142/S0217732300002358} {\bibfield  {journal} {\bibinfo  {journal} {Mod. Phys. Lett. A}\ }\textbf {\bibinfo {volume} {15}},\ \bibinfo {pages} {1941} (\bibinfo {year} {2000})},\ \Eprint {https://arxiv.org/abs/hep-ph/0010120} {arXiv:hep-ph/0010120} \BibitemShut {NoStop}%
\bibitem [{\citenamefont {Mashhoon}(1988)}]{mashhoon1988complementarity}%
  \BibitemOpen
  \bibfield  {author} {\bibinfo {author} {\bibfnamefont {B.}~\bibnamefont {Mashhoon}},\ }\bibfield  {title} {\bibinfo {title} {Complementarity of absolute and relative motion},\ }\href {https://doi.org/https://doi.org/10.1016/0375-9601(88)90799-2} {\bibfield  {journal} {\bibinfo  {journal} {Physics Letters A}\ }\textbf {\bibinfo {volume} {126}},\ \bibinfo {pages} {393} (\bibinfo {year} {1988})}\BibitemShut {NoStop}%
\bibitem [{\citenamefont {Synge}(1960)}]{synge1960relativity}%
  \BibitemOpen
  \bibfield  {author} {\bibinfo {author} {\bibfnamefont {J.~L.}\ \bibnamefont {Synge}},\ }\bibfield  {title} {\bibinfo {title} {Relativity: the general theory},\ }\href@noop {} {\  (\bibinfo {year} {1960})}\BibitemShut {NoStop}%
\bibitem [{\citenamefont {Roshan}\ and\ \citenamefont {Mashhoon}(2021)}]{MashRosh}%
  \BibitemOpen
  \bibfield  {author} {\bibinfo {author} {\bibfnamefont {M.}~\bibnamefont {Roshan}}\ and\ \bibinfo {author} {\bibfnamefont {B.}~\bibnamefont {Mashhoon}},\ }\bibfield  {title} {\bibinfo {title} {Relativistic tidal accelerations in the exterior schwarzschild spacetime},\ }\href {https://doi.org/10.1103/PhysRevD.103.064081} {\bibfield  {journal} {\bibinfo  {journal} {Phys. Rev. D}\ }\textbf {\bibinfo {volume} {103}},\ \bibinfo {pages} {064081} (\bibinfo {year} {2021})}\BibitemShut {NoStop}%
\bibitem [{\citenamefont {Singleton}\ and\ \citenamefont {Wilburn}(2011)}]{Singleton}%
  \BibitemOpen
  \bibfield  {author} {\bibinfo {author} {\bibfnamefont {D.}~\bibnamefont {Singleton}}\ and\ \bibinfo {author} {\bibfnamefont {S.}~\bibnamefont {Wilburn}},\ }\bibfield  {title} {\bibinfo {title} {Hawking radiation, unruh radiation, and the equivalence principle},\ }\href {https://doi.org/10.1103/PhysRevLett.107.081102} {\bibfield  {journal} {\bibinfo  {journal} {Phys. Rev. Lett.}\ }\textbf {\bibinfo {volume} {107}},\ \bibinfo {pages} {081102} (\bibinfo {year} {2011})}\BibitemShut {NoStop}%
\bibitem [{\citenamefont {Mukhanov}\ and\ \citenamefont {Winitzki}(2007)}]{Mukhanov}%
  \BibitemOpen
  \bibfield  {author} {\bibinfo {author} {\bibfnamefont {V.}~\bibnamefont {Mukhanov}}\ and\ \bibinfo {author} {\bibfnamefont {S.}~\bibnamefont {Winitzki}},\ }\href@noop {} {\emph {\bibinfo {title} {{Introduction to quantum effects in gravity}}}}\ (\bibinfo  {publisher} {Cambridge University Press},\ \bibinfo {year} {2007})\BibitemShut {NoStop}%
\bibitem [{\citenamefont {Bialynicki-Birula}\ and\ \citenamefont {Białynicka-Birula}(2019)}]{bialynicki2019photons}%
  \BibitemOpen
  \bibfield  {author} {\bibinfo {author} {\bibfnamefont {I.}~\bibnamefont {Bialynicki-Birula}}\ and\ \bibinfo {author} {\bibfnamefont {Z.}~\bibnamefont {Białynicka-Birula}},\ }\bibfield  {title} {\bibinfo {title} {Photons---light quanta},\ }\href {https://arxiv.org/abs/1912.07008} {\bibfield  {journal} {\bibinfo  {journal} {arXiv: Quantum Physics}\ } (\bibinfo {year} {2019})}\BibitemShut {NoStop}%
\bibitem [{\citenamefont {Exirifard}\ and\ \citenamefont {Karimi}(2022)}]{exirifard2022gravitational}%
  \BibitemOpen
  \bibfield  {author} {\bibinfo {author} {\bibfnamefont {Q.}~\bibnamefont {Exirifard}}\ and\ \bibinfo {author} {\bibfnamefont {E.}~\bibnamefont {Karimi}},\ }\bibfield  {title} {\bibinfo {title} {{Gravitational distortion on photon state at the vicinity of the Earth}},\ }\href {https://doi.org/10.1103/PhysRevD.105.084016} {\bibfield  {journal} {\bibinfo  {journal} {Phys. Rev. D}\ }\textbf {\bibinfo {volume} {105}},\ \bibinfo {pages} {084016} (\bibinfo {year} {2022})},\ \Eprint {https://arxiv.org/abs/2110.13990} {arXiv:2110.13990 [gr-qc]} \BibitemShut {NoStop}%
\bibitem [{\citenamefont {Tanaka}(1983)}]{tanaka1983detect}%
  \BibitemOpen
  \bibfield  {author} {\bibinfo {author} {\bibfnamefont {K.}~\bibnamefont {Tanaka}},\ }\bibfield  {title} {\bibinfo {title} {How to detect the gravitationally induced phase shift of electromagnetic waves by optical-fiber interferometry},\ }\href {https://doi.org/10.1103/PhysRevLett.51.378} {\bibfield  {journal} {\bibinfo  {journal} {Phys. Rev. Lett.}\ }\textbf {\bibinfo {volume} {51}},\ \bibinfo {pages} {378} (\bibinfo {year} {1983})}\BibitemShut {NoStop}%
\bibitem [{\citenamefont {Brodutch}\ \emph {et~al.}(2015)\citenamefont {Brodutch}, \citenamefont {Gilchrist}, \citenamefont {Guff}, \citenamefont {Smith},\ and\ \citenamefont {Terno}}]{Post-Newtonian}%
  \BibitemOpen
  \bibfield  {author} {\bibinfo {author} {\bibfnamefont {A.}~\bibnamefont {Brodutch}}, \bibinfo {author} {\bibfnamefont {A.}~\bibnamefont {Gilchrist}}, \bibinfo {author} {\bibfnamefont {T.}~\bibnamefont {Guff}}, \bibinfo {author} {\bibfnamefont {A.~R.~H.}\ \bibnamefont {Smith}},\ and\ \bibinfo {author} {\bibfnamefont {D.~R.}\ \bibnamefont {Terno}},\ }\bibfield  {title} {\bibinfo {title} {Post-newtonian gravitational effects in optical interferometry},\ }\href {https://doi.org/10.1103/PhysRevD.91.064041} {\bibfield  {journal} {\bibinfo  {journal} {Phys. Rev. D}\ }\textbf {\bibinfo {volume} {91}},\ \bibinfo {pages} {064041} (\bibinfo {year} {2015})}\BibitemShut {NoStop}%
\bibitem [{\citenamefont {Cohen}\ and\ \citenamefont {Mashhoon}(1993)}]{cohen1993standard}%
  \BibitemOpen
  \bibfield  {author} {\bibinfo {author} {\bibfnamefont {J.~M.}\ \bibnamefont {Cohen}}\ and\ \bibinfo {author} {\bibfnamefont {B.}~\bibnamefont {Mashhoon}},\ }\bibfield  {title} {\bibinfo {title} {Standard clocks, interferometry, and gravitomagnetism},\ }\href {https://doi.org/https://doi.org/10.1016/0375-9601(93)90387-F} {\bibfield  {journal} {\bibinfo  {journal} {Physics Letters A}\ }\textbf {\bibinfo {volume} {181}},\ \bibinfo {pages} {353} (\bibinfo {year} {1993})}\BibitemShut {NoStop}%
\bibitem [{\citenamefont {Mieling}(2022)}]{Mieling}%
  \BibitemOpen
  \bibfield  {author} {\bibinfo {author} {\bibfnamefont {T.~B.}\ \bibnamefont {Mieling}},\ }\bibfield  {title} {\bibinfo {title} {Gupta-bleuler quantization of optical fibers in weak gravitational fields},\ }\href {https://doi.org/10.1103/PhysRevA.106.063511} {\bibfield  {journal} {\bibinfo  {journal} {Phys. Rev. A}\ }\textbf {\bibinfo {volume} {106}},\ \bibinfo {pages} {063511} (\bibinfo {year} {2022})}\BibitemShut {NoStop}%
\bibitem [{\citenamefont {Gerry}\ and\ \citenamefont {Knight}(2005)}]{gerry2005introductory}%
  \BibitemOpen
  \bibfield  {author} {\bibinfo {author} {\bibfnamefont {C.}~\bibnamefont {Gerry}}\ and\ \bibinfo {author} {\bibfnamefont {P.~L.}\ \bibnamefont {Knight}},\ }\href {https://doi.org/10.1017/CBO9780511791239} {\emph {\bibinfo {title} {Introductory quantum optics}}}\ (\bibinfo  {publisher} {Cambridge university press},\ \bibinfo {year} {2005})\BibitemShut {NoStop}%
\bibitem [{\citenamefont {Englert}(1996)}]{PhysRevLett.77.2154}%
  \BibitemOpen
  \bibfield  {author} {\bibinfo {author} {\bibfnamefont {B.-G.}\ \bibnamefont {Englert}},\ }\bibfield  {title} {\bibinfo {title} {Fringe visibility and which-way information: An inequality},\ }\href {https://doi.org/10.1103/PhysRevLett.77.2154} {\bibfield  {journal} {\bibinfo  {journal} {Phys. Rev. Lett.}\ }\textbf {\bibinfo {volume} {77}},\ \bibinfo {pages} {2154} (\bibinfo {year} {1996})}\BibitemShut {NoStop}%
\bibitem [{\citenamefont {Zych}\ \emph {et~al.}(2016)\citenamefont {Zych}, \citenamefont {Pikovski}, \citenamefont {Costa},\ and\ \citenamefont {Brukner}}]{Zych2016}%
  \BibitemOpen
  \bibfield  {author} {\bibinfo {author} {\bibfnamefont {M.}~\bibnamefont {Zych}}, \bibinfo {author} {\bibfnamefont {I.}~\bibnamefont {Pikovski}}, \bibinfo {author} {\bibfnamefont {F.}~\bibnamefont {Costa}},\ and\ \bibinfo {author} {\bibfnamefont {{\v{C}}.}~\bibnamefont {Brukner}},\ }\bibfield  {title} {\bibinfo {title} {General relativistic effects in quantum interference of “clocks”},\ }in\ \href {https://doi.org/10.1088/1742-6596/723/1/012044} {\emph {\bibinfo {booktitle} {Journal of Physics: Conference Series}}},\ Vol.\ \bibinfo {volume} {723}\ (\bibinfo {organization} {IOP Publishing},\ \bibinfo {year} {2016})\ p.\ \bibinfo {pages} {012044}\BibitemShut {NoStop}%
\bibitem [{\citenamefont {Colella}\ \emph {et~al.}(1975)\citenamefont {Colella}, \citenamefont {Overhauser},\ and\ \citenamefont {Werner}}]{colella1975observation}%
  \BibitemOpen
  \bibfield  {author} {\bibinfo {author} {\bibfnamefont {R.}~\bibnamefont {Colella}}, \bibinfo {author} {\bibfnamefont {A.~W.}\ \bibnamefont {Overhauser}},\ and\ \bibinfo {author} {\bibfnamefont {S.~A.}\ \bibnamefont {Werner}},\ }\bibfield  {title} {\bibinfo {title} {Observation of gravitationally induced quantum interference},\ }\href {https://doi.org/10.1103/PhysRevLett.34.1472} {\bibfield  {journal} {\bibinfo  {journal} {Phys. Rev. Lett.}\ }\textbf {\bibinfo {volume} {34}},\ \bibinfo {pages} {1472} (\bibinfo {year} {1975})}\BibitemShut {NoStop}%
\bibitem [{\citenamefont {Rauch}\ and\ \citenamefont {Werner}(2015)}]{rauch2015neutron}%
  \BibitemOpen
  \bibfield  {author} {\bibinfo {author} {\bibfnamefont {H.}~\bibnamefont {Rauch}}\ and\ \bibinfo {author} {\bibfnamefont {S.~A.}\ \bibnamefont {Werner}},\ }\href {https://doi.org/https://doi.org/10.1093/acprof:oso/9780198712510.001.0001} {\emph {\bibinfo {title} {Neutron Interferometry: Lessons in Experimental Quantum Mechanics, Wave-Particle Duality, and Entanglement}}},\ Vol.~\bibinfo {volume} {12}\ (\bibinfo  {publisher} {Oxford University Press},\ \bibinfo {year} {2015})\BibitemShut {NoStop}%
\bibitem [{\citenamefont {Bord{\'e}}\ \emph {et~al.}(2001)\citenamefont {Bord{\'e}}, \citenamefont {Houard},\ and\ \citenamefont {Karasiewicz}}]{borde2001relativistic}%
  \BibitemOpen
  \bibfield  {author} {\bibinfo {author} {\bibfnamefont {C.~J.}\ \bibnamefont {Bord{\'e}}}, \bibinfo {author} {\bibfnamefont {J.-C.}\ \bibnamefont {Houard}},\ and\ \bibinfo {author} {\bibfnamefont {A.}~\bibnamefont {Karasiewicz}},\ }\bibfield  {title} {\bibinfo {title} {Relativistic phase shifts for dirac particles interacting with weak gravitational fields in matter---wave interferometers},\ }in\ \href {https://doi.org/https://doi.org/10.1007/3-540-40988-2_21} {\emph {\bibinfo {booktitle} {Gyros, Clocks, Interferometers...: Testing Relativistic Graviy in Space}}},\ \bibinfo {editor} {edited by\ \bibinfo {editor} {\bibfnamefont {C.}~\bibnamefont {L{\"a}mmerzahl}}, \bibinfo {editor} {\bibfnamefont {C.~W.~F.}\ \bibnamefont {Everitt}},\ and\ \bibinfo {editor} {\bibfnamefont {F.~W.}\ \bibnamefont {Hehl}}}\ (\bibinfo  {publisher} {Springer Berlin Heidelberg},\ \bibinfo {address} {Berlin, Heidelberg},\ \bibinfo {year} {2001})\ pp.\ \bibinfo {pages} {403--438}\BibitemShut {NoStop}%
\bibitem [{\citenamefont {Zych}\ \emph {et~al.}(2011)\citenamefont {Zych}, \citenamefont {Costa}, \citenamefont {Pikovski},\ and\ \citenamefont {Brukner}}]{zych2011quantum}%
  \BibitemOpen
  \bibfield  {author} {\bibinfo {author} {\bibfnamefont {M.}~\bibnamefont {Zych}}, \bibinfo {author} {\bibfnamefont {F.}~\bibnamefont {Costa}}, \bibinfo {author} {\bibfnamefont {I.}~\bibnamefont {Pikovski}},\ and\ \bibinfo {author} {\bibfnamefont {C.}~\bibnamefont {Brukner}},\ }\bibfield  {title} {\bibinfo {title} {{Quantum interferometric visibility as a witness of general relativistic proper time}},\ }\href {https://doi.org/10.1038/ncomms1498} {\bibfield  {journal} {\bibinfo  {journal} {Nature Commun.}\ }\textbf {\bibinfo {volume} {2}},\ \bibinfo {pages} {505} (\bibinfo {year} {2011})},\ \Eprint {https://arxiv.org/abs/1105.4531} {arXiv:1105.4531 [quant-ph]} \BibitemShut {NoStop}%
\bibitem [{\citenamefont {Einstein}(2013)}]{einstein2003meaning}%
  \BibitemOpen
  \bibfield  {author} {\bibinfo {author} {\bibfnamefont {A.}~\bibnamefont {Einstein}},\ }\href {https://doi.org/https://doi.org/10.1007/978-94-011-6022-3} {\emph {\bibinfo {title} {The meaning of Relativity}}}\ (\bibinfo  {publisher} {Springer Dordrecht},\ \bibinfo {year} {2013})\BibitemShut {NoStop}%
\bibitem [{\citenamefont {Bassi}\ \emph {et~al.}(2017)\citenamefont {Bassi}, \citenamefont {Großardt},\ and\ \citenamefont {Ulbricht}}]{Bassi}%
  \BibitemOpen
  \bibfield  {author} {\bibinfo {author} {\bibfnamefont {A.}~\bibnamefont {Bassi}}, \bibinfo {author} {\bibfnamefont {A.}~\bibnamefont {Großardt}},\ and\ \bibinfo {author} {\bibfnamefont {H.}~\bibnamefont {Ulbricht}},\ }\bibfield  {title} {\bibinfo {title} {Gravitational decoherence},\ }\href {https://doi.org/10.1088/1361-6382/aa864f} {\bibfield  {journal} {\bibinfo  {journal} {Classical and Quantum Gravity}\ }\textbf {\bibinfo {volume} {34}},\ \bibinfo {pages} {193002} (\bibinfo {year} {2017})}\BibitemShut {NoStop}%
\bibitem [{\citenamefont {Yuri~Bonder}(2016)}]{question}%
  \BibitemOpen
  \bibfield  {author} {\bibinfo {author} {\bibfnamefont {E.~O. . D.~S.}\ \bibnamefont {Yuri~Bonder}},\ }\bibfield  {title} {\bibinfo {title} {Questioning universal decoherence due to gravitational time dilation},\ }\bibfield  {journal} {\bibinfo  {journal} {Nature Physics}\ }\textbf {\bibinfo {volume} {12}},\ \href {https://doi.org/10.1038/nphys3573} {10.1038/nphys3573} (\bibinfo {year} {2016})\BibitemShut {NoStop}%
\bibitem [{\citenamefont {Pikovski}\ \emph {et~al.}(2016)\citenamefont {Pikovski}, \citenamefont {Zych}, \citenamefont {Costa},\ and\ \citenamefont {Brukner}}]{replyto}%
  \BibitemOpen
  \bibfield  {author} {\bibinfo {author} {\bibfnamefont {I.}~\bibnamefont {Pikovski}}, \bibinfo {author} {\bibfnamefont {M.}~\bibnamefont {Zych}}, \bibinfo {author} {\bibfnamefont {F.}~\bibnamefont {Costa}},\ and\ \bibinfo {author} {\bibfnamefont {{\v{C}}.}~\bibnamefont {Brukner}},\ }\bibfield  {title} {\bibinfo {title} {Reply to 'questioning universal decoherence due to gravitational time dilation},\ }\bibfield  {journal} {\bibinfo  {journal} {Nature Physics}\ }\textbf {\bibinfo {volume} {12}},\ \href {https://doi.org/10.1038/nphys3650} {10.1038/nphys3650} (\bibinfo {year} {2016})\BibitemShut {NoStop}%
\bibitem [{\citenamefont {Pikovski}\ \emph {et~al.}(2015)\citenamefont {Pikovski}, \citenamefont {Zych}, \citenamefont {Costa},\ and\ \citenamefont {Brukner}}]{pikovski2015universal}%
  \BibitemOpen
  \bibfield  {author} {\bibinfo {author} {\bibfnamefont {I.}~\bibnamefont {Pikovski}}, \bibinfo {author} {\bibfnamefont {M.}~\bibnamefont {Zych}}, \bibinfo {author} {\bibfnamefont {F.}~\bibnamefont {Costa}},\ and\ \bibinfo {author} {\bibfnamefont {{\v{C}}.}~\bibnamefont {Brukner}},\ }\bibfield  {title} {\bibinfo {title} {Universal decoherence due to gravitational time dilation},\ }\href {https://doi.org/10.1038/nphys3366} {\bibfield  {journal} {\bibinfo  {journal} {Nature Physics}\ }\textbf {\bibinfo {volume} {11}},\ \bibinfo {pages} {668} (\bibinfo {year} {2015})}\BibitemShut {NoStop}%
\bibitem [{\citenamefont {Smith}\ and\ \citenamefont {Raymer}(2006)}]{Smith}%
  \BibitemOpen
  \bibfield  {author} {\bibinfo {author} {\bibfnamefont {B.~J.}\ \bibnamefont {Smith}}\ and\ \bibinfo {author} {\bibfnamefont {M.~G.}\ \bibnamefont {Raymer}},\ }\bibfield  {title} {\bibinfo {title} {Two-photon wave mechanics},\ }\href {https://doi.org/10.1103/PhysRevA.74.062104} {\bibfield  {journal} {\bibinfo  {journal} {Phys. Rev. A}\ }\textbf {\bibinfo {volume} {74}},\ \bibinfo {pages} {062104} (\bibinfo {year} {2006})}\BibitemShut {NoStop}%
\bibitem [{\citenamefont {Smith}\ and\ \citenamefont {Raymer}(2007)}]{Smith_2007}%
  \BibitemOpen
  \bibfield  {author} {\bibinfo {author} {\bibfnamefont {B.~J.}\ \bibnamefont {Smith}}\ and\ \bibinfo {author} {\bibfnamefont {M.~G.}\ \bibnamefont {Raymer}},\ }\bibfield  {title} {\bibinfo {title} {Photon wave functions, wave-packet quantization of light, and coherence theory},\ }\href {https://doi.org/10.1088/1367-2630/9/11/414} {\bibfield  {journal} {\bibinfo  {journal} {New Journal of Physics}\ }\textbf {\bibinfo {volume} {9}},\ \bibinfo {pages} {414} (\bibinfo {year} {2007})}\BibitemShut {NoStop}%
\bibitem [{\citenamefont {Sakurai}\ and\ \citenamefont {Napolitano}(2020)}]{Sakurai}%
  \BibitemOpen
  \bibfield  {author} {\bibinfo {author} {\bibfnamefont {J.~J.}\ \bibnamefont {Sakurai}}\ and\ \bibinfo {author} {\bibfnamefont {J.}~\bibnamefont {Napolitano}},\ }\href {https://doi.org/10.1017/9781108587280} {\emph {\bibinfo {title} {{Modern Quantum Mechanics}}}},\ Quantum physics, quantum information and quantum computation\ (\bibinfo  {publisher} {Cambridge University Press},\ \bibinfo {year} {2020})\BibitemShut {NoStop}%
\bibitem [{\citenamefont {Roy}\ and\ \citenamefont {Sen}(2019)}]{roy2019study}%
  \BibitemOpen
  \bibfield  {author} {\bibinfo {author} {\bibfnamefont {S.}~\bibnamefont {Roy}}\ and\ \bibinfo {author} {\bibfnamefont {A.~K.}\ \bibnamefont {Sen}},\ }\bibfield  {title} {\bibinfo {title} {Study of gravitational deflection of light ray}\ }(\bibinfo  {publisher} {IOP Publishing},\ \bibinfo {year} {2019})\ p.\ \bibinfo {pages} {012002}\BibitemShut {NoStop}%
\bibitem [{\citenamefont {Lebed}(2022)}]{lebed2022breakdown}%
  \BibitemOpen
  \bibfield  {author} {\bibinfo {author} {\bibfnamefont {A.~G.}\ \bibnamefont {Lebed}},\ }\href@noop {} {\emph {\bibinfo {title} {Breakdown Of Einstein's Equivalence Principle}}}\ (\bibinfo  {publisher} {World Scientific},\ \bibinfo {year} {2022})\BibitemShut {NoStop}%
\bibitem [{\citenamefont {Brandt}\ and\ \citenamefont {Seidel}(1996)}]{IsoKerr}%
  \BibitemOpen
  \bibfield  {author} {\bibinfo {author} {\bibfnamefont {S.~R.}\ \bibnamefont {Brandt}}\ and\ \bibinfo {author} {\bibfnamefont {E.}~\bibnamefont {Seidel}},\ }\bibfield  {title} {\bibinfo {title} {Evolution of distorted rotating black holes. iii. initial data},\ }\href {https://doi.org/10.1103/PhysRevD.54.1403} {\bibfield  {journal} {\bibinfo  {journal} {Phys. Rev. D}\ }\textbf {\bibinfo {volume} {54}},\ \bibinfo {pages} {1403} (\bibinfo {year} {1996})}\BibitemShut {NoStop}%
\end{thebibliography}
\end{document}